\documentclass[amsmath,amssymb,
showpacs,twocolumn,
superscriptaddress,
prl]{revtex4-1}
\usepackage{graphicx,bm,color,subfigure}
\usepackage[T1]{fontenc}
\usepackage{mathrsfs}
\usepackage{bm}
\usepackage{amsmath}
\usepackage{amssymb}
\usepackage{graphicx}
\usepackage{esint}
\usepackage{multirow}
\usepackage{float}
\usepackage{array}
\usepackage{makecell}
\usepackage{harpoon}
\usepackage{booktabs}
\usepackage{gensymb}
\usepackage{simplewick}
\usepackage{subfigure}
\usepackage{soul}
\usepackage{makecell}


\begin{document}

\title{Critical Density-Wave Vestigial Phases of Commensurate Pair Density Wave}

\author{Chu-Tian Gao}
\affiliation{School of Physics, Beijing Institute of Technology, Beijing 100081, China}
\author{Jing Zhou}
\email{zhoujing@cqupt.edu.cn}
\affiliation{School of Electronic Science and Engineering, Chongqing University of Posts and Telecommunications, Chongqing 400065, China}
\author{Yu-Bo Liu}
\affiliation{Institute of Theoretical Physics, Chinese Academic of Science, Beijing 100080, China}
\author{Fan Yang}
\email{yangfan_blg@bit.edu.cn}
\affiliation{School of Physics, Beijing Institute of Technology, Beijing 100081, China}

\begin{abstract}
The pair-density-wave (PDW) is an exotic pairing state hosting a spatially modulated pairing order parameter, which has attracted great interest. Due to its simultaneously breaking U(1)-gauge and translational symmetries, intriguing vestigial phases which restore only one broken symmetry can emerge at an intermediate temperature regime. Previously, investigations on the vestigial phases of PDW were mainly focused on incommensurate PDW. However, the experimentally observed PDW is usually commensurate, whose vestigial phases have not been systematically investigated. Here we study the vestigial phases of 2D commensurate PDW with $n$-times expanded unit vectors, hosting different numbers of wave vectors. Based on the Ginzburg-Landau theory, we get the low energy effective model Hamiltonian. Subsequent renormalization group (RG) and Monte-Carlo (MC) studies are conducted to obtain the phase diagram and spatial dependent correlation functions. Our RG and MC calculations consistently yield the following result. For $n\le 4$, besides the charge-4e/2e superconductivity, there exists the translational symmetry broken charge-density-wave (CDW) vetigial phase. Intriguingly, for $n\ge 5$, the restore of the translational symmetry with increasing temperature is realized through two successive Berezinskii-Kosterlitz-Thouless transitions. Such a two-step process leads into two critical vestigial phases, i.e. the critical-PDW and the critical-CDW phases, in which the discrete translational symmetry is quasily broken, leading into a power-law decaying density-density correlation even at 2D. Our work appeals for experimental verifications.
\end{abstract}

\maketitle

{\bf Introduction:} The pair density wave (PDW) is an unconventional superconducting state in which Cooper pairs carry a non-zero center-of-mass momentum, characterized by a spatially modulated pairing order~\cite{FF1964,LO1965,Radzihovsky2009,Agterberg,Zhai2010,Cho2012,Lee2014,Maciejko2014,Jian2015,Jian2017,Jian2020,Han2020,Agterberg2020,Berg_2009,Wang,Wang1,Jin2022,Yao2025,Yang2025}. Since the PDW is irrelevant in the weak-coupling renormalization group (RG) analysis~\cite{Shankar1994}, its emergence suggests strong correlations. It has been identified in various correlated electron materials, exampled by Cuprate superconductors~\cite{Bi_PDW,Ruan2018,Du2020,Li2021,Bi_PDW1}, kagome lattice superconductors~\cite{CsV3Sb5_PDW,han2024}, transition-metal dichalcogenide~\cite{S_PDW}, Iron-based superconducting materials~\cite{Zhao2023, Liu2023}, Uranium-based heavy-fermion superconducting compounds~\cite{Gu2023} and may also exist in cold atomic systems~\cite{Liao2010,Vitali2022}. These research holds significant importance for understanding superconductivity (SC) in correlated electron systems.

The PDW state exhibits spontaneously breaking both translational and $U(1)$-gauge symmetry. As these two symmetries are generally not restored simultaneously, widespread interest has been ignited in characterizing the resultant vestigial phases~\cite{Agterberg,Berg2009,Agterberg1,You2012,Babaev2004,WHKo,Herland,FFSong,PLi,LFZhang2024,SZhou,Rampp,YYu,Curtis,Poduval,Zeng2024,Jian2021,Liang_Fu,MHecker,Wuyiming2023,Song2022,Varma2023}.  For 2D incommensurate PDW, the ground state symmetry breaking is $U(1)\times U(1)$~\cite{Berg2009} or $U(1)\times U(1) \times U(1)$~\cite{Agterberg, Agterberg1}, leading into two or three independent low-energy fluctuating phase modes. These low-energy phase modes can be rearranged into the total phase and the relative phase(s). When the relative phase(s) is(are) locked and the total phase is disordered, it implies that the $U(1)$ symmetry is restored while the translation symmetry is broken, leading to the charge density wave (CDW) phase. In contrast, when the total phase is locked and the relative phase(s) is (are) disordered, the translation symmetry is restored, whereas the $U(1)$ symmetry is broken. This results in the emergence of charge-2e/4e/6e SC. The higher charge superconductivity with fractional flux quantization have been reported~\cite{Ge2024,Pan2024,Lin2025,Song2025}. 

While in real materials, the incommensurate PDW can be viewed as an approximation as a commensurate one with long period, most of the presently identified PDW states are commensurate one with short period. For example, in copper-based materials, the PDW period is 4$a_0$~\cite{Bi_PDW} and 8$a_0$~\cite{Du2020,Bi_PDW1}, while in $CsV_3Sb_5$, the PDW period is $2a_0\times 2a_0$~\cite{CsV3Sb5_PDW,han2024,Deng2024}. The vestigial phases of commensurate PDW has not been systematically investigated yet, which will be the focus of our work. In particular, we shall address the following questions. What is the phase diagram for commensurate PDW? Are these phase diagrams depend on the period of the PDW? Are there any exotic phases in the phase diagrams which are unconventional for 2D electronic system?   

In this paper, we explore the vestigial phases of commensurate PDW with periodicity $na_0\times na_0$. Based on the Ginzburg-Landau (GL) theory, we derive the low-energy effective Hamiltonian which describes the pairing phase fluctuations of the commensurate PDW state. In particular, the relative phase fluctuates between its $n$ saddle points, which is described by a $n$-state clock model~\cite{Jose1977,Tobochnik1982,Challa1986,Surungan2019,Ziqian_Li2020,Hao_Chen2020,Miyajima2021,Yu_Bo_Liu2023,Yu_Bo_Liu2024}. We determine the phase diagrams and phase transitions by combined RG and Monte-Carlo (MC) studies, which consistently yield the following results. For $n\le 4$, the previously known charge-2e/4e/6e SC or long-range CDW state can be the intermediate vestigial phase between the low-temperature PDW ground state and the high-temperature normal metal (MT) state. For $n\ge 5$, two additional critical vestigial phases emerge, i.e. the critical-PDW (C-PDW) and the critical-CDW (C-CDW), characterized by power-law decaying CDW correlation. Remarkably, the discrete translational symmetry is quasily broken even in 2D in these intriguing critical phases. More over, due to the presence of these critical phases, all the phase transitions in the phase diagram are Berezinskii-Kosterlitz-Thouless (BKT) transitions. This $n$-dependent result is related to the properties of the $n$-state clock model. Our results are highly anticipated for experimental verifications.

\begin{figure}[h]
	\centering
	\includegraphics[width=0.45\textwidth]{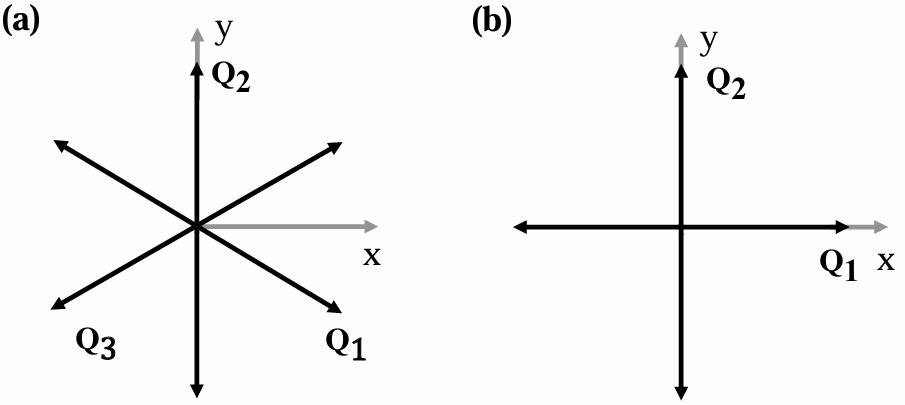}
	\caption{Directions of $\mathbf{Q}_{\alpha}$, with (a) for $3Q$ PDW, (b) for $2Q$ PDW, respectively.} \label{pdw}
\end{figure}

{\bf Model:} Consider the PDW order parameter with multiple symmetry-related degenerate wave vectors $\{\mathbf{Q}_{\alpha}\}$:
\begin{equation}\label{gap_function}
\Delta(\mathbf{r})=\sum_{\alpha}(\Delta_{\mathbf{Q}_{\alpha}}e^{i\mathbf{Q}_{\alpha}\cdot \mathbf{r}}+\Delta_{-\mathbf{Q}_{\alpha}}e^{-i\mathbf{Q}_{\alpha}\cdot\mathbf{r}}).
\end{equation}
As shown in Fig.~\ref{pdw}, the hexagonal $3Q$ PDW host three pairs of wave vectors $\pm \mathbf{Q_1},\pm \mathbf{Q_2},\pm \mathbf{Q_3}$ satisfying $\mathbf{Q}_{1}+\mathbf{Q}_{2}+\mathbf{Q}_{3}=0$, and the $2Q$ PDW host two pairs of wave vectors $\pm \mathbf{Q_1},\pm \mathbf{Q_2}$. Let us start from the ground states $\{ \Delta_{\mathbf{Q}_{1}}$, $\Delta_{\mathbf{Q}_{2}}$, $\Delta_{\mathbf{Q}_{3}}$, $\Delta_{-\mathbf{Q}_{1}}$, $\Delta_{-\mathbf{Q}_{2}}$, $\Delta_{-\mathbf{Q}_{3}}\}=\Delta_{0}e^{i\theta}\{e^{i\phi_1}$, $e^{i\phi_2}$, $e^{i\phi_3}$, $e^{-i\phi_1}$, $e^{-i\phi_2}$, $e^{-i\phi_3}\}$$\equiv$$\Delta_{0}\{e^{i\theta_1}$, $e^{i\theta_2}$, $e^{i\theta_3}$, $e^{i\theta^{'}_1}$, $e^{i\theta^{'}_2}$, $e^{i\theta^{'}_3}\}$~\cite{Agterberg1} and $\{ \Delta_{\mathbf{Q}_{1}}$, $\Delta_{\mathbf{Q}_{2}}$, $\Delta_{-\mathbf{Q}_{1}}$, $\Delta_{-\mathbf{Q}_{2}}\}$=$\Delta_{0}e^{i\theta}\{e^{i\phi_1}$, $ e^{i\phi_2}$, $e^{-i\phi_1}, e^{-i\phi_2}\}$$\equiv$$\Delta_{0}\{e^{i\theta_1}$, $ e^{i\theta_2}$, $e^{i\theta^{'}_1}, e^{i\theta^{'}_2}\}$ ($\Delta_{0}>0$)~\cite{Agterberg} for the two PDW states, respectively. Here $\theta$ and $\phi_\alpha$ are free parameters, and we have $\theta_\alpha=\theta+\phi_\alpha$, and $\theta^\prime_\alpha=\theta-\phi_\alpha$. Note that in the $3Q$ PDW state, the relative phases additionally satisfy $\phi_1 + \phi_2 + \phi_3 = 0$~\cite{Agterberg1}.

At finite temperature, the thermal fluctuations lead to low-energy phase fluctuations of the PDW order parameter, so that these free parameters become phase fields:  $\Delta_{\mathbf{Q}_{\alpha}}=\Delta_{0}e^{i \theta_{\alpha}(\mathbf{r})}$, $\Delta_{-\mathbf{Q}_{\alpha}}=\Delta_{0}e^{i \theta^{'}_{\alpha}(\mathbf{r})}$, with $\theta_{\alpha}(\mathbf{r}) =\theta(\mathbf{r}) + \phi_{\alpha}(\mathbf{r})$ and $\theta^{'}_{\alpha}(\mathbf{r}) =\theta(\mathbf{r}) - \phi_{\alpha}(\mathbf{r})$. The physical $\theta_{\alpha}(\mathbf{r})$ and $\theta^{'}_{\alpha}(\mathbf{r})$ phase fields should host only integer vortices to ensure the single-valuedness of the superconducting order parameters. This restricts  the allowed vortices in the $\theta$ and $\phi_\alpha$ fields for different PDW states. For the $3Q$ PDW groundstate, only integer vortex and dislocation charge is allowed because $\phi_1 + \phi_2 + \phi_3 = 0$~\cite{Agterberg1}. For the $2Q$ PDW case, since $\phi_1$ and $\phi_2$ are independent, $\theta$ and $\phi_{1,2}$ can simultaneously host integer or half-integer vortices~\cite{Agterberg}, respectively. This is the ``kinematics constraint'' between the $\theta$ and $\phi_\alpha$ fields~\cite{Yu_Bo_Liu2023,Yu_Bo_Liu2024}. 

Let us derive the low-energy effective Hamiltonian for the phase fluctuations. Firstly, the spatial variation of the phase fields $\theta$ and $\phi_{\alpha}$ costs the following energy $H_0$ which depends on the gradient of the fields, 
\begin{eqnarray}\label{Hamiltonian_r}
H_{0}&=&\int d^{2}\mathbf{r}\Big( \frac{\rho}{2}|\nabla\theta|^{2}+\frac{\mu}{2}\sum_{\alpha}|\nabla \phi_\alpha|^{2}\Big).
\end{eqnarray}
Here $\rho$ and $\mu$ are stiffness parameters for the total and relative phases, respectively. For the incommensurate PDW, Eq. (\ref{Hamiltonian_r}) is already the full Hamiltonian~\cite{Agterberg,Agterberg1}. However, for the commensurate PDW, there can be the following additional anisotropic term in the Hamiltonian. For a 2D commensurate PDW with the periodicity $na_{0}$$\times na_{0}$, the wave vector $\mathbf{Q}_{\alpha}$ can be $\mathbf{Q}_{\alpha}$=$\frac{m_{1\alpha}}{n}\mathbf{b}_{1}$ +$\frac{m_{2\alpha}}{n}\mathbf{b}_{2}$, where $\mathbf{a}_{1,2}$ are the unit cell vectors, $\mathbf{b}_{1,2}$ are the reciprocal lattice vectors, $ m_{i\alpha}$ $(i=1,2)$ and $n$ are coprime integers. Under the translation $\mathbf{a}_{i}$, we have
\begin{equation}\label{translation operation}
\Delta_{\pm \mathbf{Q}_{\alpha}}(\mathbf{r})\rightarrow\Delta_{\pm \mathbf{Q}_{\alpha}}(\mathbf{r}-\mathbf{a}_{i})e^{\mp i\mathbf{Q}_{\alpha}\cdot \mathbf{a}_{i}},
\end{equation}
where $\mathbf{Q}_{\alpha} \cdot \mathbf{a}_{i}=2\pi \frac{m_{i\alpha}}{n}$. The invariance of the free energy under this translation and the $U(1)$-gauge operation allows for the following anisotropic term:
\begin{eqnarray}\label{anisotropic}
F_{ani}&=&A_0\sum_{\alpha}(\Delta_{\mathbf{Q}_{\alpha}}^{n\ast}(\mathbf{r})\Delta_{-\mathbf{Q}_{\alpha}}^{n}(\mathbf{r})+c.c)  \nonumber\\
&=& A\sum_{\alpha}\cos{(2n\phi_{\alpha})}.
\end{eqnarray}
Here $A$ is proportional to $|\Delta_{0}|^{2n}$. Finally, the total low-energy effective Hamiltonian is:
\begin{eqnarray}\label{Hamiltonian_r}
H=H_0+A\int d^{2}\mathbf{r}\sum_{\alpha}\cos(2n\phi_{\alpha}).
\end{eqnarray}

Eq.~(\ref{Hamiltonian_r}) shows that while the $\theta$ field follows a conventional XY model that experiences a BKT transition when $T$ increases, the $\phi_{\alpha}$ field behaves like a XY model with $q$-fold ($q=n$) anisotropy, resembling the symmetry of the $q$-state clock model. It should be noted that the states described by $(\theta(\mathbf{r}),\phi_{\alpha}(\mathbf{r})+\pi)$ and $(\theta(\mathbf{r}),\phi_{\alpha}(\mathbf{r}))$ are gauge equivalent, as their corresponding physical configurations $\theta_{\alpha}\left(\mathbf{r}\right)(\theta{'}_{\alpha}\left(\mathbf{r}\right))$ differ only by a global constant $\pi$~\cite{Yu_Bo_Liu2023,Yu_Bo_Liu2024}. Consequently, while Eq.~(\ref{Hamiltonian_r}) appears $2n$ saddle points for each $\phi_{\alpha}$ field, these actually correspond to just $n$ physically distinct states, resulting in $n$-fold anisotropy.

{\bf RG study:} We use the RG to study the Hamiltonian Eq.~(\ref{Hamiltonian_r}). In particular, for the 2D PDW state, only two of the $\phi_\alpha$ fields are independent. We introduce two phonon fields $\mathbf{u} = (u_x, u_y)$ \cite{Agterberg1} to simultaneously describe the $2Q$ and $3Q$ PDW states through the relation $\phi_\alpha = \mathbf{Q}_{\alpha} \cdot \mathbf{u}$, as for the $3Q$ state $\sum_\alpha\mathbf{Q}_{\alpha}=0\to \sum_\alpha\phi_\alpha =0$. Then the Hamiltonian Eq.~(\ref{Hamiltonian_r}) can be rewritten as: 
\begin{eqnarray}
H_{eff}&=&\int d^{2}r\Big( \frac{\rho}{2}|\nabla\theta|^{2}+\frac{\mu}{2}(\frac{2\pi}{a})^{2}|\nabla u_{x}|^{2} \nonumber \\
&&+\frac{\mu}{2}(\frac{2\pi}{a})^{2}|\nabla u_{y}|^{2}+A\sum_{\alpha}\cos(2n\phi_{\alpha})\Big).
\end{eqnarray}
As shown in Table~\ref{tabcharge}, we have the vortex and dislocation charges of the topological excitations of the $3Q$ PDW and $2Q$ PDW phases by $\{ \theta,u_{x},u_{y} \}$.
The corresponding action function of the multiple components Sine-Gordon model in dual space is:
\begin{eqnarray}\label{eqn:action-SG}
S_{D}&=&\int d^{2}x_{D}\Big( \frac{T}{2\rho}|\nabla\widetilde{\theta}|^{2}+\frac{T}{2\mu}|\nabla\widetilde{u}_{x}|^{2}+\frac{T}{2\mu}|\nabla\widetilde{u}_{y}|^{2} \nonumber\\
&+&g_{\theta}\cos2\pi\widetilde{\theta}+g_{x}\cos(\frac{2\pi a_{1x}\widetilde{u}_{x}}{a})\cos(\frac{2\pi a_{1y}\widetilde{u}_{y}}{a}) \nonumber\\
&+&g_{y}\cos(\frac{2\pi a_{2x}\widetilde{u}_{x}}{a})\cos(\frac{2\pi a_{2y}\widetilde{u}_{y}}{a})+\sum_{\alpha}g_{\alpha}\cos(2n\phi_{\alpha})
\nonumber\\
&+&g_{\frac{1}{2}}^{x}\cos\pi\theta\cos\pi \widetilde{u}_{x}+g_{\frac{1}{2}}^{y}\cos\pi\theta\cos\pi \widetilde{u}_{y}
\Big),
\end{eqnarray}
where we have replaced $\frac{2\pi}{a}\widetilde{u}_{i}$ by $\widetilde{u}_{i}$ ($i=x,y$).  The Wigner-Seitz vector $\mathbf{a_{1}}=(a,0)$ and $\mathbf{a_{2}}=(\frac{a}{2},\frac{\sqrt{3}}{2}a)$ for $3Q$ PDW, $\mathbf{a}_{1}=(a,0)$, and $\mathbf{a}_{2}=(0,a)$ for $2Q$ PDW.The dual bosonic field $\widetilde{\theta}$ and $\widetilde{u}_{x},\widetilde{u}_{y}$ describe the vortice fields of $\theta$ and $u_{x},u_{y}$. $g_{\theta}$ and $g_{x},g_{y}$ are proportional to the fugacities parameters of different integer vortex excitations. $g_{\frac{1}{2}}^{x}$, $g_{\frac{1}{2}}^{y}$ describe coupling parameters of half-half vortices. In the 3Q PDW state, only integer vortices are proliferated ($g_{\frac{1}{2}}^{x/y}=0$).


\begin{table}
\begin{tabular}{|c|c|c|}
  \hline
  Phase & Vortex Charge & Dislocation Charge\\
  \hline
  $3Q$ PDW &$\frac{1}{2\pi}\oint d\theta =n$ & $\oint d{\bf u}=l_1\mathbf{a_1}$$+l_2\mathbf{a_2}$\\
  $2Q$ PDW & $\frac{1}{2\pi}\oint d\theta =\frac{1}{2}(n_\alpha+n_{\alpha^{'}})$ & $\oint du_{x/y}= \frac{a}{2}(n_\alpha-n_{\alpha^{'}})$\\
  \hline
\end{tabular}
\caption{The vortex and dislocation charges of the topological excitations of the $3Q$ PDW and $2Q$ PDW phases ($n_\alpha$, $n_{\alpha^{'}}$ and $l_i$ are integers). } \label{tabcharge}
\end{table}

In Table~\ref{tab:1}, we present fixed points of the tree level RG flow equations and corresponding phases in the vestigial phases of the commensurate PDW state. To simplify the expression, the half vortices coupling parameter $g_{\frac{1}{2}}^{x/y}$ is set as zero for $3Q$ PDW state. If $g_{\theta}$ or $g_{\frac{1}{2}}^{x/y}$ is relevant, suggesting the integer type votices or the half vortices are proliferated in the $\theta$ field, the superconductivity is killed. If $g_{x/y}$ or $g_{\frac{1}{2}}^{x/y}$ is relevant, suggesting the integer vortices or the half votices are proliferated in the $u_{x/y}$ field, the CDW is killed, and the translation symmetry is recovered. When $g_{x/y}$ and $g_{\frac{1}{2}}^{x/y}$ are irrelevant, if $g_{\alpha}$ is relevant, the translation symmetry is broken; if $g_{\alpha}$ are irrelevant, the system enters critical phase.

\begin{table}[!h]
\centering
\caption{Fixed points of the coupling parameters under RG, and the corresponding phases for the commensurate PDW. The abbreviations denote: 2e/4e SC is charge 2e SC and charge 4e SC; MT is normal metal; PDW is pair density wave; CDW is charge density wave;  C-PDW is critical pair density wave;  C-CDW is critical charge density wave.}\label{tab:1}
\begin{tabular}{|c|c|c|c|c|}
  \hline\hline
  $g_{\theta}$ & $g_{x/y}$ & $g_{\frac{1}{2}}^{x/y}$ & $g_{\alpha}$ & phase \\
  \hline
  $\infty$ & $\infty$ & 0 & 0 &  MT \\
  \hline
  $\infty$ & 0 & $\infty$ & 0 &  MT \\
  \hline
  $\infty$ & 0 & 0 & $\infty$ &  CDW \\
  \hline
  $\infty$ & 0 & 0 & 0 & C-CDW \\
  \hline
  0 & 0 & 0 & $\infty$ & PDW \\
  \hline
  0 & 0 & 0 & 0 & C-PDW \\
  \hline
  0 & $\infty$ & 0 & 0 & charge-2e/4e SC \\
  \hline\hline
\end{tabular}
\end{table}

\begin{figure}[h]
	\centering
	\includegraphics[width=0.45\textwidth]{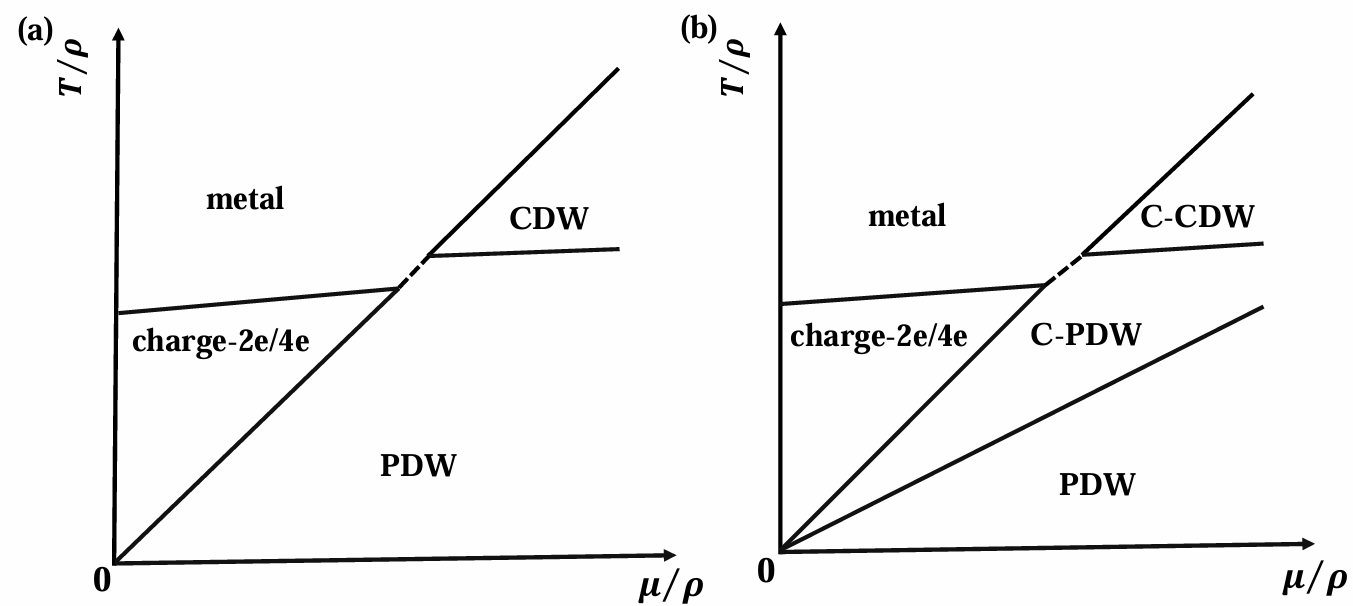}
	\caption{Schematic phase diagrams of the commensurate PDW with (a) for $2 \le n \le 4$ and (b) for $n\ge5$. $\rho$, $\mu$ and $T$ denote the superfluid stiffness, the CDW elastic constant and temperature, respectively. The black lines mark the phase boundaries. The dashed line represents a point for the 3Q PDW, while it represents for a solid line for the 2Q PDW.} \label{schematicphasediagram}
\end{figure}

The phase diagram for $2\le n\le 4$ is shown in Fig.~\ref{schematicphasediagram}(a). At low enough $T$, the fugacity parameter $g_{\alpha}$ is relevant while $g_{\theta},g_{x/y},g_{\frac{1}{2}}^{x/y}$ are all irrelevant, suggesting $\theta$ and $u_{x/y}$ are ordered and no vortices are proliferated, leading to the PDW phase. Enhancing $T$, when $\mu/\rho$ is small, translation symmetry is restored by proliferating integer vortices in dislocation while superconductivity still exists, leading to charge-$2e$ ($3Q$ PDW) or charge-$4e$ ($2Q$ PDW) phase; when $\mu/\rho$ is large, $g_{\theta}$ becomes relevant by proliferating vortices in $\theta$ field while translation symmetry is still broken, leading to the CDW phase. For large enough $T$, the fugacity parameters $g_{\theta}$ and $g_{x/y}$ (or $g_{\frac{1}{2}}^{x/y}$) become all relevant by proliferating vortices in $\theta$ and $u_{x/y}$ field, resulting into the normal MT phase. For intermediate regime of $\mu/\rho$, there are two triple points for $2Q$ PDW case in the phase diagram, and the dashed line indicates a solid line which divides PDW phase and normal MT phase, suggesting the direct phase transition and all the orders are killed by half-half vortices proliferated ($g_{\frac{1}{2}}^{x/y}$ is relevant). For $3Q$ PDW case, only integer vortices are allowed to exist in $\theta$ or $u_{x/y}$ channel, so the dashed line decays into a quadruple point.

The phase diagram for $n\ge5$ is shown in Fig.~\ref{schematicphasediagram}(b), wherein two new phases emerge: the C-PDW and the C-CDW. For these critical phases, $g_{x/y}$, $g_{\frac{1}{2}}^{x/y}$ and $g_{\alpha}$ are all irrelevant, suggesting that the translation symmetry is quasi broken. At low enough $T$, the system is PDW. When $T$ increases, the systems will first enter the C-PDW phase in which all fugacity parameters are irrelevant and then enter different phases depending on the value of $\mu/\rho$. If $\mu/\rho$ is small, increasing $T$ further, $g_{x/y}$ becomes relevant, suggesting the charge-2e/4e phase. If $\mu/\rho$ is large, increasing $T$ further, the system will enter the C-CDW phase wherein integer vortices proliferate in the $\theta$ field. For high enough $T$, the system enters the normal MT phase. There are also two triple points connected by a dashed line. For $2Q$ PDW case, the dashed line indicates a solid line separating the C-PDW and the normal MT phase. For $3Q$ PDW case, the dashed line decays into a quadruple point.  

The difference between the phase diagrams for $n\le4$ and $n\ge5$ can be understood from the $n-$ state clock model~\cite{Jose1977} describing the $u_{x,y}$ fields. For this model, for $n\le4$, the system will experience a second-order phase transition from the low-$T$ ordered phase to high-$T$ disordered phase; for $n\ge5$, the system will experience two subsequent BKT transitions first from the low-$T$ ordered phase to the intermediate-$T$ critical phase and then to the high-$T$ disordered phase. The intermediate-$T$ critical phase for $n\ge5$ just correspond to the C-PDW and the C-CDW phases obtained here.

{\bf MC study:} To perform the MC study, we discretize the continuous Hamiltonian (\ref{Hamiltonian_r}) on the square lattice as,
\begin{eqnarray}\label{Hamiltonian_d32}
H &=& -\kappa\sum_{\langle ij\rangle} \cos(m\theta(\mathbf{r}_{i})-m\theta(\mathbf{r}_{j}))  \nonumber\\
&&- \lambda \sum_{\langle ij\rangle, \alpha}  \cos(m\phi_\alpha(\mathbf{r}_{i})-m\phi_\alpha(\mathbf{r}_{j}))  \nonumber\\
&&- \gamma \sum_{\langle ij\rangle, \alpha} [\cos(\theta(\mathbf{r}_{i})+\phi_{\alpha}(\mathbf{r}_{i})-\theta(\mathbf{r}_{j})-\phi_{\alpha}(\mathbf{r}_{j})) \nonumber\\
&&+\cos(\theta(\mathbf{r}_{i})-\phi_{\alpha}(\mathbf{r}_{i})-\theta(\mathbf{r}_{j})+\phi_{\alpha}(\mathbf{r}_{j}))] \nonumber\\
&&+ A\sum_{i, \alpha} \cos(2n\phi_{\alpha}(\mathbf{r}_{i})).
\end{eqnarray}
Here $\langle ij\rangle$ represents nearest-neighbor bonding and the coefficients $\kappa$, $\lambda$ and $\gamma$ satisfy
\begin{eqnarray}\label{relation}
\kappa =\frac{\rho-4\gamma}{m^2},~~~~~~~~~~~ \lambda = \frac{\mu-2\gamma}{m^2},
\end{eqnarray}
with $m=1, \gamma=0$ for the $3Q$ PDW, and $m=2, \gamma>0$ for the $2Q$ PDW. The positive coefficients $\kappa$, $\lambda$ and $\gamma$ ensure the discretized Hamiltonian~ (\ref{Hamiltonian_d32}) match the continuous Hamiltonian~(\ref{Hamiltonian_r}) in the thermodynamic limit, respectively. Note that we have set $m=2$ and finite $\gamma>0$ for the case of $2Q$ PDW due to the following reason. Firstly, $m=2$ energetically allows for integer and half-integer vortices of the $\theta$ and $\phi_{\alpha}$ fields, as required by Table~\ref{tabcharge}. Secondly, the $\gamma$ term with $\gamma>0$ energetically only allows for integer $\theta_{\alpha}$ and $\theta^\prime_{\alpha}$ vortices, imposing the kinematic constraint between the $\theta$ and $\phi_{\alpha}$ fields: they can either both host integer vortices or both host half-integer vortices, as implied by Table~\ref{tabcharge}. This  ensures the correct topology of the phase diagram for the $2Q$ PDW~\cite{Yu_Bo_Liu2023,Yu_Bo_Liu2024}. In the following MC studies, we set $\gamma=\frac{1}{4}\rho\mu/(\rho+\mu)$. 


\begin{table}[!h]
	\centering
	\caption{The correlation functions $\eta_\theta$ and $\eta_{\phi_{\alpha}}$ decay for all possible phases in Fig.~\ref{schematicphasediagram}.}\label{tab:3}
	\begin{tabular}{|c|c|c|c|c|c|c|}
		\hline\hline
		Phase & $\eta_\theta$ & $\eta_{\phi_{\alpha}}$ \\
		\hline
		2e/4e SC & ~~$r^{-\sigma}$~~ & ~~$e^{-r/\xi}$~~ \\
		\hline
		MT & ~~$e^{-r/\xi_1}$~~ & ~~$e^{-r/\xi_2}$~~ \\
		\hline
		PDW & ~~$r^{-\sigma}$ & ~~$ const $~~ \\
		\hline
		CDW & ~~$e^{-r/\xi}$~~ & ~~$const$~~ \\
            \hline
		C-PDW & ~~$r^{-\sigma_{1}}$~~ & ~~$r^{-\sigma_{2}}$~~ \\
		\hline
		C-CDW & ~~$e^{-r/\xi}$~~ & ~~$r^{-\sigma}$~~ \\
		\hline\hline
	\end{tabular}
\end{table}

The nature of all phases can be characterized by the correlation function $\eta_{\theta/\phi_{\alpha}}$ for the $\theta$ and $\phi_{\alpha}$ fields. The dependence of these functions on $\Delta r$ ($\equiv |\Delta \mathbf{r}|$) for each phase is summarized in Table~\ref{tab:3}. See their formulas in the Supplementary Material (SM)~\cite{SM}. An exponential decay in $\eta_{\theta}$, signaling $U(1)$ gauge symmetry restoration, destroys superconductivity. In contrast, an exponential decay in $\eta_{\phi_{\alpha}}$ restores translational symmetry and leads to a spatially uniform state, distinct from a density wave. When $\eta_{\phi_{\alpha}}$ exhibits power-law decay, and $\eta_{\theta}$ exhibits either power-law or exponential decay, they correspond to C-PDW or C-CDW, respectively. The phase diagrams for different $n$ from our MC simulations qualitatively agree with the RG results, as schematically illustrated in Fig.~\ref{schematicphasediagram}.

For $n \le 4$, as depicted in Fig.~\ref{schematicphasediagram}(a), when the temperature is low enough, $\eta_{\theta}$ follows a power law decay with $\Delta r$, $\eta_{\phi_{\alpha}}$ saturates to a finite value as $\Delta r \to \infty$, reflecting the PDW. If $\mu/\rho$ is small, increasing the temperature, while $\eta_{\theta}$ exhibits a power-law decay with $\Delta r$, $\eta_{\phi_{\alpha}}$ shows exponential decay with $\Delta r$, reflecting the charge-$2e/4e$ SC; if $\mu/\rho$ is large, raising the temperature, whereas $\eta_{\theta}$ decays exponentially with $\Delta r$, $\eta_{\phi_{\alpha}}$ saturates to a finite value as $\Delta r \to \infty$, reflecting the CDW. When the temperature is high enough, both $\eta_{\theta}$ and $\eta_{\phi_{\alpha}}$ decay exponentially with $\Delta r$, reflecting the MT. When $\mu/\rho$ is moderate, a dashed line appears in the phase diagram. However, its physical meaning differs: it marks a solid-phase boundary for the $2Q$ PDW, whereas it physically degenerates to a single point for the $3Q$ PDW.

\begin{figure}[h]
	\centering
	\includegraphics[width=0.45\textwidth]{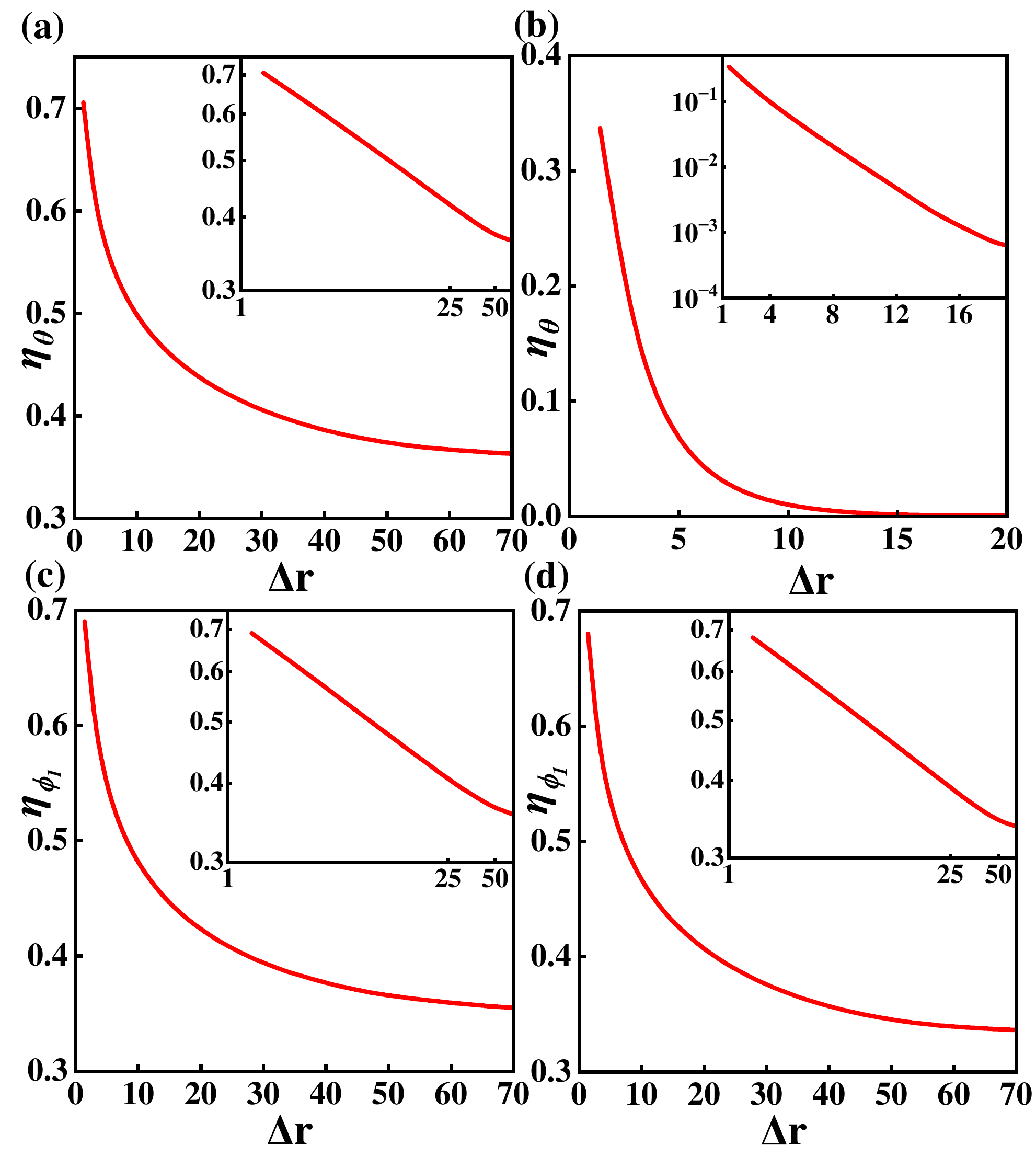}
	\caption{(Color online) The correlation functions $\eta_{\theta/\phi_{1}}$ for the $3Q$ PDW are shown for the C-PDW phase in panels (a) and (c), and for the C-CDW phase in panels (b) and (d). Insets of (a, c, d) the log-log plot, and (b) only the y-axis is logarithmic. Further details are provided in the SM~\cite{SM}.}\label{CE}
\end{figure}

For $n \ge 5$, as depicted in Fig.~\ref{schematicphasediagram}(b), in addition to the phases mentioned above, there are two other phases that require particular emphasis in our study. We provide the correlation functions in the Fig.~\ref{CE} for the two critical phases of the $3Q$ PDW. The correlation functions in the two critical phases of the 2Q PDW follow the same pattern. As the temperature increases, both $\eta_{\theta}$ and $\eta_{\phi_{\alpha}}$ exhibit power-law decay with $\Delta r$, see Fig.~\ref{CE}(a) and (c). The system transitions from the PDW state to the C-PDW state. As the temperature is further increased, the system enters a charge $2e/4e$ SC phase for small $\mu/\rho$, a MT phase for intermediate $\mu/\rho$, and a C-CDW phase for large $\mu/\rho$. For the C-CDW phase, $\eta_{\theta}$ shows exponential decay with $\Delta r$ and $\eta_{\phi_{\alpha}}$ exhibits a power-law decay with $\Delta r$, see Fig.~\ref{CE}(b) and (d).

Notably, we have discovered two novel critical phases, C-PDW and C-CDW, which emerge only for the commensurate PDW with periodicity $n \ge 5$ and are absent for $n \le 4$. In these phases, the correlation $\eta_{\phi_{\alpha}}(\Delta r)$ exhibits quasi-long-range order (QLRO), which can be regarded as the remarkable ``quasi-breaking'' of translation symmetry. We provide the detailed results of the hexagonal $3Q$ PDW and the $2Q$ PDW in the SM~\cite{SM}. 

To investigate the nature of the phase transitions, we present the thermodynamic quantities as functions of temperature for different lattice sizes $L$, with details provided in the SM~\cite{SM}. These quantities include the specific heat $C_v$, the susceptibilities of the $\theta$ and $\phi_{\alpha}$ fields, the Binder cumulant $3U_\theta-1$ and $3U_{\phi_{\alpha}}-1$, the stiffness $S$ of $\theta$ field, the Ising order parameter $I_{\alpha}$ of $\phi_{\alpha}$ field. As shown in Fig.~\ref{schematicphasediagram}(a), for systems with $n \le 4$, we find that for small $\mu/\rho$, the enhancement of $T$ first leads to a second-order phase transition in the $\phi_{\alpha}$ field, followed by a BKT transition in the $\theta$ field. Conversely, for large $\mu/\rho$, a BKT transition first occurs in the $\theta$ field, after which the $\phi_{\alpha}$ field undergoes a second-order phase transition. For systems with $n \ge 5$, as illustrated in Fig.~\ref{schematicphasediagram}(b), when $\mu/\rho$ is small, three successive BKT transitions occur: the first two are driven by the $\phi_{\alpha}$ field, and the third is driven by the $\theta$ field. When $\mu/\rho$ is large, the system similarly exhibits three successive BKT transitions: first, the $\phi_{\alpha}$ field undergoes a BKT transition, then the $\theta$ field undergoes a BKT transition, and finally, the $\phi_{\alpha}$ field undergoes a second BKT transition. To conclude, for $n \le 4$, the $\phi_{\alpha}$ field always undergoes a single second-order phase transition from long-range order (LRO) to disorder, while for $n \ge 5$, it experiences two successive BKT transitions: first from LRO to QLRO, and then from QLRO to disorder. This behavior is consistent with the $n$-state clock model.

{\bf Conclusion and Discussion:} 
We have also studied the 1$Q$ PDW, i.e. the unidirectional PDW with wave vectors $\pm \mathbf{Q} $, which is described by two complex gap functions $\Delta_{\pm \mathbf{Q}}(\mathbf{r})$~\cite{Berg2009}. In the incommensurate case, previous study~\cite{Berg2009} has yielded the charge-4e SC and CDW as vestigial phases. Here we studied the commensurate case with period $na_0$. In comparison with the incommensurate case, the free energy for the commensurate PDW acquires an additional term, $ (\Delta_{\mathbf{+Q}}^{n\ast}(\mathbf{r})\Delta_{-\mathbf{Q}}^{n}(\mathbf{r})+c.c)$, as detailed in the SM~\cite{SM}. Considering the anisotropy in the x and y directions, our analysis therefore employs the MC simulations rather than the RG method. We find that the main results are qualitatively the same as that of the $2Q$ PDW shown in Fig.~\ref{schematicphasediagram}. In particular, for $n\ge 5$, the C-PDW and C-CDW phases also emerge.

In conclusion, we have conducted a systematic investigation on the vestigial phases of 2D commensurate PDW in the form of $3Q$, $2Q$ and $1Q$. We find that the resultant phase diagrams differ for different periods $n$ of the PDW with expanded unit cell $na_0 \times na_0$.  For $2 \le n\le 4$, the systems exhibit charge-2e/4e SC and CDW vestigial phases, which are known previously. The new discovery here is that for $n\ge 5$, the translational symmetry undergoes a two-step transition first from LRO to QLRO and then from QLRO to disorder through two successive BKT transitions, leading into the two intermediate-temperature critical phases, C-PDW and C-CDW, as secondary orders. Our findings provide insights into vestigial phases in commensurate PDW.

\section*{Acknowledgements:} 
F. Y. is supported by the National Natural Science Foundation of China under the Grant Nos. 12574141, 12234016, 12074031. J. Zhou is supported by the Scientific Research Program from Science and Technology Bureau of Chongqing City (Grant No. CSTB2025NSCQ-GPX1303).

\bibliography{references}
~~~~~~~~~~~~~~~~~~~~~~~~~~~~~~~~~~~~~

\newpage

\renewcommand{\thefigure}{S\arabic{figure}}
\setcounter{figure}{0}
\renewcommand{\thetable}{S\arabic{table}}
\setcounter{table}{0}
\begin{widetext}

\appendix

\section{Ginzburg-Landau analysis for the hexagonal $3Q$ PDW}
We consider 3$\mathbf{Q}$ PDW state:
\begin{equation}\label{gap_function}
\Delta(\mathbf{r})=\sum_{i=1,2,3}(\Delta_{\mathbf{Q}_{i}}e^{i\mathbf{Q}_{i}\cdot \mathbf{r}}+\Delta_{-\mathbf{Q}_{i}}e^{-i\mathbf{Q}_{i}\cdot\mathbf{r}}),
\end{equation}
with $\mathbf{Q}_{1}+\mathbf{Q}_{2}+\mathbf{Q}_{3}=0$. The origin Hamiltonian is invariant under the rotation group $P6m$: $G=\{E,C_{6},C_{6}^{2},C_{6}^{3},C_{6}^{4},C_{6}^{5},m_{x},m_{y}\}$ and translation operation $P$, time reversal operation $T$. To get the Ginzburg-Landau free energy, we find the symmetry properties of the order parameter. Under translation operation $P$:
\begin{equation}\label{translation operation}
\Delta_{\pm \mathbf{Q}_{i}}(\mathbf{r})\rightarrow\Delta_{\pm \mathbf{Q}_{i}}(\mathbf{r}-\mathbf{a})e^{\mp i\mathbf{Q}_{i}\cdot \mathbf{a}}.
\end{equation}
Under time reversal operation $T$,
\begin{equation}\label{time reversal operation}
\Delta_{\pm \mathbf{Q}_{i}}(\mathbf{r})\rightarrow\Delta^{\ast}_{\mp \mathbf{Q}_{i}}(\mathbf{r}).
\end{equation}
Furthermore, under $C_{6}$ rotation, we have:
\begin{eqnarray}\label{rotation operation}
C_{6}:(\Delta_{\mathbf{Q}_{1}},\Delta_{\mathbf{Q}_{2}},\Delta_{\mathbf{Q}_{3}},\Delta_{-\mathbf{Q}_{1}},\Delta_{-\mathbf{Q}_{2}},\Delta_{-\mathbf{Q}_{3}})\rightarrow(\Delta_{-\mathbf{Q}_{3}},\Delta_{-\mathbf{Q}_{1}},\Delta_{-\mathbf{Q}_{2}},\Delta_{\mathbf{Q}_{3}},\Delta_{\mathbf{Q}_{1}},\Delta_{\mathbf{Q}_{2}}).
\end{eqnarray}
Taking $m_{x}$ and $m_{y}$ operation, we get:
\begin{eqnarray}\label{mx and my operation}
m_{x}:(\Delta_{\mathbf{Q}_{1}},\Delta_{\mathbf{Q}_{2}},\Delta_{\mathbf{Q}_{3}},\Delta_{-\mathbf{Q}_{1}},\Delta_{-\mathbf{Q}_{2}},\Delta_{-\mathbf{Q}_{3}})\rightarrow (\Delta_{\mathbf{Q}_{3}},\Delta_{\mathbf{Q}_{2}},\Delta_{\mathbf{Q}_{1}},\Delta_{-\mathbf{Q}_{3}},\Delta_{-\mathbf{Q}_{2}},\Delta_{-\mathbf{Q}_{1}}), \\
m_{y}:(\Delta_{\mathbf{Q}_{1}},\Delta_{\mathbf{Q}_{2}},\Delta_{\mathbf{Q}_{3}},\Delta_{-\mathbf{Q}_{1}},\Delta_{-\mathbf{Q}_{2}},\Delta_{-\mathbf{Q}_{3}})\rightarrow(\Delta_{-\mathbf{Q}_{3}},\Delta_{-\mathbf{Q}_{2}},\Delta_{-\mathbf{Q}_{1}},\Delta_{\mathbf{Q}_{3}},\Delta_{\mathbf{Q}_{2}},\Delta_{\mathbf{Q}_{1}}).
\end{eqnarray}
The free energy should be invariant under point group $P6m$. We write out the free energy density of our model with the symmetry above:
\begin{eqnarray}\label{free energy}
F^{(2)}&=&-\alpha\sum_{i}(|\Delta_{\mathbf{Q}_{i}}|^{2}+|\Delta_{-\mathbf{Q}_{i}}|^{2}), \\
F^{(4)}&=&\beta_{1}\sum_{i\neq j}\Delta_{\mathbf{Q}_{i}}^{\ast}\Delta_{-\mathbf{Q}_{i}}^{\ast}\Delta_{\mathbf{Q}_{j}}\Delta_{-\mathbf{Q}_{j}}+\beta_{2}\sum_{i\neq j}|\Delta_{\mathbf{Q}_{i}}|^{2}|\Delta_{\mathbf{Q}_{j}}|^{2}+\beta_{3}(\sum_{i}|\Delta_{\mathbf{Q}_{i}}|^{2})^{2} \nonumber \\
&+&\beta_{4}\sum_{i}|\Delta_{\mathbf{Q}_{i}}|^{2}|\Delta_{-\mathbf{Q}_{i}}|^{2} 
+\beta_{5}\sum_{i\neq j}(|\Delta_{\mathbf{Q}_{i}}|^{2}|\Delta_{\mathbf{Q}_{j}}|^{2}+|\Delta_{-\mathbf{Q}_{i}}|^{2}|\Delta_{-\mathbf{Q}_{j}}|^{2}).
\end{eqnarray}
The part induced by the phase fluctuation of the order parameters takes the following form:
\begin{eqnarray}\label{kinetic term}
F_{flu}&=&(|\Delta_{\mathbf{Q}_{1}}(\mathbf{k})|^{2}+|\Delta_{-\mathbf{Q}_{1}}(\mathbf{k})|^{2})(\alpha_{1}\mathbf{k}_{+}^{2}+\beta_{1}\mathbf{k}_{-}^{2}+\gamma_{1}\mathbf{k}_{+}\mathbf{k}_{-}) 
+(|\Delta_{\mathbf{Q}_{2}}(\mathbf{k})|^{2}+|\Delta_{-\mathbf{Q}_{2}}(\mathbf{k})|^{2})(\alpha_{2}\mathbf{k}_{+}^{2}+\beta_{1}\mathbf{k}_{-}^{2}+\gamma_{2}\mathbf{k}_{+}\mathbf{k}_{-}) \nonumber \\
&+&(|\Delta_{\mathbf{Q}_{3}}(\mathbf{k})|^{2}+|\Delta_{-\mathbf{Q}_{3}}(\mathbf{k})|^{2})(\alpha_{3}\mathbf{k}_{+}^{2}+\beta_{3}\mathbf{k}_{-}^{2}+\gamma_{3}\mathbf{k}_{+}\mathbf{k}_{-}),
\end{eqnarray}
with the definition $\mathbf{k}_{\pm}=k_x \pm i k_y$.
$C_{6}$ rotation symmetry requires $\alpha_{3}=\alpha_{1}e^{i2\pi/3},\beta_{3}=\beta_{1}e^{-i2\pi/3}, \gamma_{3}=\gamma_{1}$, and $\alpha_{2}=\alpha_{3}e^{i2\pi/3},\beta_{2}=\beta_{3}e^{-i2\pi/3}, \gamma_{2}=\gamma_{3}$. Finally, we get the fluctuation part of the free energy density:
\begin{eqnarray}\label{F_kin}
F_{flu}&=&\kappa_{1}\sum_{i}|\nabla\Delta_{\mathbf{Q}_{i}}|^{2}+\kappa_{2}(|\nabla_{+}\Delta_{\mathbf{Q}_{1}}|^{2}+|\nabla_{+}\Delta_{-\mathbf{Q}_{1}}|^{2} 
+\nu(|\nabla_{+}\Delta_{\mathbf{Q}_{2}}|^{2}+|\nabla_{+}\Delta_{-\mathbf{Q}_{2}}|^{2})  \nonumber \\
&+&\nu^{2}(|\nabla_{+}\Delta_{\mathbf{Q}_{3}}|^{2}+|\nabla_{+}\Delta_{-\mathbf{Q}_{3}}|^{2})+h.c.).
\end{eqnarray} 
Here, $\nu=e^{2i\pi/3}$ and $\nabla_{\pm}=\nabla_{x} \pm i \nabla_{y}$.

\section{Ginzburg-Landau analysis for the $2Q$ PDW}
We consider 2$\mathbf{Q}$ PDW state:
\begin{equation}\label{gap_function}
\Delta(\mathbf{r})=\sum_{i=1,2}(\Delta_{\mathbf{Q}_{i}}e^{i\mathbf{Q}_{i}\cdot \mathbf{r}}+\Delta_{-\mathbf{Q}_{i}}e^{-i\mathbf{Q}_{i}\cdot\mathbf{r}}).
\end{equation}
The origin Hamiltonian is invariant under the symmetry: U(1) gauge symmetry, $C_4^1, m_{x}, m_{y}$, time reversal symmetry (TRS), translation symmetry. To get the Ginzburg-Landau free energy, we find the symmetry properties of the order parameter,
\begin{eqnarray}\label{symmetry2}
&&\text{(1) U(1)-gauge :~~~~} \Delta_{\mathbf{Q}_{i}}\rightarrow e^{i\theta}\Delta_{\mathbf{Q}_{i}}\nonumber\\
&&\text{(2) C}_4^1\text{-rotation :~~~~} (\Delta_{\mathbf{Q}_{1}},\Delta_{\mathbf{Q}_{2}},\Delta_{-\mathbf{Q}_{1}},\Delta_{-\mathbf{Q}_{2}})\rightarrow(\Delta_{\mathbf{Q}_{2}},\Delta_{-\mathbf{Q}_{1}},\Delta_{-\mathbf{Q}_{2}},\Delta_{\mathbf{Q}_{1}})\nonumber\\
&&\text{(3) } m_x\text{-mirror :~~~~}(\Delta_{\mathbf{Q}_{1}},\Delta_{\mathbf{Q}_{2}},\Delta_{-\mathbf{Q}_{1}},\Delta_{-\mathbf{Q}_{2}})\rightarrow (\Delta_{\mathbf{Q}_{1}},\Delta_{-\mathbf{Q}_{2}},\Delta_{-\mathbf{Q}_{1}},\Delta_{\mathbf{Q}_{2}}) \nonumber\\
&&\text{(4) } m_y\text{-mirror :~~~~}(\Delta_{\mathbf{Q}_{1}},\Delta_{\mathbf{Q}_{2}},\Delta_{-\mathbf{Q}_{1}},\Delta_{-\mathbf{Q}_{2}})\rightarrow(\Delta_{-\mathbf{Q}_{1}},\Delta_{\mathbf{Q}_{2}},\Delta_{\mathbf{Q}_{1}},\Delta_{-\mathbf{Q}_{2}})\nonumber\\
&&\text{(5)}~~\text{TRS :~~~~}
\Delta_{\pm \mathbf{Q}_{i}}(\mathbf{r})\rightarrow\Delta^{\ast}_{\mp \mathbf{Q}_{i}}(\mathbf{r}) \nonumber\\
&&\text{(6)}~~\text{Translation symmetry:~~~~}\Delta_{\pm \mathbf{Q}_{i}}(\mathbf{r})\rightarrow\Delta_{\pm \mathbf{Q}_{i}}(\mathbf{r}-\mathbf{a})e^{\mp i\mathbf{Q}_{i}\cdot \mathbf{a}}.
\end{eqnarray}

The free energy should be invariant under all the symmetries mentioned above. The free energy density takes the following form:
\begin{eqnarray}
F&=&\alpha\sum_{i}|\Delta_{\mathbf{Q}_{i}}|^2+\beta_1(\sum_{i}|\Delta_{\mathbf{Q}_{i}}|)^2 +
\beta_2\sum_{i<j}|\Delta_{\mathbf{Q}_{i}}|^2|\Delta_{\mathbf{Q}_{j}}|^2+\beta_3\sum_{i}|\Delta_{\mathbf{Q}_{i}}|^2|\Delta_{-\mathbf{Q}_{i}}|^2 \nonumber\\
&&+\beta_4(\Delta_{\mathbf{Q}_{1}}\Delta_{-\mathbf{Q}_{1}}\Delta_{\mathbf{Q}_{2}}^{\ast}\Delta_{-\mathbf{Q}_{2}}^{\ast}+\Delta_{\mathbf{Q}_{1}}^{\ast}\Delta_{-\mathbf{Q}_{1}}^{\ast}\Delta_{\mathbf{Q}_{2}}\Delta_{-\mathbf{Q}_{2}}).
\end{eqnarray}

\section{Ginzburg-Landau analysis and effective Hamiltonian for the unidirectional PDW}

In this section, we derive the effective Hamiltonian appearing in the Eq.~(\ref{Hamiltonian_r}) of the main text by the Ginzburg-Landau (G-L) theory of the commensurate unidirectional PDW.

\subsection{Symmetry}

The G-L free energy is constructed by requiring invariance under the translation symmetry, the mirror reflection symmetry, the time-reversal symmetry (TRS) and U(1)-gauge symmetry. We set
$\Delta_{+ \mathbf{Q_1}}(\mathbf{r})=e^{i\theta_{1}(\mathbf{r})}\Delta_{0}$, $\Delta_{-\mathbf{Q_1}}(\mathbf{r})=e^{i\theta^{'}_{1}(\mathbf{r})}\Delta_{0}$, with $\theta_{1}(\mathbf{r})=\theta(\mathbf{r})+\phi(\mathbf{r})$, $\theta^{'}_{1}(\mathbf{r})=\theta(\mathbf{r})-\phi(\mathbf{r})$ and $\mathbf{Q_1}$ along the $x$ direction. Here we focus on the low-energy phase fluctuations, and have set the global amplitude $\Delta_0>0$ as a constant. The phase fields $\theta(\mathbf{r})$ and $\phi(\mathbf{r})$ are slowly-varying of the coarse-grained position $\mathbf{r}$.
Under the translation symmetry, the spatial dependent pairing amplitudes change to:
\begin{eqnarray}
\Delta_{\mathbf{+Q_1}}(\mathbf{r})\to{\Delta}_{\mathbf{+Q_1}}(\mathbf{r-a}) e^{-i\mathbf{Q_1} \cdot \mathbf{a}},\qquad \Delta_{\mathbf{-Q_1}}(\mathbf{r})\to{\Delta}_{\mathbf{-Q_1}}(\mathbf{r-a}) e^{i\mathbf{Q_1} \cdot \mathbf{a}}.
\end{eqnarray}
Under the mirror reflection symmetry $\sigma_{yz}$, the spatial dependent pairing amplitudes change to:
\begin{eqnarray}
\Delta_{\mathbf{+Q_1}}(x,y)\to{\Delta}_{\mathbf{-Q_1}}(-x,y),\qquad \Delta_{\mathbf{-Q_1}}(x,y)\to{\Delta}_{\mathbf{+Q_1}}(-x,y).
\end{eqnarray}
Under the mirror reflection symmetry $\sigma_{xz}$, the spatial dependent pairing amplitudes is not change:
\begin{eqnarray}
\Delta_{\mathbf{+Q_1}}({x,y})\to{\Delta}_{\mathbf{+Q_1}}({x,-y}),\qquad \Delta_{\mathbf{-Q_1}}({x,y})\to{\Delta}_{\mathbf{-Q_1}}({x,-y}).
\end{eqnarray}
Under TRS, the spatial dependent pairing amplitudes change to:
\begin{eqnarray}
\Delta_{\mathbf{-Q_1}}(\mathbf{r})\to{\Delta}_{\mathbf{+Q_1}}^{\ast}(\mathbf{r}),\qquad \Delta_{\mathbf{+Q_1}}(\mathbf{r})\to{\Delta}_{\mathbf{-Q_1}}^{\ast}(\mathbf{r}).
\end{eqnarray}
Under U(1)-gauge, the spatial dependent pairing amplitudes change to:
\begin{eqnarray}\label{symmetry}
\text{ U(1)-gauge :~~~~} \Delta_{\pm \mathbf{Q_1}}(\mathbf{r})\rightarrow e^{i\theta}\Delta_{\pm \mathbf{Q_1}}(\mathbf{r}).
\end{eqnarray}

It is convenient to rewrite the above transformation in the $\mathbf{k}-$space
\begin{eqnarray}
&&\text{ translation symmetry :~~~~}\Delta_{\pm \mathbf{Q_1}}(\mathbf{k})\to{\Delta}_{\pm \mathbf{Q_1}}(\mathbf{k}) e^{\mp i\mathbf{Q_1} \cdot \mathbf{a}} e^{-i\mathbf{k} \cdot \mathbf{a}},\nonumber\\
&&\text{  mirror reflection symmetry $\sigma_{yz}$ :~~~~}\Delta_{\pm \mathbf{Q_1}}(\mathbf{k}_{x},\mathbf{k}_{y})\to{\Delta}_{\mp \mathbf{Q_1}}(-\mathbf{k}_{x},\mathbf{k}_{y}),\nonumber\\
&&\text{  mirror reflection symmetry $\sigma_{xz}$ :~~~~}\Delta_{\pm \mathbf{Q_1}}(\mathbf{k}_{x},\mathbf{k}_{y})\to{\Delta}_{\pm \mathbf{Q_1}}(\mathbf{k}_{x},-\mathbf{k}_{y}),\nonumber\\
&&\text{ TRS :~~~~}\Delta_{\mp \mathbf{Q_1}}(\mathbf{k})\to{\Delta}_{\pm \mathbf{Q_1}}^{\ast}(\mathbf{-k}),\nonumber\\
&&\text{ U(1)-gauge :~~~~}\Delta_{\pm \mathbf{Q_1}}(\mathbf{k})\rightarrow e^{i\theta}\Delta_{\pm \mathbf{Q_1}}(\mathbf{k}).
\end{eqnarray}

In order to derive the free energy as an explicit function of the $\theta$ and $\phi$ fields, we need to expand the free energy up to $2n$-th order of its argument $\Delta_{\pm \mathbf{Q_1}}$. The symmetry-allowed $2n$-th order term in the free energy is
\begin{eqnarray}
F^{(2n)}=A_{0}(\Delta_{+\mathbf{Q_1}}(\mathbf{r})^{n}\Delta_{-\mathbf{Q_1}}(\mathbf{r})^{n*}+\Delta_{+\mathbf{Q_1}}(\mathbf{r})^{n*}\Delta_{-\mathbf{Q_1}}(\mathbf{r})^{n}).
\end{eqnarray}
This term contributes to the anisotropy-field part $F^{(2n)}=2A_{0}\Delta_{0}^{2n}\cos(2n\phi)$ in the low-energy classical Hamiltonian.

\subsection{The second-order G-L expansion}

Up to $O(\Delta_{\pm \mathbf{Q_1}}^2)$, the differential term in G-L free energy has the following general form in the $\mathbf{k}-$ space:
\begin{eqnarray}
F_{0}^{(2)}& = &\sum_{\mathbf{k}}\Delta_{+\mathbf{Q_1}}^{\ast}(\mathbf{k})\Delta_{+\mathbf{Q_1}}(\mathbf{k})(a_{1}\mathbf{k}_{x}^2+b_{1}\mathbf{k}_{y}^2+c_{1}\mathbf{k}_{x}\mathbf{k}_{y}) \nonumber\\
& + &\sum_{\mathbf{k}}\Delta_{+\mathbf{Q_1}}^{\ast}(\mathbf{k})\Delta_{-\mathbf{Q_1}}(\mathbf{k})(a_{2}\mathbf{k}_{x}^2+b_{2}\mathbf{k}_{y}^2+c_{2}\mathbf{k}_{x}\mathbf{k}_{y}) \nonumber\\
& + &\sum_{\mathbf{k}}\Delta_{-\mathbf{Q_1}}^{\ast}(\mathbf{k})\Delta_{+\mathbf{Q_1}}(\mathbf{k})(a_{3}\mathbf{k}_{x}^2+b_{3}\mathbf{k}_{y}^2+c_{3}\mathbf{k}_{x}\mathbf{k}_{y}) \nonumber\\
& + &\sum_{\mathbf{k}}\Delta_{-\mathbf{Q_1}}^{\ast}(\mathbf{k})\Delta_{-\mathbf{Q_1}}(\mathbf{k})(a_{4}\mathbf{k}_{x}^2+b_{4}\mathbf{k}_{y}^2+c_{4}\mathbf{k}_{x}\mathbf{k}_{y}).
\end{eqnarray}
Under the translation symmetry, the invariance of $F_{0}^{(2)}$ requires only $a_{1},b_{1},c_{1},a_{4},b_{4},c_{4}\neq0$  while all the other coefficients keep zero. Further more, $c_{1}=c_{4}=0$ is required by the mirror-reflection symmetry $\sigma_{xz}$ and $a_{1}=a_{4}=A$, $b_{1}=b_{4}=B$ is required by TRS. Changing back to the real space, the resulting $F_{0}^{(2)}$ is:
\begin{eqnarray}
\label{eqn:F2}
F_{0}^{(2)}&=&A\int d^{2}\mathbf{r}[(\nabla_{x}\Delta_{+\mathbf{Q_1}}^{\ast})\cdot(\nabla_{x}\Delta_{+\mathbf{Q_1}})+(\nabla_{x}\Delta_{-\mathbf{Q_1}}^{\ast})\cdot(\nabla_{x}\Delta_{-\mathbf{Q_1}})] \nonumber\\
&+&B\int d^{2}\mathbf{r}[(\nabla_{y}\Delta_{+\mathbf{Q_1}}^{\ast})\cdot(\nabla_{y}\Delta_{+\mathbf{Q_1}})+(\nabla_{y}\Delta_{-\mathbf{Q_1}}^{\ast})\cdot(\nabla_{y}\Delta_{-\mathbf{Q_1}})] \nonumber\\
&=&A\Delta_{0}^{2}\int d^{2}\mathbf{r}[\nabla_{x}(e^{-i(\theta+\phi)})\cdot\nabla_{x}(e^{i(\theta+\phi)})+\nabla_{x}(e^{-i(\theta-\phi)})\cdot\nabla_{x}(e^{i(\theta-\phi)})] \nonumber\\
&+&B\Delta_{0}^{2}\int d^{2}\mathbf{r}[\nabla_{y}(e^{-i(\theta+\phi)})\cdot\nabla_{y}(e^{i(\theta+\phi)})+\nabla_{y}(e^{-i(\theta-\phi)})\cdot\nabla_{y}(e^{i(\theta-\phi)})]\nonumber\\
&=&2A\Delta_{0}^{2}\int d^{2}\mathbf{r}(|\nabla_{x}\theta|^{2}+|\nabla_{x}\phi|^{2})+2B\Delta_{0}^{2}\int d^{2}\mathbf{r}(|\nabla_{y}\theta|^{2}+|\nabla_{y}\phi|^{2}).
\end{eqnarray}.

\subsection{The fourth-order G-L expansion}

According to the second order expansion of the differential term in the G-L free energy, the coefficients before $\theta$ and $\phi$ are the same in the x and y direction, respectively. To get different coefficients, considering the fourth order with the general form as of $F_{0}$:
\begin{eqnarray}
F_{0}^{(4)}&=&\sum_{\mathbf{k}_{1},\mathbf{k}_{2},\mathbf{k}_{3},\mathbf{k}_{4}}\Delta_{\alpha}^{\ast}(\mathbf{k}_{1})\Delta_{\beta}^{\ast}(\mathbf{k}_{2})\Delta_{\gamma}(\mathbf{k}_{3})\Delta_{\nu}(\mathbf{k}_{4})
(\sum_{i,j=1}^{4}\alpha_{ij}\mathbf{k}_{ix}\cdot\mathbf{k}_{jx}+\beta_{ij}\mathbf{k}_{iy}\cdot\mathbf{k}_{jy}+\gamma_{ij}\mathbf{k}_{ix}\cdot\mathbf{k}_{jy}+\nu_{ij}\mathbf{k}_{iy}\cdot\mathbf{k}_{jx}), \nonumber\\
\end{eqnarray}
where $\alpha,\beta,\gamma,\nu=\pm \mathbf{Q_1}$. Obviously, $\alpha\beta\gamma\nu=\left\{++++,+-+-,----\right\} \mathbf{Q_1}$ is required by the translation symmetry. $\gamma_{ij}=\nu_{ij}=0$ is required by the mirror-reflection symmetry $\sigma_{xz}$. The $F_{0}^{(4)}$ can be rewrited:
\begin{eqnarray}
\label{eqn:genergal_F4}
F_{0}^{(4)}&=&\sum_{\mathbf{k}_{1},\mathbf{k}_{2},\mathbf{k}_{3},\mathbf{k}_{4}}\Delta_{+\mathbf{Q_1}}^{\ast}(\mathbf{k}_{1})\Delta_{+\mathbf{Q_1}}^{\ast}(\mathbf{k}_{2})\Delta_{+\mathbf{Q_1}}(\mathbf{k}_{3})\Delta_{+\mathbf{Q_1}}(\mathbf{k}_{4})(\sum_{i,j=1}^{4}\alpha^{(1)}_{ij}\mathbf{k}_{ix}\cdot \mathbf{k}_{jx}+\beta^{(1)}_{ij}\mathbf{k}_{iy}\cdot \mathbf{k}_{jy}) \nonumber\\
&+&\sum_{\mathbf{k}_{1},\mathbf{k}_{2},\mathbf{k}_{3},\mathbf{k}_{4}}\Delta_{+\mathbf{Q_1}}^{\ast}(\mathbf{k}_{1})\Delta_{-\mathbf{Q_1}}^{\ast}(\mathbf{k}_{2})\Delta_{+\mathbf{Q_1}}(\mathbf{k}_{3})\Delta_{-\mathbf{Q_1}}(\mathbf{k}_{4})(\sum_{i,j=1}^{4}\alpha^{(2)}_{ij}\mathbf{k}_{ix}\cdot \mathbf{k}_{jx}+\beta^{(2)}_{ij}\mathbf{k}_{iy}\cdot \mathbf{k}_{jy}) \nonumber\\
&+&\sum_{\mathbf{k}_{1},\mathbf{k}_{2},\mathbf{k}_{3},\mathbf{k}_{4}}\Delta_{-\mathbf{Q_1}}^{\ast}(\mathbf{k}_{1})\Delta_{-\mathbf{Q_1}}^{\ast}(\mathbf{k}_{2})\Delta_{-\mathbf{Q_1}}(\mathbf{k}_{3})\Delta_{-\mathbf{Q_1}}(\mathbf{k}_{4})(\sum_{i,j=1}^{4}\alpha^{(3)}_{ij}\mathbf{k}_{ix}\cdot \mathbf{k}_{jx}+\beta^{(3)}_{ij}\mathbf{k}_{iy}\cdot \mathbf{k}_{jx}).
\end{eqnarray}

To simplify the derivation, we only consider x-direction, and it is the same to y-direction. Since the first and the third term in the general form of $F_{0}^{(4)}$ in terms of only $\Delta_{+\mathbf{Q_1}}$ or $\Delta_{-\mathbf{Q_1}}$ and recall that all the transformation relation, the form of equation (\ref{eqn:genergal_F4}) becomes:
\begin{eqnarray}
\label{eqn:F4_13}
F^{(4)}_{0(1,3)x}=\sum_{\mathbf{k}_{1},\mathbf{k}_{2},\mathbf{k}_{3},\mathbf{k}_{4}}[\Delta_{+\mathbf{Q_1}}^{\ast}(\mathbf{k}_{1})\Delta_{+\mathbf{Q_1}}^{\ast}(\mathbf{k}_{2})\Delta_{+\mathbf{Q_1}}(\mathbf{k}_{3})\Delta_{+\mathbf{Q_1}}(\mathbf{k}_{4})  +\Delta_{-\mathbf{Q_1}}^{\ast}(\mathbf{k}_{1})\Delta_{-\mathbf{Q_1}}^{\ast}(\mathbf{k}_{2})\Delta_{-\mathbf{Q_1}}(\mathbf{k}_{3})\Delta_{-\mathbf{Q_1}}(\mathbf{k}_{4})] \nonumber\\
\cdot[a(\mathbf{k}_{1x}^{2}+\mathbf{k}_{2x}^{2}+\mathbf{k}_{3x}^{2}+\mathbf{k}_{4x}^{2})+b(\mathbf{k}_{1x}\cdot \mathbf{k}_{2x}+\mathbf{k}_{3x}\cdot \mathbf{k}_{4x})+c(\mathbf{k}_{1x}+\mathbf{k}_{2x})\cdot(\mathbf{k}_{3x}+\mathbf{k}_{4x})].
\end{eqnarray}
Let’s consider the conservation of momentum in x-direction is $(\mathbf{k}_{1x}+\mathbf{k}_{2x}-\mathbf{k}_{3x}-\mathbf{k}_{4x})^{2}=0$. From this constraint, we have:
\begin{equation}
\sum_{i=1}^{4}\mathbf{k}_{ix}^{2}=2(\mathbf{k}_{1x}+\mathbf{k}_{2x})\cdot(\mathbf{k}_{3x}+\mathbf{k}_{4x})-2(\mathbf{k}_{1x}\cdot \mathbf{k}_{2x}+\mathbf{k}_{3x}\cdot \mathbf{k}_{4x}).
\end{equation}
The first and third term can be writed:
\begin{eqnarray}
F^{(4)}_{0(1,3)x}=\sum_{\mathbf{k}_{1},\mathbf{k}_{2},\mathbf{k}_{3},\mathbf{k}_{4}}[\Delta_{+\mathbf{Q_1}}^{\ast}(\mathbf{k}_{1})\Delta_{+\mathbf{Q_1}}^{\ast}(\mathbf{k}_{2})\Delta_{+\mathbf{Q_1}}(\mathbf{k}_{3})\Delta_{+\mathbf{Q_1}}(\mathbf{k}_{4})  +\Delta_{-\mathbf{Q_1}}^{\ast}(\mathbf{k}_{1})\Delta_{-\mathbf{Q_1}}^{\ast}(\mathbf{k}_{2})\Delta_{-\mathbf{Q_1}}(\mathbf{k}_{3})\Delta_{-\mathbf{Q_1}}(\mathbf{k}_{4})] \nonumber\\
\cdot[ (b-2a)\cdot(\mathbf{k}_{1x}\cdot \mathbf{k}_{2x}+\mathbf{k}_{3x}\cdot \mathbf{k}_{4x}) + (c+2a)\cdot(\mathbf{k}_{1x}+\mathbf{k}_{2x})\cdot(\mathbf{k}_{3x}+\mathbf{k}_{4x}) ] .
\end{eqnarray}
By the same method, the second term of the fourth order expansion of the differential term in G-L free energy is
\begin{eqnarray}
\label{eqn:F4_2}
F^{(4)}_{0(2)x}&=&\sum_{\mathbf{k}_{1},\mathbf{k}_{2},\mathbf{k}_{3},\mathbf{k}_{4}}\Delta_{+\mathbf{Q_1}}^{\ast}(\mathbf{k}_{1})\Delta_{-\mathbf{Q_1}}^{\ast}(\mathbf{k}_{2})\Delta_{+\mathbf{Q_1}}(\mathbf{k}_{3})\Delta_{-\mathbf{Q_1}}(\mathbf{k}_{4}) \nonumber\\
&\cdot&[a'(\mathbf{k}_{1x}^{2}+\mathbf{k}_{2x}^{2}+\mathbf{k}_{3x}^{2}+\mathbf{k}_{4x}^{2})+b'(\mathbf{k}_{1x}\cdot\mathbf{k}_{2x}+\mathbf{k}_{3x}\cdot \mathbf{k}_{4x})
+c'(\mathbf{k}_{1x}\cdot\mathbf{k}_{3x}+\mathbf{k}_{2x}\cdot \mathbf{k}_{4x})+d'(\mathbf{k}_{1x}\cdot\mathbf{k}_{4x}+\mathbf{k}_{2x}\cdot \mathbf{k}_{3x}) ]. \nonumber\\
\end{eqnarray}
After the Fourier transform, the total form of $F_{0x}^{(4)}$ in the real space,
\begin{eqnarray}
F^{(4)}_{0x}&=&-(b-2a)\int d^{2}\mathbf{r}[ (\nabla_{x}\Delta_{+\mathbf{Q_1}}^{\ast})^{2}\Delta_{+\mathbf{Q_1}}^{2}+(\Delta_{+\mathbf{Q_1}}^{\ast})^{2}(\nabla_{x}\Delta_{+\mathbf{Q_1}})^{2}+(\nabla_{x}\Delta_{-\mathbf{Q_1}}^{\ast})^{2}\Delta_{-\mathbf{Q_1}}^{2}+(\Delta_{-\mathbf{Q_1}}^{\ast})^{2}(\nabla_{x}\Delta_{-\mathbf{Q_1}})^{2}] \nonumber\\
&+&(c+2a)\int d^{2}\mathbf{r}[ \nabla_{x}(\Delta_{+\mathbf{Q_1}}^{\ast 2})\cdot\nabla_{x}(\Delta_{+\mathbf{Q_1}}^{2})+\nabla_{x}(\Delta_{-\mathbf{Q_1}}^{\ast 2})\cdot\nabla_{x}(\Delta_{-\mathbf{Q_1}}^{2}) ] \nonumber\\
&-&(b'-2a')\int d^{2}\mathbf{r}[ (\nabla_{x}\Delta_{+\mathbf{Q_1}}^{\ast})\cdot(\nabla_{x}\Delta_{-\mathbf{Q_1}}^{\ast})\Delta_{+\mathbf{Q_1}}\Delta_{-\mathbf{Q_1}}+\Delta_{+\mathbf{Q_1}}^{\ast}\Delta_{-\mathbf{Q_1}}^{\ast}(\nabla_{x}\Delta_{+\mathbf{Q_1}})\cdot(\nabla_{x}\Delta_{-\mathbf{Q_1}})]\nonumber\\
&+&(c'+2a')\int d^{2}\mathbf{r}[ \nabla_{x}\Delta_{+\mathbf{Q_1}}^{\ast}\cdot\nabla_{x}\Delta_{+\mathbf{Q_1}}|\Delta_{-\mathbf{Q_1}}|^{2}+|\Delta_{+\mathbf{Q_1}}|^{2}\nabla_{x}\Delta_{-\mathbf{Q_1}}^{\ast}\cdot\nabla_{x}\Delta_{-\mathbf{Q_1}} ]\nonumber\\
&+&(d'+2a')\int d^{2}\mathbf{r} [ \nabla_{x}\Delta_{+\mathbf{Q_1}}^{\ast}\cdot\nabla_{x}\Delta_{-\mathbf{Q_1}}\Delta_{-\mathbf{Q_1}}^{\ast}\Delta_{+\mathbf{Q_1}} + \nabla_{x}\Delta_{-\mathbf{Q_1}}^{\ast}\cdot\nabla_{x}\Delta_{+\mathbf{Q_1}}\Delta_{+\mathbf{Q_1}}^{\ast}\Delta_{-\mathbf{Q_1}} ] \nonumber\\
&=&(8a+4b+8c+4a'+2b'+2c'+2d')\Delta_{0}^{4} \int d^{2}\mathbf{r}|\nabla_{x}\theta|^{2}  \nonumber\\
&+&(8a+4b+8c+4a'-2b'+2c'-2d')\Delta_{0}^{4}\int d^{2}\mathbf{r}|\nabla_{x}\phi|^{2}.
\end{eqnarray}
We can get the stiffness parameters $\rho$ and $\mu$ in the text:
\begin{eqnarray}
\rho&=&4(A+B)\Delta_{0}^{2}+2(8a+4b+8c+4a'+2b'+2c'+2d')\Delta_{0}^{4}, \\
\mu&=&4(A+B)\Delta_{0}^{2}+2(8a+4b+8c+4a'-2b'+2c'-2d')\Delta_{0}^{4}.
\end{eqnarray}
By the same method, in the y-direction,
\begin{eqnarray}
F_{0y}^{(4)}&=&(8a_{1}+4b_{1}+8c_{1}+4a_{1}'+2b_{1}'+2c_{1}'+2d_{1}')\Delta_{0}^{4} \int d^{2}\mathbf{r}|\nabla_{y}\theta|^{2}  \nonumber\\
&+&(8a_{1}+4b_{1}+8c_{1}+4a_{1}'-2b_{1}'+2c_{1}'-2d_{1}')\Delta_{0}^{4}\int d^{2}\mathbf{r}|\nabla_{y}\phi|^{2}.
\end{eqnarray}
We introduce the anisotropy coefficients $\alpha_{1}$ and $\alpha_{2}$ to rewrite the coefficients,
\begin{eqnarray}
\rho\alpha_{1}&=&4(A+B)\Delta_{0}^{2}+2(8a_{1}+4b_{1}+8c_{1}+4a_{1}'+2b_{1}'+2c_{1}'+2d_{1}')\Delta_{0}^{4}, \\
\mu\alpha_{2}&=&4(A+B)\Delta_{0}^{2}+2(8a_{1}+4b_{1}+8c_{1}+4a_{1}'-2b_{1}'+2c_{1}'-2d_{1}')\Delta_{0}^{4}.
\end{eqnarray}

Thus, the Hamiltonian for the melting system is given by:
\begin{equation}
H=\int d^{2}\mathbf{r}\left( \frac{\rho}{2}(|\nabla_{x}\theta|^{2}+\alpha_{1}|\nabla_{y}\theta|^{2})+\frac{\mu}{2}(|\nabla_{x}\phi|^{2}+\alpha_{2}|\nabla_{y}\phi|^{2})+ A\cos(2n\phi) \right),
\end{equation}
here $A=2A_{0}|\Delta_{0}|^{2n}.$

\section{More details Results about the RG study for the hexagonal $3Q$ PDW and the $2Q$ PDW}

\noindent With the standard Renormalization Group analysis for the hexagonal $3Q$ PDW, we have the tree level RG flow equation:
\begin{eqnarray}\label{fow equation}
\frac{dg_{\theta}}{dln b}&=&(2-\pi\rho^{'})g_{\theta} \nonumber\\
\frac{dg_{x}}{dln b}&=&(2-\pi\mu^{'})g_{x} \nonumber\\
\frac{dg_{y}}{dln b}&=&(2-\pi\mu^{'})g_{y} \nonumber\\
\frac{dg_{1}}{dln b}&=&0.5(2-\frac{n^{2}Q_{1x}^{2}}{\pi\mu}-\frac{n^{2}Q_{1y}^{2}}{\pi\mu})g_{1} \nonumber \\
\frac{dg_{2}}{dln b}&=&0.5(2-\frac{n^{2}Q_{2x}^{2}}{\pi\mu}-\frac{n^{2}Q_{2y}^{2}}{\pi\mu})g_{2} \nonumber \\
\frac{dg_{3}}{dln b}&=&0.5(2-\frac{n^{2}Q_{3x}^{2}}{\pi\mu}-\frac{n^{2}Q_{3y}^{2}}{\pi\mu})g_{3} 
\end{eqnarray}
Here, $b$ represents the renormalization scale, $g_{\theta}$, $g_{x}$, and $g_{y}$ represent the coupling strength of different types of integer vortices. We have replace $\rho/T$ by $\rho^{'}$, $\mu/T$ by $\mu^{'}$ to simplify the formula. $\mathbf{Q}_{1}=(1,-\frac{1}{\sqrt{3}})$ and $\mathbf{Q}_{2}=(0,\frac{2}{\sqrt{3}})$ and 
$\mathbf{Q}_{3}=-\mathbf{Q}_{1}-\mathbf{Q}_{2}$.

For the $2Q$ PDW, the corresponding RG equation is as follow:
\begin{eqnarray}\label{RG_4Q}
\frac{g_{\theta}}{dlnb}&=&(2-\pi\rho^{'})g_{\theta} \nonumber \\
\frac{g_{x}}{dlnb}&=&(2-\pi\mu^{'})g_{x} \nonumber \\
\frac{g_{y}}{dlnb}&=&(2-\pi\mu^{'})g_{y}  \nonumber \\
\frac{g_{\frac{1}{2}}^{x}}{dlnb}&=&\Big(2-\frac{\pi}{4}(\rho^{'}+\mu^{'})\Big)g_{\frac{1}{2}}^{x}  \nonumber \\
\frac{g_{\frac{1}{2}}^{y}}{dlnb}&=&\Big(2-\frac{\pi}{4}(\rho^{'}+\mu^{'})\Big)g_{\frac{1}{2}}^{y}  \nonumber \\
\frac{g_{1}}{dlnb}&=&(2-\frac{n^{2}Q_{1x}^{2}}{\pi\mu}-\frac{n^{2}Q_{1y}^{2}}{\pi\mu})g_{1} \nonumber \\
\frac{g_{2}}{dlnb}&=&(2-\frac{n^{2}Q_{2x}^{2}}{\pi\mu}-\frac{n^{2}Q_{2y}^{2}}{\pi\mu})g_{2} 
\end{eqnarray}
Here, $g_{\frac{1}{2}}^{x}$ and $g_{\frac{1}{2}}^{y}$ are coupling parameters of half vortices, $\mathbf{Q}_{1}=(1,0)$, $\mathbf{Q}_{2}=(0,1)$.

\begin{figure}[h]
	\centering
	\includegraphics[width=0.45\textwidth]{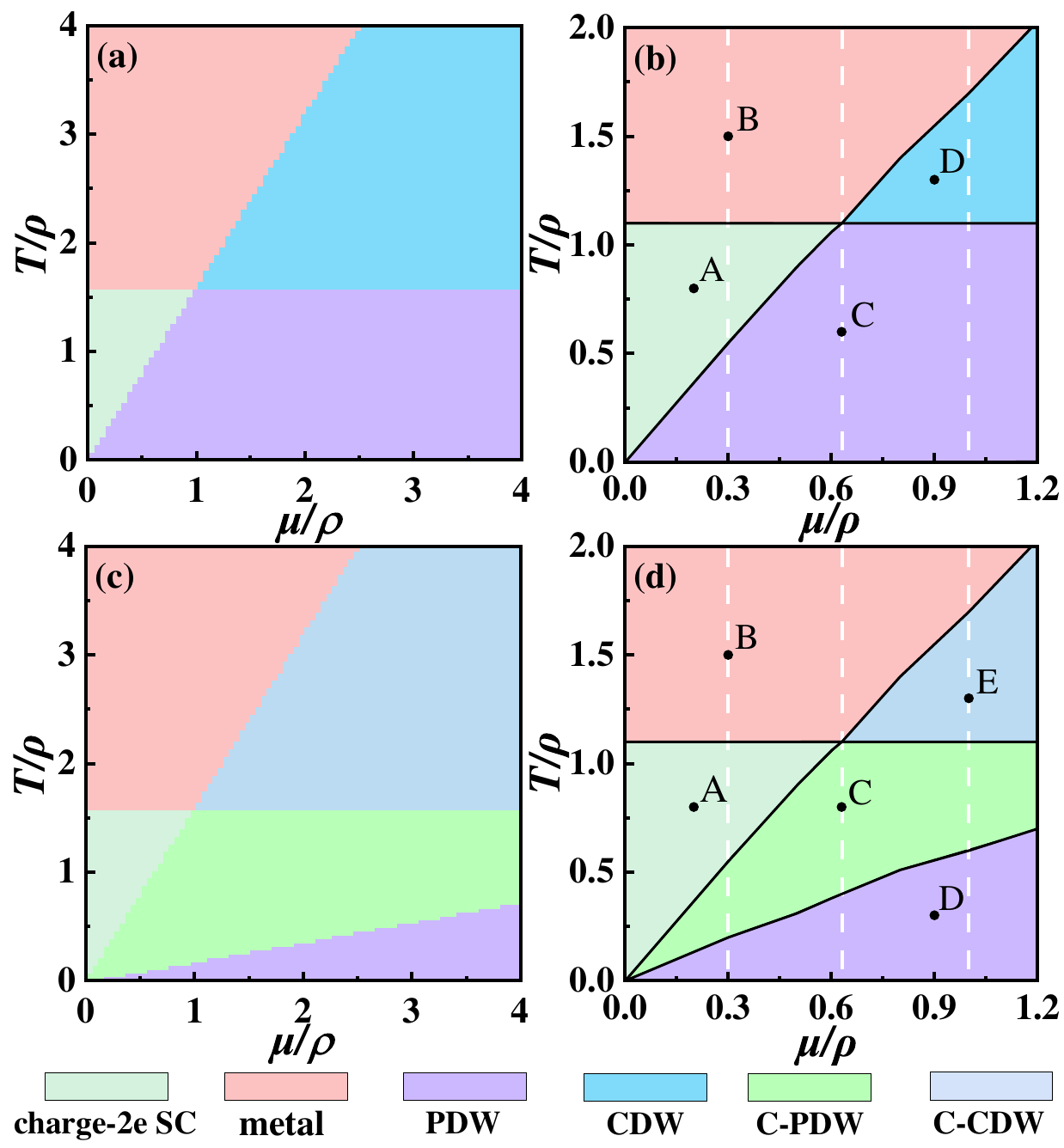}
	\caption{(Color online) Phase diagrams provided by (a,c) the RG study and (b,d) the MC study, with (a)-(b) for $n=2$ and (c)-(d) for $n=5$ (the $3Q$ PDW state). The white dashed lines in (b,d) mark $\mu/\rho=0.3, 0.63$ and $1$, respectively. The initial values of the coupling parameters in (a,c) are $g_{\theta}=0.1$, $g_{x}=g_{y}=0.1$, $g_\alpha=0.1$ in Eq. (\ref{eqn:action-SG}), and in (b,d) are $A=0.22\rho$ in Eq. (\ref{Hamiltonian_d}).}\label{phase_diagram}
\end{figure}

\begin{figure}[h]
	\centering
	\includegraphics[width=0.45\textwidth]{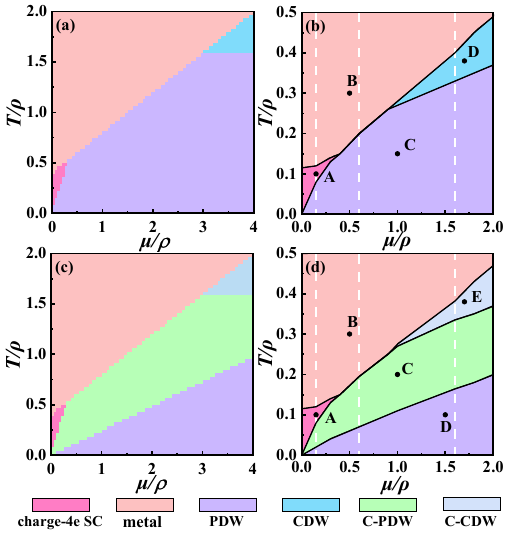}
	\caption{(Color online) Phase diagrams provided by (a,c) the RG study and (b,d) the MC study, with (a)-(b) for $n=2$ and (c)-(d) for $n=5$ (the $2Q$ PDW state). The white dashed lines in (b,d) mark $\mu/\rho=0.15, 0.6$ and $1.6$, respectively. The initial values of the coupling parameters in (a,c) are $g_{\theta}=0.1$, $g_{x}=g_{y}=g_{\frac{1}{2}}^{x}=g_{\frac{1}{2}}^{y}=0.1$, $g_\alpha=0.1$ in Eq. (\ref{eqn:action-SG}), and in (b,d) are $A=0.02\rho$ in Eq. (\ref{Hamiltonian_d4Q}).}\label{phase_diagram_2Q}
\end{figure}

We present the phase diagrams for the $3Q$ and $2Q$ PDW states in Fig.~\ref{phase_diagram} and Fig.~\ref{phase_diagram_2Q}, respectively. Specifically, panels (a) and (c) of Fig.~\ref{phase_diagram}, along with panels (a) and (c) of Fig.~\ref{phase_diagram_2Q}, are obtained using the RG approach, while panels (b) and (d) in both figures are derived from the MC simulations. All these results consistently support the schematic phase diagram Fig.~\ref{schematicphasediagram} illustrated in the main text.

\section{More details Results about the MC study for the hexagonal $3Q$ PDW}

We can discretize the Hamiltonian (\ref{Hamiltonian_r}) on the square lattice to employ the Monte-Carlo (MC) study. For the $3Q$ PDW, we choose $\theta_{1}, \theta_{2}, \theta^{'}_{3}$ as the dynamical variables, which suggests $\theta = \theta_{1} +\theta_{2} -\theta_{3}^{'}, \phi_{1}=\theta_{3}^{'} - \theta_{2}, \phi_{2}=\theta_{3}^{'} - \theta_{1}$, and $\phi_3 = \theta_{1} +\theta_{2} -2\theta_{3}^{'}$. The discretized Hamiltonian is
\begin{eqnarray}\label{Hamiltonian_d}
H &=& -\rho\sum_{\langle ij\rangle} \cos(\theta(\mathbf{r}_{i})-\theta(\mathbf{r}_{j})) - \mu \sum_{\langle ij\rangle, \alpha} \cos(\phi_{\alpha}(\mathbf{r}_{i})-\phi_{\alpha}(\mathbf{r}_{j}))+ A\sum_{i, \alpha} \cos(2n\phi_{\alpha}(\mathbf{r}_{i})).
\end{eqnarray}
The positive coefficients $\rho, \mu$ ensure the discretized Hamiltonian~(\ref{Hamiltonian_d}) match the continuous Hamiltonian~(\ref{Hamiltonian_r}) in the thermodynamic limit.

Observables are calculated based on the following formulas.
The specific heat is defined as
\begin{eqnarray}\label{Cv2}
C_v=\dfrac{\left\langle H^2\right\rangle -\left\langle H\right\rangle^2}{NT^2},
\end{eqnarray}
where $N$ is the site number.

The phase stiffness characterizing the quasi-long-range order of the $\theta$ field and superconducting phase is \cite{Zeng2024}
\begin{eqnarray}\label{ss}
	S=\frac{1}{N}(<H_x>-\beta<I_x^2>)
\end{eqnarray}
with
\begin{eqnarray}\label{Hamiltonian_p1}
H_{x} &=& \rho\sum_{\langle ij\rangle_x} \cos(\theta(\mathbf{r}_{i})-\theta(\mathbf{r}_{j})),  \nonumber\\
I_{x} &=& \rho\sum_{\langle ij\rangle_x} \sin(\theta(\mathbf{r}_{i})-\theta(\mathbf{r}_{j})),  
\end{eqnarray}
where $\beta=1/k_BT$.

The Ising order parameter characterizing the $\phi_{\alpha}$ field ordering is,
\begin{eqnarray}
I_{\alpha}\equiv\frac{1}{N^2}\sum_{ij}\left\langle\sin(\phi_{\alpha}(\mathbf{r}_{i}))\cdot \sin(\phi_{\alpha}(\mathbf{r}_{j}))\right\rangle.
\end{eqnarray}

The susceptibility $\chi$ and Binder cumulant $U$ of $\theta$ and $\phi_{\alpha}$ fields are given as~\cite{Challa1986}
\begin{eqnarray}\label{Cv2}
\chi=\dfrac{N(\left\langle m^2\right\rangle-\left\langle m\right\rangle^2)}{K_BT},~~~~ U=1-\dfrac{\left\langle m^4\right\rangle}{3\left\langle m^2\right\rangle^2},
\end{eqnarray}
where $m_{\theta}=\frac{1}{N}\sum_ie^{i \theta(\mathbf{r}_{i})}$ for the $\theta$-field or $m_{\phi_{\alpha}}=\frac{1}{N}\sum_ie^{i \phi_{\alpha}(\mathbf{r}_{i})}$ for the $\phi_{\alpha}$-field.

The $\theta$ and $\phi_{\alpha}$ fields correlation functions are defined as
\begin{eqnarray}
\eta_{\theta}(\Delta \mathbf{r})&=&\frac{1}{N}\sum_{\mathbf{r}}\left\langle e^{i[\theta(\mathbf{r})- \theta(\mathbf{r}+\Delta \mathbf{r})]}\right\rangle,  \nonumber\\
\eta_{\phi_{\alpha}}(\Delta \mathbf{r})&=&\frac{1}{N}\sum_{\mathbf{r}}\left\langle e^{i[\phi_{\alpha}(\mathbf{r})-\phi_{\alpha}(\mathbf{r}+\Delta \mathbf{r})]}\right\rangle.   
\end{eqnarray}

\subsection{$n=2$}

\begin{figure}[h]
	\centering
	\includegraphics[width=0.45\textwidth]{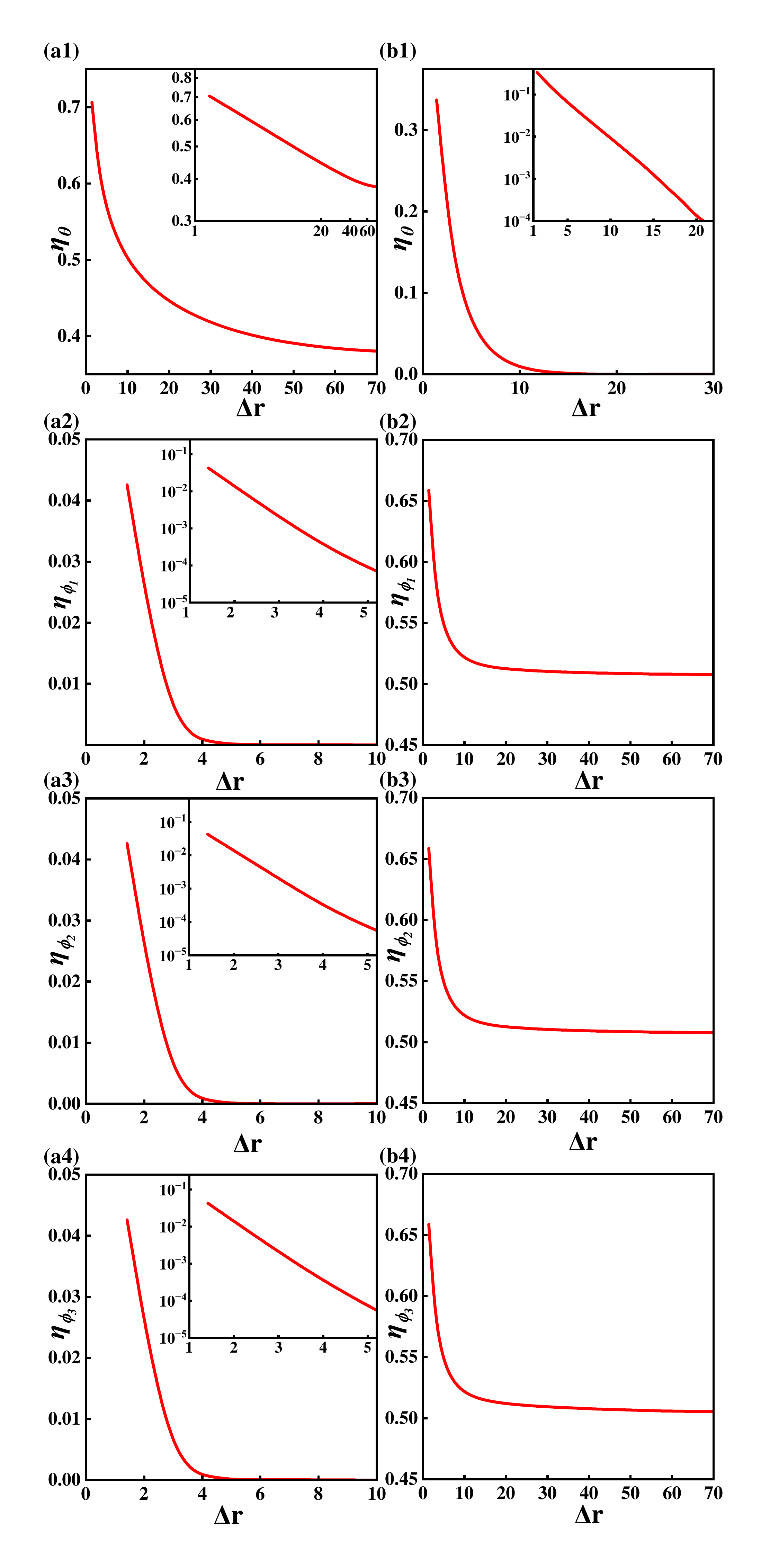}
	\caption{(Color online) The correlation function $\eta_{\theta/\phi}$ for (a1-a4) for the point $\mathbf{A}$ ($\mu=0.2\rho, T=0.8\rho$), for (b1-b4) for the point $\mathbf{D}$ ($\mu=0.9\rho, T=1.3\rho$) marked in Fig.~\ref{phase_diagram}(b). Insets of (a1) the log-log plot, and (a2-a4, b1) only the y-axis is logarithmic.}\label{ADall}
\end{figure}

\begin{figure}[h]
	\centering
	\includegraphics[width=0.45\textwidth]{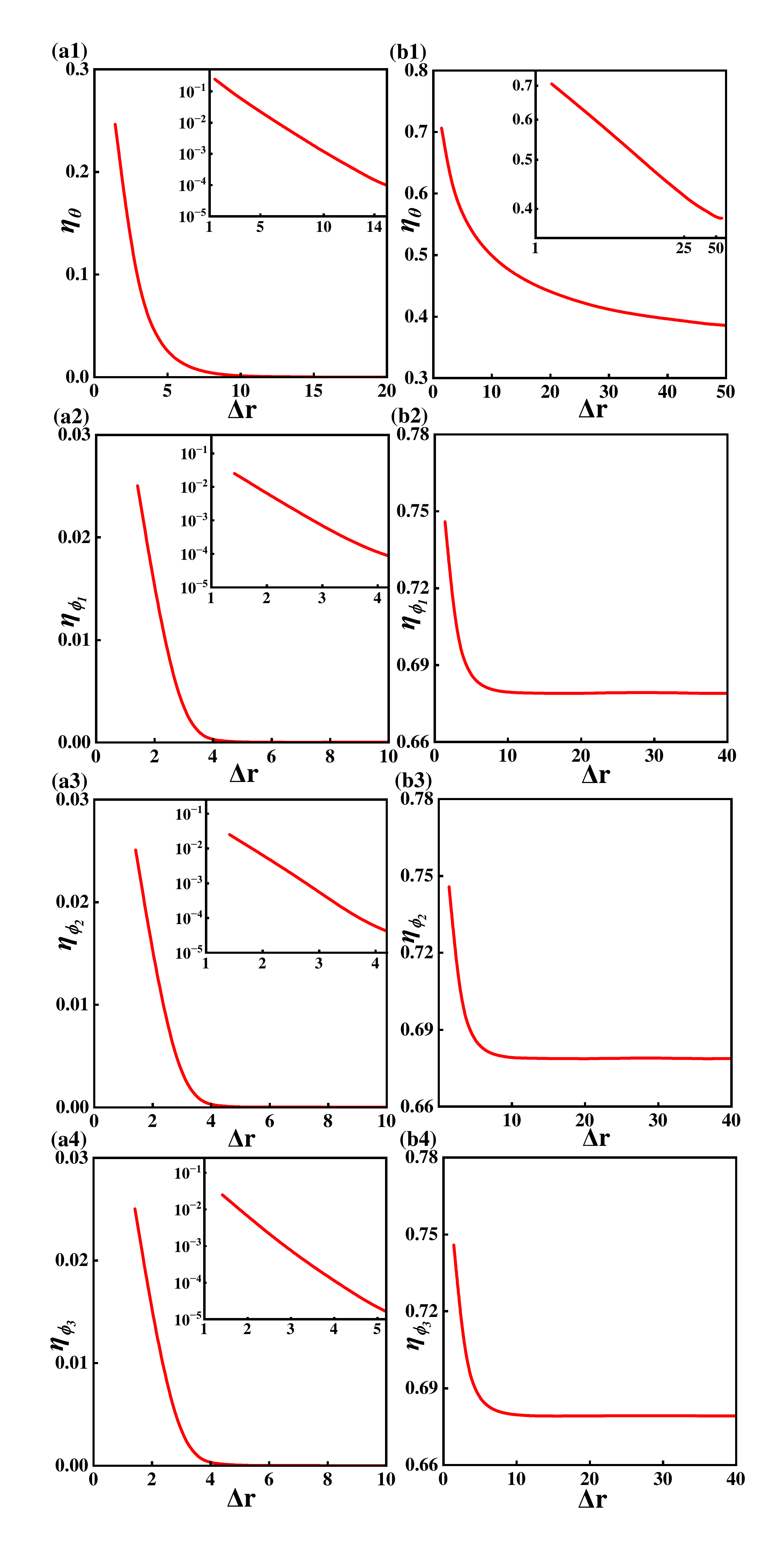}
	\caption{(Color online) The correlation function $\eta_{\theta/\phi}$ for (a1-a4) for the point $\mathbf{B}$ ($\mu=0.3\rho, T=1.5\rho$), for (b1-b4) for the point $\mathbf{C}$ ($\mu=0.63\rho, T=0.6\rho$) marked in Fig.~\ref{phase_diagram}(b). Insets of (a1-a4) only the y-axis is logarithmic, (b1) the log-log plot.}\label{BC}
\end{figure}

The phase diagram Fig.~\ref{phase_diagram}(b) is determined based on the decay characteristics of correlation functions $\eta_{\theta/\phi_{\alpha}}$, as systematically summarized in Table~\ref{tab:3}. In the main text, we present the $\eta_{\theta/\phi_{1}}$ for the representative A (2e SC) and D (CDW) points marked in the MC phase diagram Fig.~\ref{phase_diagram}(b), and their decaying behaviors are consistent with Table~\ref{tab:3}. As supplements, Fig.~\ref{ADall}(a1)-(a4) show $\Delta r$ ($\equiv |\Delta \mathbf{r}|$)-dependence of $\eta_{\theta}$ and $\eta_{\phi_{\alpha}} (\alpha=1,2,3)$ for the typical point A marked in Fig.~\ref{phase_diagram}(b): while $\eta_{\theta}$ shows a power-law decay, $\eta_{\phi_{\alpha}}$ decays exponentially, which is characteristic of the charge-$2e$ superconducting phase. For the typical point D indicated in Fig.~\ref{phase_diagram}(b), as shown in Fig.~\ref{ADall}(b1)--(b4), $\eta_{\theta}$ decays exponentially with $\Delta r$, whereas $\eta_{\phi_{\alpha}}$ saturates to a finite value as $\Delta r \to \infty$, reflecting CDW order. At point B, presented in Fig.~\ref{BC}(a1)--(a4), both $\eta_{\theta}$ and $\eta_{\phi_{\alpha}}$ exhibit exponential decay with $\Delta r$, indicating the MT phase. Finally, at point C, shown in Fig.~\ref{BC}(b1)--(b4), $\eta_{\theta}$ decays power-law with $\Delta r$, and $\eta_{\phi_{\alpha}}$ saturates to a finite value as $\Delta r \to \infty$, consistent with PDW behavior.

\begin{figure}[h]
	\centering
	\includegraphics[width=0.5\textwidth]{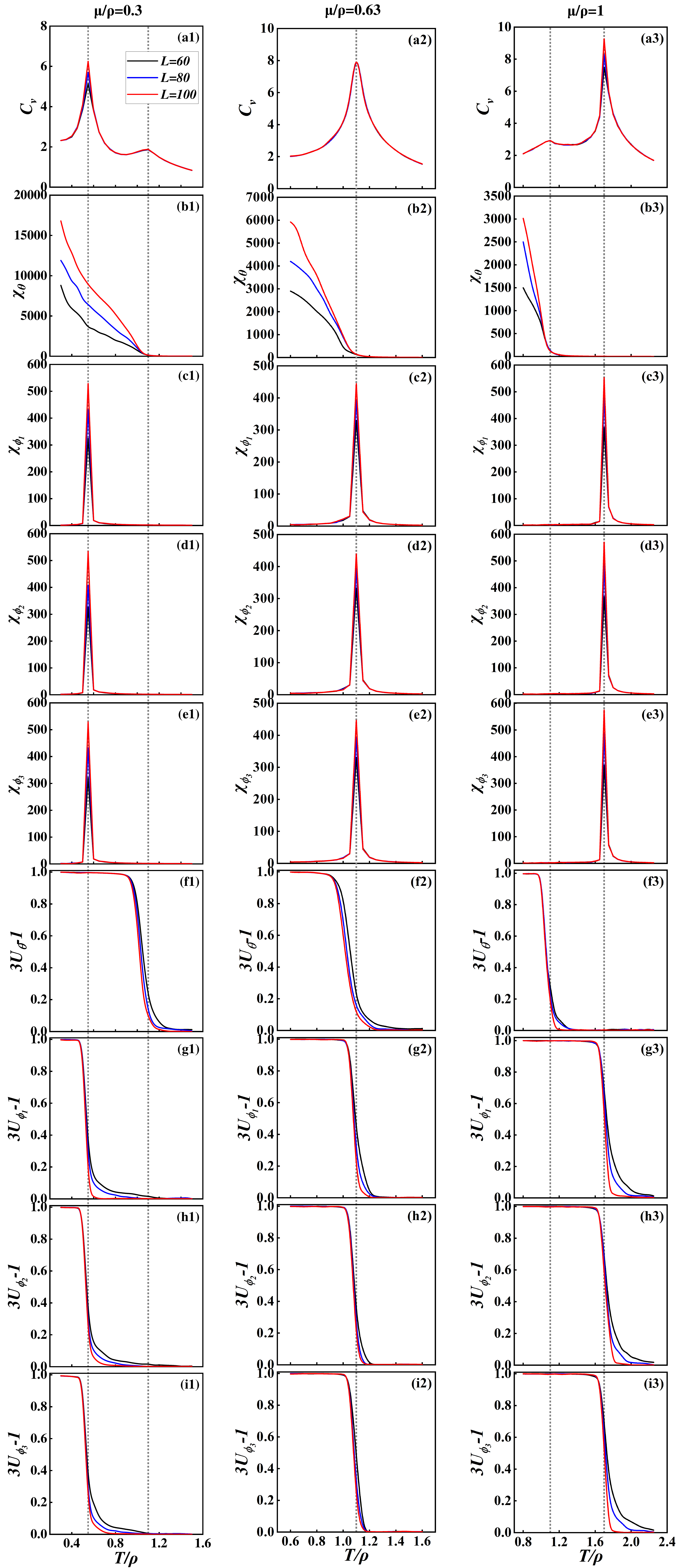}
	\caption{(Color online) Temperature evolution of thermodynamic quantities for $\mu/\rho=0.3$ (a1,b1,...,i1), $\mu/\rho=0.63$ (a2,b2,...,i2) and $\mu/\rho=1$ (a3,b3,...,i3) for $n=2$ (the $3Q$ PDW state). The scaling in all figures is $L=$ 60 (black line), 80 (blue line), and 100 (red line). (a1-a3) The specific heat $C_v$. (b1-b3) The susceptibilities $\chi_{\theta}$ of $\theta$. (c1-c3) The susceptibilities $\chi_{\phi_{1}}$ of $\phi_{1}$. (d1-d3) The susceptibilities $\chi_{\phi_{2}}$ of $\phi_{2}$. (e1-e3) The susceptibilities $\chi_{\phi_{3}}$ of $\phi_{3}$. (f1-f3) $3U_{\theta}-1$, where $U_{\theta}$ is the Binder cumulant of the $\theta$-field. (g1-g3) $3U_{\phi_{1}}-1$, where $U_{\phi_{1}}$ is the Binder cumulant of the $\phi_{1}$-field. (h1-h3) $3U_{\phi_{2}}-1$, where $U_{\phi_{2}}$ is the Binder cumulant of the $\phi_{2}$-field. (i1-i3) $3U_{\phi_{3}}-1$, where $U_{\phi_{3}}$ is the Binder cumulant of the $\phi_{3}$-field. The grey dotted lines in (a1)-(j3) mark the phase transitions.}\label{odp3}
\end{figure}

\begin{figure}[h]
	\centering
	\includegraphics[width=0.7\textwidth]{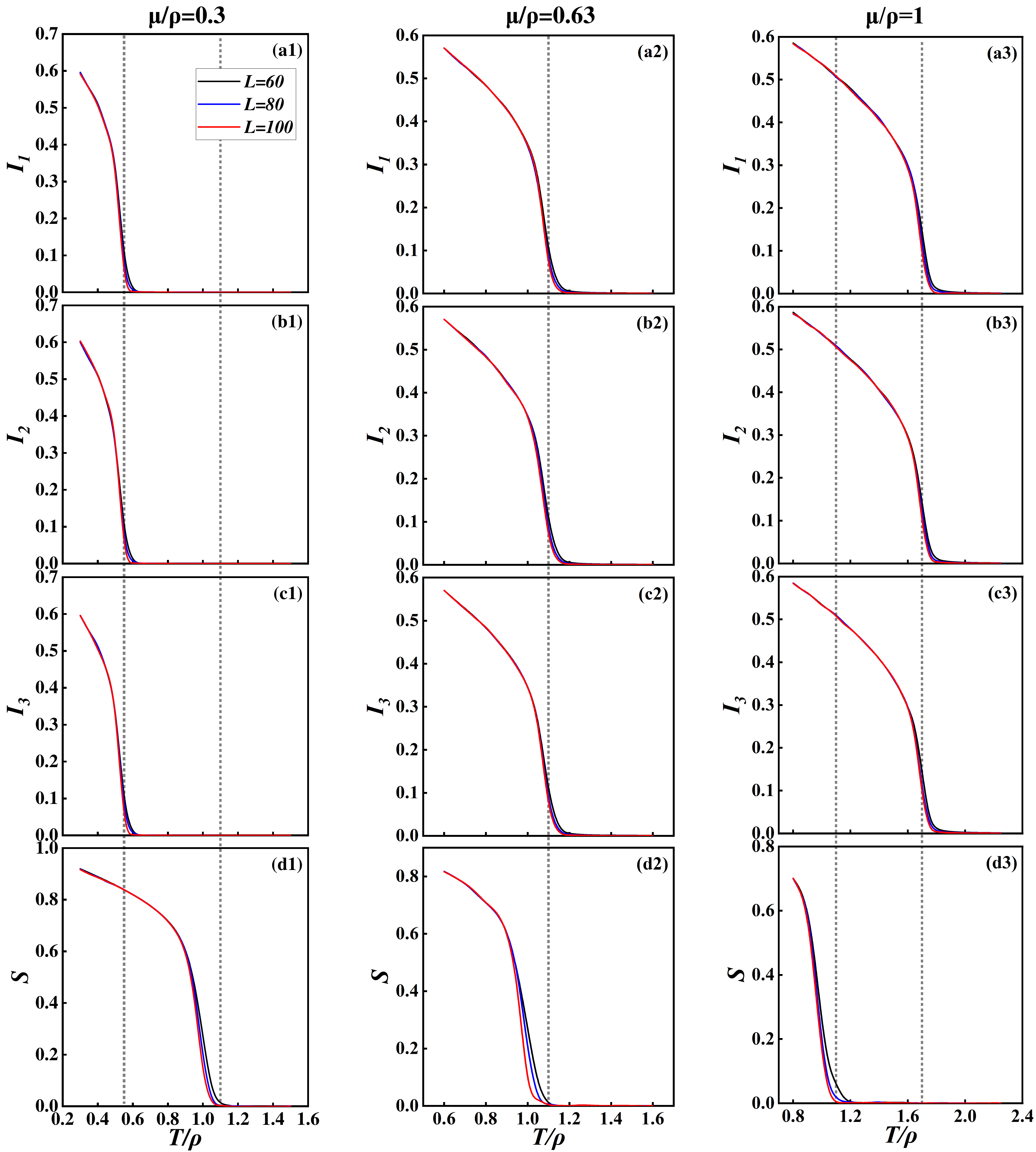}
	\caption{(Color online) Temperature evolution of thermodynamic quantities for $\mu/\rho=0.3$ (a1,b1,c1,d1), $\mu/\rho=0.63$ (a2,b2,c2,d2) and $\mu/\rho=1$ (a3,b3,c3,d3) for $n=2$ (the $3Q$ PDW state). The scaling in all figures is $L=$ 60 (black line), 80 (blue line), and 100 (red line). (a1-a3) Ising order parameter $\mathit{I_{1}}$ of $\phi_{1}$. (b1-b3) Ising order parameter $\mathit{I_{2}}$ of $\phi_{2}$. (c1-c3) Ising order parameter $\mathit{I_{3}}$ of $\phi_{3}$. (d1-d3) The phase stiffness $S$ of $\theta$. The grey dotted lines in (a1)-(d3) mark the phase transitions.}\label{odp4}
\end{figure}

Figures~\ref{odp3} and~\ref{odp4} display the temperature dependence of thermodynamic quantities for various lattice sizes at $\mu/\rho=0.3,0.63$ and $1$. More specifically, Fig.~\ref{odp3}(a1-a3) display the specific heat $C_v$, Fig.~\ref{odp3}(b1-b3), (c1-c3), (d1-d3) and (e1-e3) display the susceptibility $\chi_\theta$ and the susceptibility $\chi_{\phi_{a}}$, Fig.~\ref{odp3}(f1-f3), (g1-g3), (h1-h3) and (i1-i3) display the Binder cumulant $3U_\theta-1$ and $3U_{\phi_{\alpha}}-1$, Fig.~\ref{odp4}(a1-a3), (b1-b3) and (c1-c3) display the Ising order parameter $I_{\alpha}$ of $\phi_{\alpha}$, Fig.~\ref{odp4}(d1-d3) display the stiffness $S$ of $\theta$, respectively.

For $\mu/\rho=0.3$ (Fig.~\ref{odp3}(a1,b1,...,i1) and Fig.~\ref{odp4}(a1,b1,c1,d1)), we observe two phase transitions. First, around $T/\rho\approx0.55$, the specific heat exhibits a sharp peak and diverges upon $L$, and the susceptibility $\chi_{\phi_{\alpha}}$ switches from finite to divergence, the Binder cumulant $3U_{\phi_{\alpha}}-1$ and the Ising order parameter $I_{\alpha}$ drop to zero, suggesting that an Ising phase transition occurs, in which the $\phi_{\alpha}$-field transitions from long-range order to disorder. The system enters the normal 2e-SC phase. Subsequently, around $T/\rho\approx1.1$, the specific heat exhibits a finite broad bump, the susceptibility $\chi_{\theta}$ changes from divergence to finite, the Binder cumulant $3U_{\theta}-1$ and the stiffness $S$ rapidly drop to zero, suggesting a BKT phase transition in which the $\theta$-field transitions from quasi-long-range order to disorder. The system reaches the MT phase.

For $\mu/\rho=0.63$ (Fig.~\ref{odp3}(a2,b2,...,i2) and Fig.~\ref{odp4}(a2,b2,c2,d2)), we observe one phase transition. Around $T/\rho\approx 1.1$, the specific heat displays a peak, the susceptibility $\chi_{\theta}$ evolves from divergence to finite, the Binder cumulant $3U_{\theta}-1$ and $3U_{\phi_{\alpha}}-1$, the Ising order parameter $I_{\alpha}$ and the stiffness $S$ drop to zero, and the susceptibility $\chi_{\phi_{\alpha}}$ changes from finite to divergence, implying both $\theta$- and $\phi_{\alpha}$-fields becoming disorder. The system enters the MT phase.

For $\mu/\rho=1$ (Fig.~\ref{odp3}(a3,b3,...,i3) and Fig.~\ref{odp4}(a3,b3,c3,d3)), we observe two phase transitions. First, around $T/\rho\approx1.1$, the specific heat is a finite broad bump, the susceptibility $\chi_{\theta}$ changes from divergent to finite, the Binder cumulant $3U_{\theta}-1$ and the stiffness $S$ rapidly drop to zero, indicting that a BKT phase transition takes place in which the $\theta$-field undergoes a transition from quasi-long-range to disorder. The system enters the CDW phase. Subsequently, around $T/\rho\approx1.7$, the specific heat exhibits a sharp peak and diverges upon $L$, the susceptibility $\chi_{\phi_{\alpha}}$ evolves from finite to divergence, the Binder cumulant $3U_{\phi_{\alpha}}-1$ and the Ising order parameter $I_{\alpha}$ rapidly drop to zero. These results suggest that an Ising phase transition in which the $\phi_{\alpha}$-field undergoes a transition from long-range order to disorder, driving the system into the MT phase.

\subsection{$n=5$}

\begin{figure}[h]
	\centering
	\includegraphics[width=0.45\textwidth]{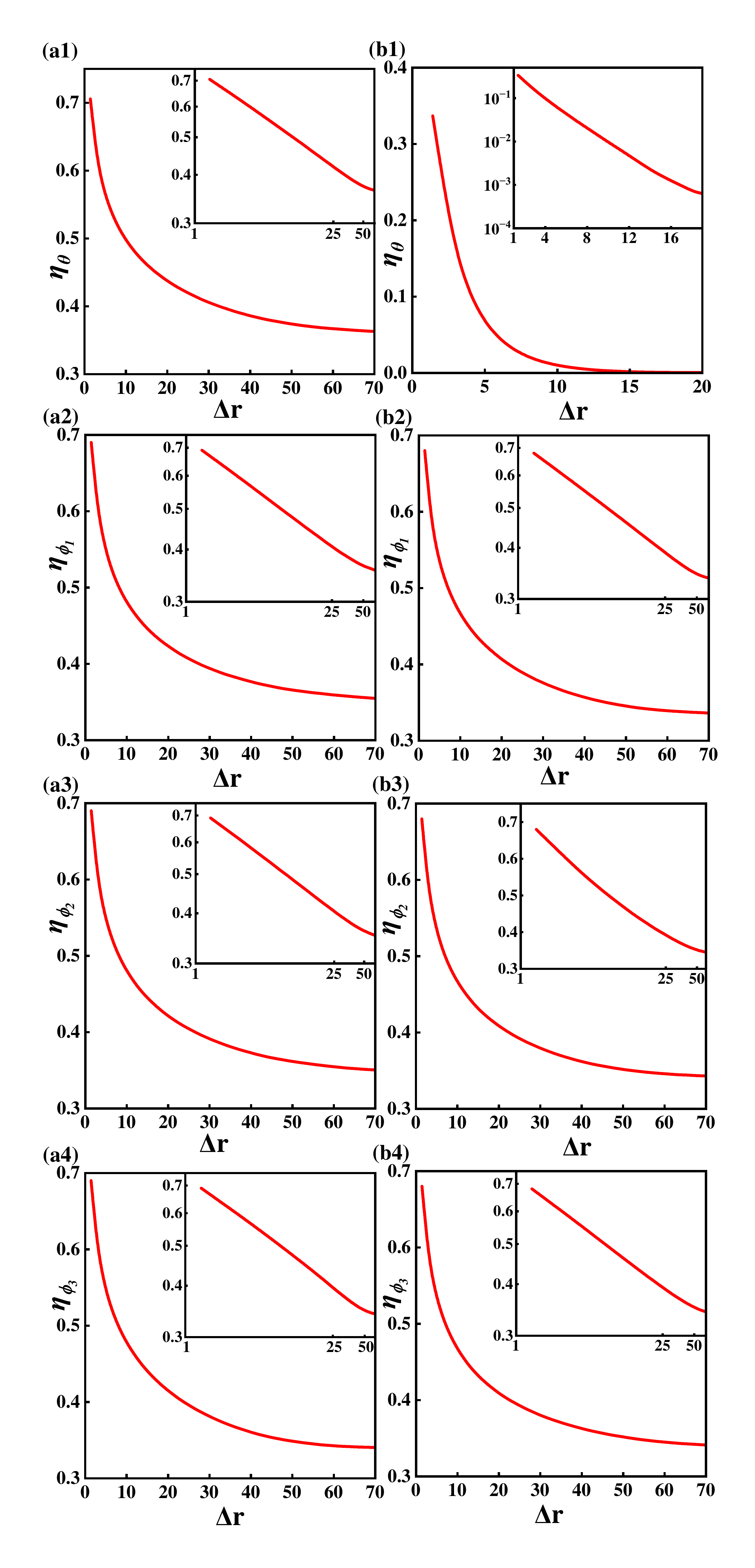}
	\caption{(Color online) The correlation function $\eta_{\theta/\phi}$ for (a1-a4) for the point $\mathbf{C}$ ($\mu=0.63\rho, T=0.8\rho$), for (b1-b4) for  the point $\mathbf{E}$ ($\mu=1\rho, T=1.3\rho$) marked in Fig.~\ref{phase_diagram}(d). Insets of (a1-a4) and (b2-b4) the log-log plot, and (b1) only the y-axis is logarithmic.}\label{CEall}
\end{figure}

\begin{figure}[h]
	\centering
	\includegraphics[width=0.7\textwidth]{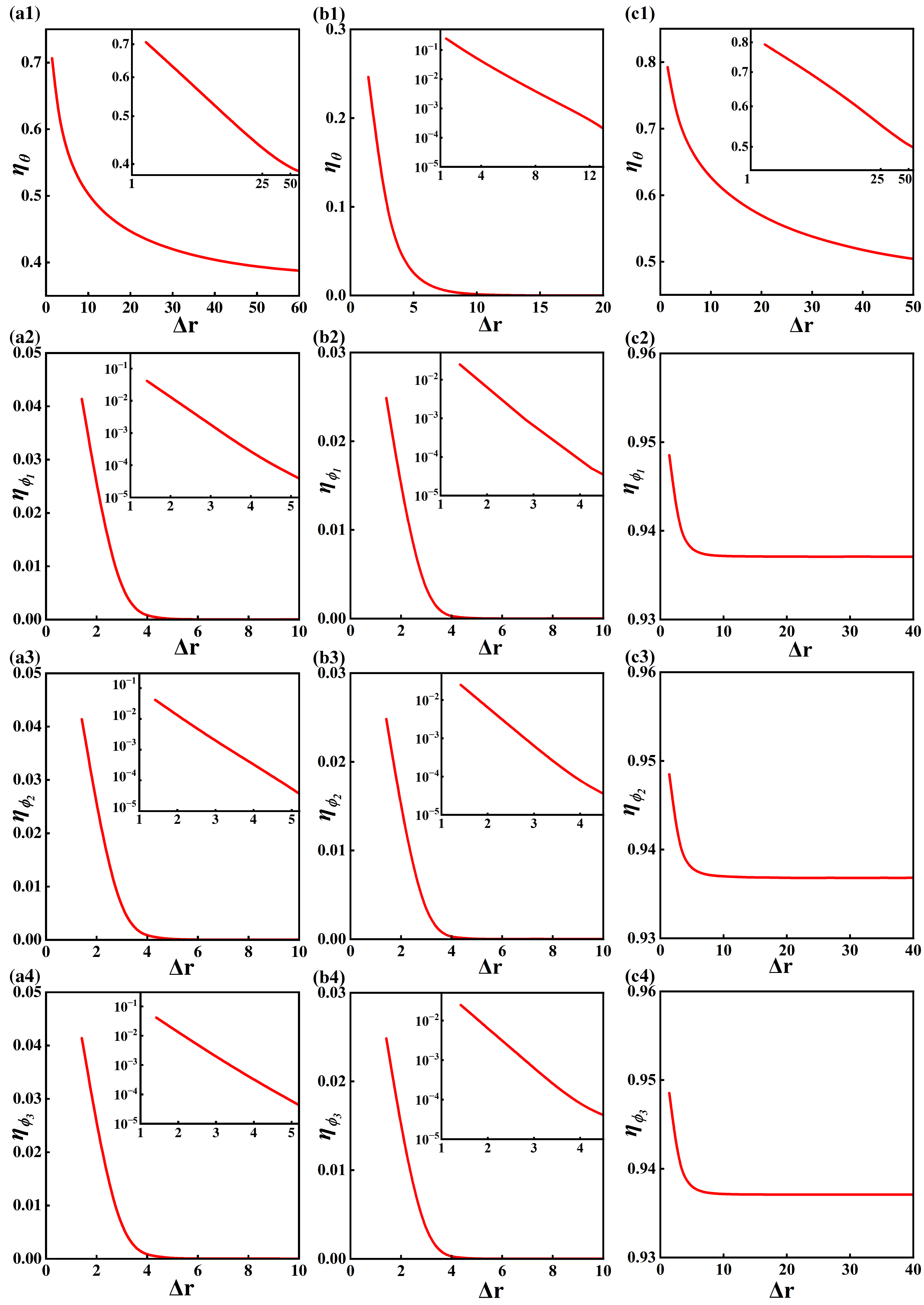}
	\caption{(Color online) The correlation function $\eta_{\theta/\phi}$ for (a1-a4) for the point $\mathbf{A}$ ($\mu=0.2\rho, T=0.8\rho $), for (b1-b4) for the point $\mathbf{B}$ ($\mu=0.3\rho, T=1.5\rho$), for (c1-c4) for the point $\mathbf{D}$ ($\mu=0.9\rho, T=0.3\rho$) marked in Fig.~\ref{phase_diagram}(d). Insets of (a1,c1) the log-log plot, and (a2-a4, b1-b4) only the y-axis is logarithmic.}\label{ABD}
\end{figure}

 The phase diagram Fig.~\ref{phase_diagram}(d) can be determined by analyzing the decaying behavior of the correlation functions $\eta_{\theta/\phi}$. As summarized in Table~\ref{tab:3}, each phase exhibits distinct decay patterns for $\eta_{\theta/\phi}$. The representative points C (C-PDW) and E (C-CDW) are highlighted in the MC phase diagram Fig.~\ref{phase_diagram}(d) in the main text and their decaying behaviors are consistent with the Table~\ref{tab:3}. Moreover, the spatial correlation functions $\eta_{\theta}$ and $\eta_{\phi_{\alpha}} (\alpha=1,2,3)$ are shown in Fig.~\ref{CEall}(a1)-(a4) for the typical point C marked in Fig.~\ref{phase_diagram}(d): both $\eta_{\theta}$ and $\eta_{\phi_{\alpha}}$ decay in power law with $\Delta r$, consistent with the C-PDW. Fig.~\ref{CEall}(b1)-(b4) are for the typical point E marked in Fig.~\ref{phase_diagram} (d): while $\eta_{\theta}$ decays exponentially with $\Delta r$, $\eta_{\phi_{\alpha}}$ decays in power law with $\Delta r$, consistent with the C-CDW. Additionally, Fig.~\ref{ABD}(a1)-(a4) show the spatial correlation functions $\eta_{\theta}$ and $\eta_{\phi_{\alpha}}$ for the typical point A marked in Fig.~\ref{phase_diagram}(d): while $\eta_{\theta}$ power-law decays with $\Delta r$, $\eta_{\phi_{\alpha}}$ decays exponentially with $\Delta r$, consistent with the 2e-SC. Fig.~\ref{ABD}(b1)-(b4) are for the typical point B marked in Fig.~\ref{phase_diagram}(d): both $\eta_{\theta}$ and $\eta_{\phi_{\alpha}}$ decay exponentially with $\Delta r$, consistent with the MT. Fig.~\ref{ABD}(c1)-(c4) are for the typical point D marked in Fig.~\ref{phase_diagram}(d): $\eta_{\theta}$ decays in power law with $\Delta r$, $\eta_{\phi_{\alpha}}$ saturates to a finite value as $\Delta r \to \infty$, consistent with the PDW.

\begin{figure}[h]
	\centering
	\includegraphics[width=0.43\textwidth]{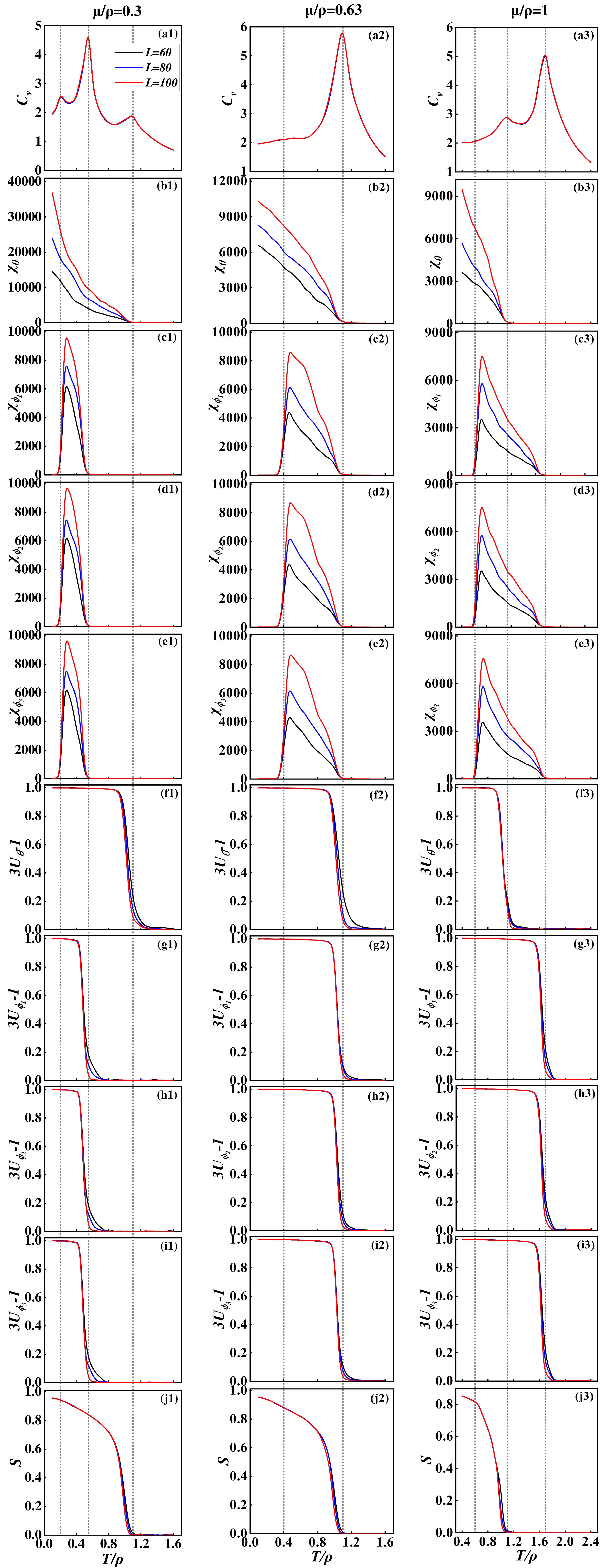}
	\caption{(Color online) Temperature evolution of thermodynamic quantities for $\mu/\rho=0.3$ (a1,b1,...,j1), $\mu/\rho=0.63$ (a2,b2,...,j2) and $\mu/\rho=1$ (a3,b3,...,j3) for $n=5$ (the $3Q$ PDW state). The scaling in all figures is $L=$ 60 (black line), 80 (blue line), and 100 (red line). (a1-a3) The specific heat $C_v$. (b1-b3) The susceptibilities $\chi_{\theta}$ of $\theta$. (c1-c3) The susceptibilities $\chi_{\phi_{1}}$ of $\phi_{1}$. (d1-d3) The susceptibilities $\chi_{\phi_{2}}$ of $\phi_{2}$. (e1-e3) The susceptibilities $\chi_{\phi_{3}}$ of $\phi_{3}$. (f1-f3) $3U_{\theta}-1$, where $U_{\theta}$ is the Binder cumulant of the $\theta$-field. (g1-g3) $3U_{\phi_{1}}-1$, where $U_{\phi_{1}}$ is the Binder cumulant of the $\phi_{1}$-field. (h1-h3) $3U_{\phi_{2}}-1$, where $U_{\phi_{2}}$ is the Binder cumulant of the $\phi_{2}$-field. (i1-i3) $3U_{\phi_{3}}-1$, where $U_{\phi_{3}}$ is the Binder cumulant of the $\phi_{3}$-field. (j1-j3) The phase stiffness $S$ of $\theta$. The grey dotted lines in (a1)-(j3) mark the phase transitions.}\label{odp5}
\end{figure}

We present detailed thermodynamic quantities for different lattice sizes at $\mu/\rho=0.3,0.63$ and $1$ in Fig.~\ref{odp5}, where: Fig.~\ref{odp5}(a1-a3) illustrate the specific heat $C_v$, Fig.~\ref{odp5}(b1-b3), (c1-c3), (d1-d3) and (e1-e3) illustrate the susceptibility $\chi_\theta$ and the susceptibility $\chi_{\phi_{\alpha}}$, Fig.~\ref{odp5}(f1-f3), (g1-g3), (h1-h3) and (i1-i3) illustrate the Binder cumulant $3U_\theta-1$ and $3U_{\phi_{\alpha}}-1$, Fig.~\ref{odp5}(j1-j3) illustrate the stiffness $S$ of $\theta$, respectively.

For $\mu/\rho=0.3$ (Fig.~\ref{odp5}(a1-j1)), we identify three phase transitions. The first occurs near $T/\rho\approx0.2$, where the specific heat exhibits a finite broad bump, and the susceptibility $\chi_{\phi_{\alpha}}$ evolves from finite to divergent—a signature of a BKT transition that drives the $\phi_{\alpha}$-field into quasi-long-range order and the system into the C-PDW phase. The second transition, around $T/\rho\approx0.55$, is marked by another finite broad bump in the specific heat, accompanied by the susceptibility $\chi_{\phi_{\alpha}}$ becoming finite and the cumulant $3U_{\phi_{\alpha}}-1$ sharply dropping to zero. These characteristics indicate that the $\phi_{\alpha}$-field transitions from quasi-long-range order to disorder and the system enters the normal 2e-SC phase. The third transition, near $T/\rho\approx1.1$, is characterized by a finite broad bump in specific heat, the susceptibility $\chi_\theta$ switching from divergence to finiteness, and the cumulant $3U_\theta-1$ and the stiffness $S$ collapsing to zero, signifying a BKT transition in the $\theta$-field and driving the system into the normal MT phase.

For $\mu/\rho=0.63$ (Fig.~\ref{odp5}(a2-j2)), we identify two phase transitions.
The first occurs near $T/\rho\approx0.4$, where the specific heat is very smooth, and the susceptibility $\chi_{\phi_{\alpha}}$ changes from finite to divergent—a signature of a BKT transition in which the $\phi_{\alpha}$-field transitions from long-range order to quasi-long-range order, driving the system into the C-PDW. The second transition, near $T/\rho\approx1.1$, where the specific heat presents a finite broad bump, the susceptibility $\chi_\theta$ and $\chi_{\phi_{\alpha}}$ changes from divergent to finite, and the cumulant $3U_\theta-1$, $3U_{\phi_{\alpha}}-1$ and the stiffness $S$ rapidly drop to zero. These results imply that a BKT phase transition occurs, in which the $\theta$- and $\phi_{\alpha}$-fields evolves from quasi-long-range order to disorder, thereby driving the system into the normal MT phase.

For $\mu/\rho=1$ (Fig.~\ref{odp5}(a3-j3)), we identify three phase transitions. The first occurs near $T/\rho\approx0.6$, where the specific heat is very smooth, and the susceptibility $\chi_{\phi_{\alpha}}$ changes from finite to divergent. These properties imply a BKT phase transition in which the $\phi_{\alpha}$-field transitions from long-range order to quasi-long-range order. The system enters the C-PDW phase. The second transition, near $T/\rho\approx1.1$, where the specific heat presents a finite broad bump, the susceptibility $\chi_\theta$ changes from divergent to finite, the Binder cumulant $3U_\theta-1$ and the stiffness $S$ rapidly drop to zero. These properties indicate a BKT phase transition which the $\theta$-field experiences a BKT phase transition from quasi-long-range order to disorder, driving the system into the C-CDW phase. The third transition, near $T/\rho\approx1.7$, where the specific heat presents a finite broad bump, the susceptibility $\chi_{\phi_{\alpha}}$ changes from divergent to finite, and the Binder cumulant $3U_{\phi_{\alpha}}-1$ rapidly drops to zero. These properties suggest a BKT phase transition takes place, in which the $\phi_{\alpha}$-field evolves from quasi-long-range order to disorder, driving the system into the normal MT phase.

\section{More details Results about the MC study for the $2Q$ PDW}

For the $2Q$ PDW, we have $\theta=\frac{\theta_1+\theta^{'}_1}{2}=\frac{\theta_2+\theta^{'}_2}{2}$, $\phi_1=\frac{\theta_1-\theta^{'}_1}{2}$ and $\phi_2=\frac{\theta_2-\theta^{'}_2}{2}$. We can use $\theta_1, \theta^{'}_1, \theta_2, \theta^{'}_2$ as the dynamical variables and the Hamiltonian is
\begin{eqnarray}\label{Hamiltonian_d4Q}
H &=& -\kappa\sum_{\langle ij\rangle, \alpha} \cos(\theta_\alpha(\mathbf{r}_{i})+\theta^{'}_\alpha(\mathbf{r}_{i})-\theta_\alpha(\mathbf{r}_{j})-\theta^{'}_\alpha(\mathbf{r}_{j}))  \nonumber\\
&&- \lambda \sum_{\langle ij\rangle, \alpha}  \cos(\theta_\alpha(\mathbf{r}_{i})-\theta^{'}_\alpha(\mathbf{r}_{i})-\theta_\alpha(\mathbf{r}_{j})+\theta^{'}_\alpha(\mathbf{r}_{j}))  \nonumber\\
&&- \gamma \sum_{\langle ij\rangle, \alpha} \cos(\theta_{\alpha}(\mathbf{r}_{i})-\theta_{\alpha}(\mathbf{r}_{j}))+\cos(\theta^{'}_{\alpha}(\mathbf{r}_{i})-\theta^{'}_{\alpha}(\mathbf{r}_{j})) \nonumber\\
&&+ A\sum_{i, \alpha} \cos(n\theta_{\alpha}(\mathbf{r}_{i})-n\theta^{'}_{\alpha}(\mathbf{r}_{i})).
\end{eqnarray}
Here $\langle ij\rangle$ represents nearest-neighbor bonding and the coefficients $\kappa$, $\lambda$ and $\gamma$ satisfy
\begin{eqnarray}\label{relation}
\kappa =\frac{\rho-4\gamma}{8},~~~~~~~~~~~ \lambda = \frac{\mu-2\gamma}{4}.
\end{eqnarray}
The positive coefficients $\kappa$, $\lambda$ and $\gamma$ ensure the discretized Hamiltonian~(\ref{Hamiltonian_d4Q}) match the continuous Hamiltonian~(\ref{Hamiltonian_r}) in the thermodynamic limit, respectively. 

The definition of specific heat is similar to the previous one. To access the superfluid response of the melting systems, we calculate the phase stiffness of the total-phase for the x-direction defined by

\begin{eqnarray}\label{ss}
	S_{x}=\frac{1}{N}(<H_{x}>-\beta<I_{x}^2>),
\end{eqnarray}
with
\begin{eqnarray}\label{Hamiltonian_p1x}
H_x &=& 4\kappa\sum_{<ij>_x, \alpha}\cos[\theta_{\alpha}(\mathbf{r}_i)+\theta^{'}_{\alpha}(\mathbf{r}_i)-\theta_{\alpha}(\mathbf{r}_j)-\theta^{'}_{\alpha}(\mathbf{r}_j)]\nonumber\\
&+&\gamma\sum_{<ij>_x,\alpha}\cos[\theta_{\alpha}(\mathbf{r}_{i})-\theta_{\alpha}(\mathbf{r}_{j})]+\cos[\theta^{'}_{\alpha}(\mathbf{r}_{i})-\theta^{'}_{\alpha}(\mathbf{r}_{j})], \nonumber\\
I_x &=& 2\kappa\sum_{<ij>_x,\alpha}\sin[\theta_{\alpha}(\mathbf{r}_i)+\theta^{'}_{\alpha}(\mathbf{r}_i)-\theta_{\alpha}(\mathbf{r}_j)-\theta^{'}_{\alpha}(\mathbf{r}_j)]\nonumber\\
&+&\gamma\sum_{<ij>_x,\alpha}\sin[\theta_{\alpha}(\mathbf{r}_{i})-\theta_{\alpha}(\mathbf{r}_{j})]+\sin[\theta^{'}_{\alpha}(\mathbf{r}_{i})-\theta^{'}_{\alpha}(\mathbf{r}_{j}) ].\nonumber\\
\end{eqnarray}

To characterize the relative-phase ordering, the Ising order parameter is,
\begin{equation}
I_\alpha \equiv\frac{1}{N^2}\sum_{ij}\left\langle\sin[\theta_{\alpha}(\mathbf{r}_{i})-\theta^{'}_{\alpha}(\mathbf{r}_{i})]\cdot \sin[\theta_{\alpha}(\mathbf{r}_{j})-\theta^{'}_{\alpha}(\mathbf{r}_{j})]\right\rangle.
\end{equation}

The susceptibility $\chi$ and Binder cumulant $U$ of $\theta$ and $\phi_{\alpha}$ are given as
\begin{eqnarray}\label{Cv2}
\chi=\dfrac{N(\left\langle m^2\right\rangle-\left\langle m\right\rangle^2)}{K_BT},~~~~ U=1-\dfrac{\left\langle m^4\right\rangle}{3\left\langle m^2\right\rangle^2},
\end{eqnarray}
where $m_{\theta}=\frac{1}{N}\sum_ie^{i(\theta_1(\mathbf{r}_{i})+\theta^{'}_1(\mathbf{r}_{i}))}$ for the $\theta$-field or $m_{\phi_\alpha}=\frac{1}{N}\sum_ie^{i(\theta_\alpha(\mathbf{r}_{i})-\theta^{'}_\alpha(\mathbf{r}_{i}))}$ for the $\phi_{\alpha}$-field. 

The $\theta$ and $\phi_{\alpha}$ fields correlation functions are defined as
\begin{eqnarray}
\eta_{\theta}(\Delta \mathbf{r})\equiv\frac{1}{N}\sum_{\mathbf{r}}\left\langle e^{i[\theta_1(\mathbf{r})+ \theta^{'}_1(\mathbf{r})-\theta_1(\mathbf{r}+\Delta \mathbf{r}) -\theta^{'}_1(\mathbf{r}+\Delta \mathbf{r})]}\right\rangle,  \nonumber\\
\eta_{\phi_\alpha}(\Delta \mathbf{r})\equiv\frac{1}{N}\sum_{\mathbf{r}}\left\langle e^{i[\theta_\alpha(\mathbf{r}) - \theta^{'}_\alpha(\mathbf{r})-\theta_\alpha(\mathbf{r}+\Delta \mathbf{r}) + \theta^{'}_\alpha(\mathbf{r}+\Delta \mathbf{r})]}\right\rangle. \nonumber\\
\end{eqnarray}

\subsection{$n=2$}

\begin{figure}[h]
	\centering
	\includegraphics[width=0.7\textwidth]{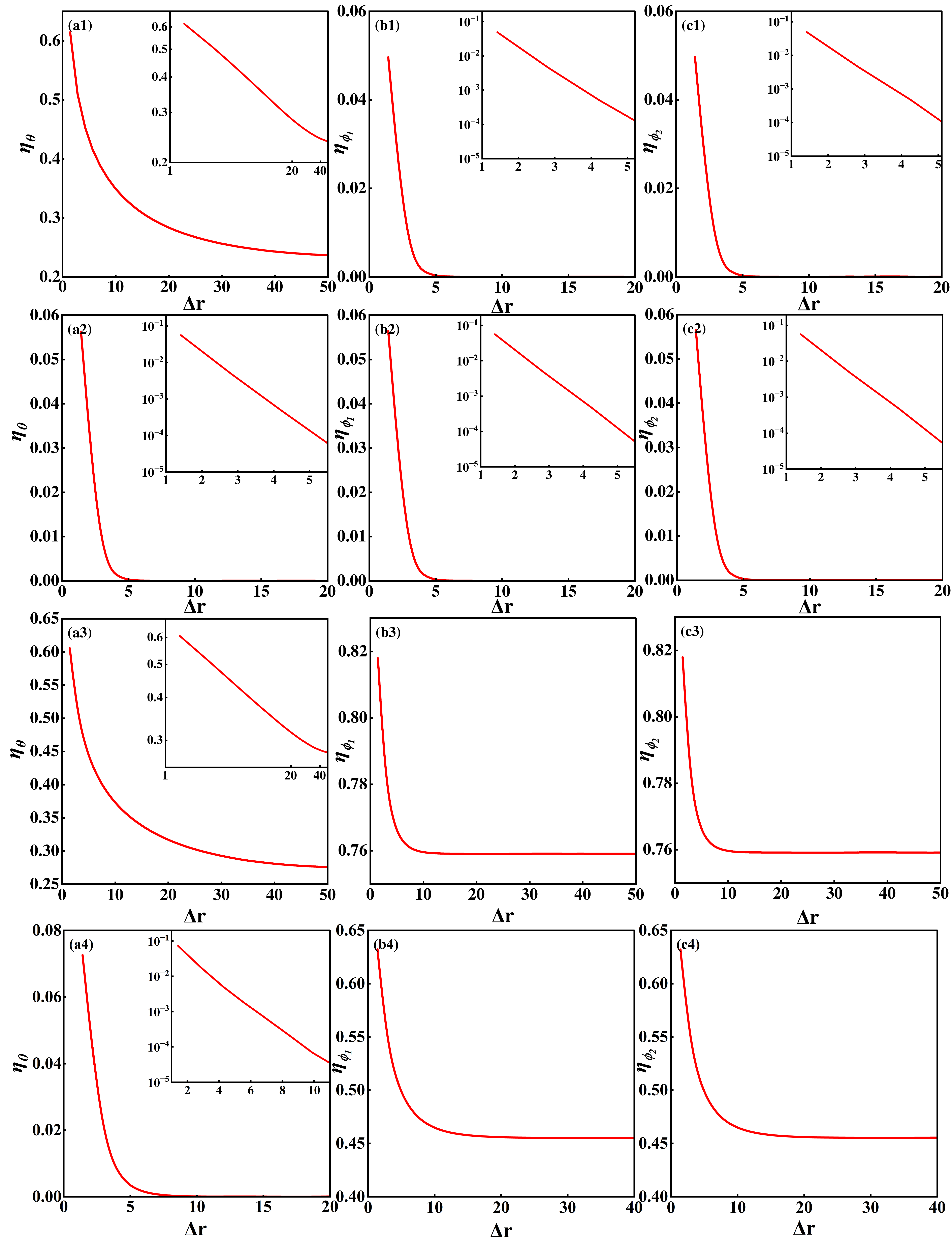}
	\caption{(Color online) The correlation function $\eta_{\theta/\phi_{\alpha}}$ for (a1)-(c1) for the point $\mathbf{A}$ ($\mu=0.15\rho, T=0.1\rho$), for (a2)-(c2) for the point $\mathbf{B}$ ($\mu=0.5\rho, T=0.3\rho$), for (a3)-(c3) for the point $\mathbf{C}$ ($\mu=1\rho, T=0.15\rho$), for (a4)-(c4) for the point $\mathbf{D}$ ($\mu=1.7\rho, T=0.38\rho$) labeled in Fig.~\ref{phase_diagram_2Q}(b). Insets of (a1) and (a3) the log-log plot, and (b1)-(c1), (a2)-(c2), (a4) only the y-axis is logarithmic.}\label{ABCD}
\end{figure}

The phase diagram in Fig.~\ref{phase_diagram_2Q}(b) is derived from the decay behavior of the correlation functions $\eta_{\theta/\phi_{\alpha}}$. At point A, indicated in Fig.~\ref{phase_diagram_2Q}(b), $\eta_{\theta}$ decays as a power law with $\Delta r$ ($\equiv |\Delta \mathbf{r}|$), whereas $\eta_{\phi_{\alpha}} (\alpha=1,2)$ exhibits exponential decay, consistent with a charge 4e superconducting (4e SC) state (Fig.~\ref{ABCD}(a1)-(c1)). In contrast, at point B, both $\eta_{\theta}$ and $\eta_{\phi_{\alpha}}$ decay exponentially (Fig.~\ref{ABCD}(a2)-(c2)), signaling a metallic (MT) phase. For point C, $\eta_{\theta}$ follows a power-law decay, while $\eta_{\phi_{\alpha}}$ saturates to a finite value at large $\Delta r$ (Fig.~\ref{ABCD}(a3)-(c3)), indicative of a pair-density-wave (PDW) state. Finally, at point D, $\eta_{\theta}$ shows exponential decay, but $\eta_{\phi_{\alpha}}$ saturates to a nonzero value (Fig.~\ref{ABCD}(a4)-(c4)), reflecting a charge-density-wave (CDW) phase.

\begin{figure}[h]
	\centering
	\includegraphics[width=0.45\textwidth]{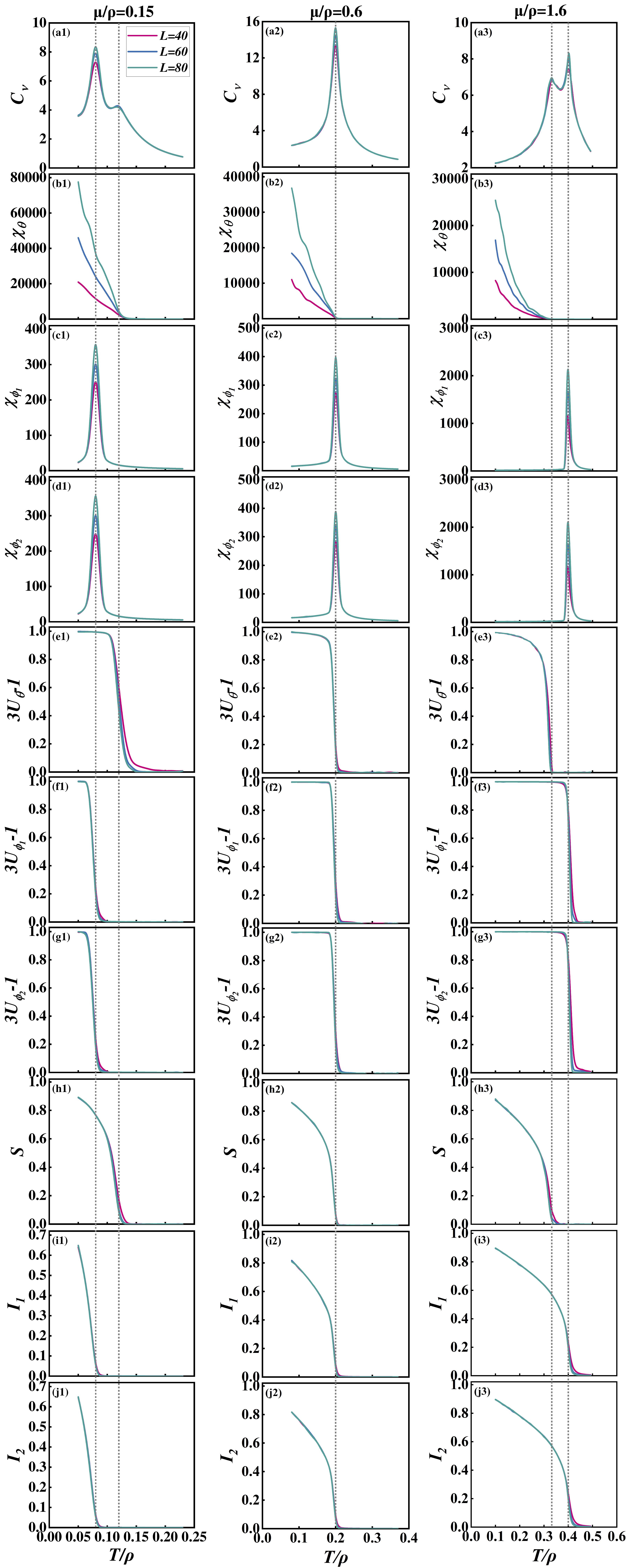}
	\caption{(Color online) The quantities as functions of temperature for $\mu/\rho=0.15$ (a1,b1,...,j1), $\mu/\rho=0.6$ (a2,b2,...,j2) and $\mu/\rho=1.6$ (a3,b3,...,j3) for $n=2$ (the $2Q$ PDW state). The scaling in all figures is $L=$ 40 (red line), 60 (blue line), and 80 (green line). (a1-a3) The specific heat $C_v$. (b1-b3) The susceptibilities $\chi_{\theta}$ of $\theta$. (c1-c3) The susceptibilities $\chi_{\phi_1}$ of $\phi_1$. (d1-d3) The susceptibilities $\chi_{\phi_2}$ of $\phi_2$. (e1-e3) $3U_{\theta}-1$, where $U_{\theta}$ is the Binder cumulant of the $\theta$-field. (f1-f3) $3U_{\phi_1}-1$, where $U_{\phi_1}$ is the Binder cumulant  of the $\phi_1$-field. (g1-g3) $3U_{\phi_2}-1$, where $U_{\phi_2}$ is the Binder cumulant  of the $\phi_2$-field. (h1-h3) The phase stiffness $S$ of $\theta$. (i1-i3) Ising order parameter $\mathit{I_1}$ of the $\phi_1$-field. (j1-j3) Ising order parameter $\mathit{I_2}$ of the $\phi_2$-field. The grey dotted lines represent the phase transitions in (a1)-(j3).}\label{ob_2Q_2}
\end{figure}

The temperature evolution of thermodynamic properties across different system sizes is shown in Fig.~\ref{ob_2Q_2} for $\mu/\rho=0.15,0.6$ and $1.6$. More specifically, Fig.~\ref{ob_2Q_2}(a1-a3) display the specific heat $C_v$, Fig.~\ref{ob_2Q_2}(b1-b3), (c1-c3), (d1-d3) display the susceptibility $\chi_\theta$ and the susceptibility $\chi_{\phi_{a}}$, Fig.~\ref{ob_2Q_2}(e1-e3), (f1-f3), (g1-g3) display the Binder cumulant $3U_\theta-1$ and $3U_{\phi_{\alpha}}-1$, 
Fig.~\ref{ob_2Q_2}(h1-h3) display the stiffness $S$ of $\theta$, Fig.~\ref{ob_2Q_2}(i1-i3), (j1-j3) display the Ising order parameter $I_{\alpha}$ of $\phi_{\alpha}$, respectively.

For $\mu/\rho=0.15$ (Fig.~\ref{ob_2Q_2}(a1,b1,...,j1), the system exhibits two distinct transitions. The first, at $T/\rho \approx 0.08$, is marked by a sharp, size-divergent peak in specific heat, accompanied by a shift in $\chi_{\phi_{\alpha}}$ from finite to divergent values. Simultaneously, the Binder cumulant $3U_{\phi_{\alpha}}-1$ and the Ising order parameter $I_{\alpha}$ drop to zero, signaling an Ising transition where the $\phi_{\alpha}$-field loses long-range order, driving the system into the 4e-SC phase. At higher temperature ($T/\rho \approx 1.2$), a second transition occurs, characterized by a broad finite peak in specific heat, $\chi_{\theta}$ switching from divergence to finiteness, and the collapse of $3U_{\theta}-1$ and stiffness $S$ to zero. These features indicate a BKT transition disordering the $\theta$-field, ultimately leading to a metallic phase.

For $\mu/\rho=0.6$ (Fig.~\ref{ob_2Q_2}(a2,b2,...,j2), a single transition emerges near $T/\rho \approx 0.2$. The specific heat peaks while $\chi_{\theta}$ transitions from divergent to finite, and $\chi_{\phi_{\alpha}}$ changes from finite to divergent. Concurrently, $3U_{\theta}-1$, $3U_{\phi_{\alpha}}-1$, $I_{\alpha}$, and $S$ all drop to zero, implying simultaneous disordering of both $\theta$- and $\phi_{\alpha}$-fields and a direct transition to the metallic phase.

For $\mu/\rho=1.6$ (Fig.~\ref{ob_2Q_2}(a3,b3,...,j3), two transitions are observed. The first, at $T/\rho \approx 0.33$, features a broad finite specific heat peak, $\chi_{\theta}$ becoming finite, and $3U_{\theta}-1$ and $S$ dropping sharply to zero—a BKT transition that disorders the $\theta$-field and stabilizes the CDW phase. The second transition ($T/\rho \approx 0.4$) displays a sharp, size-divergent specific heat peak, $\chi_{\phi_{\alpha}}$ turning divergent, and $3U_{\phi_{\alpha}}-1$ and $I_{\alpha}$ collapsing to zero, confirming an Ising transition that disorders the $\phi_{\alpha}$-field and establishes the metallic state.

\subsection{$n=5$}

\begin{figure}[h]
	\centering
	\includegraphics[width=0.7\textwidth]{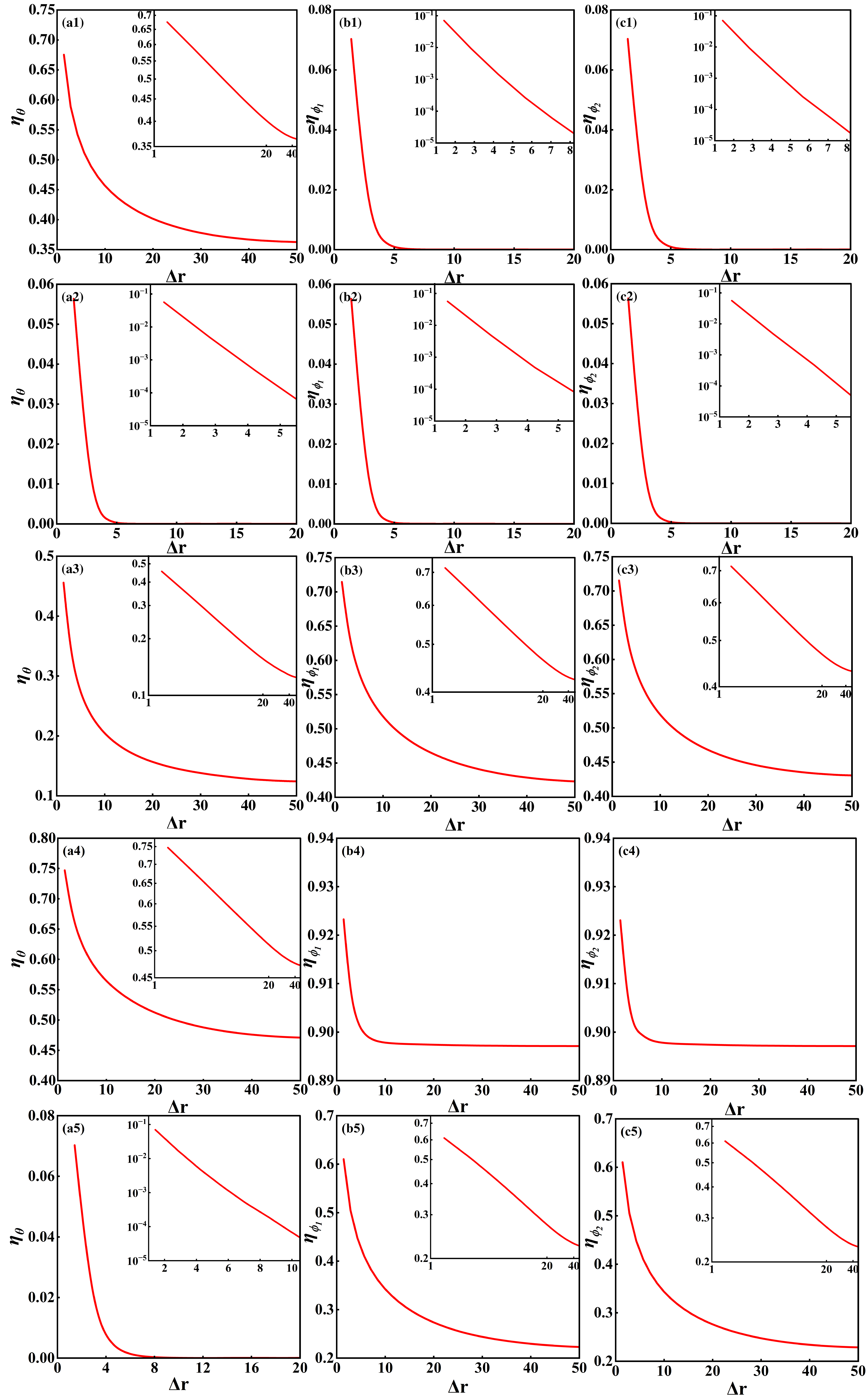}
	\caption{(Color online) The correlation function $\eta_{\theta/\phi_{\alpha}}$ for (a1)-(c1) for the point $\mathbf{A}$ ($\mu=0.15\rho, T=0.1\rho$), for (a2)-(c2) for the point $\mathbf{B}$ ($\mu=0.5\rho, T=0.3\rho$), for (a3)-(c3) for the point $\mathbf{C}$ ($\mu=1\rho, T=0.2\rho$), for (a4)-(c4) for the point $\mathbf{D}$ ($\mu=1.5\rho, T=0.1\rho$), for (a5)-(c5) for the point $\mathbf{E}$ ($\mu=1.7\rho, T=0.38\rho$) labeled in Fig.~\ref{phase_diagram_2Q}(d). Insets of (a1), (a3)-(c3), (a4), (b5)-(c5) the log-log plot, and (b1)-(c1), (a2)-(c2), (a5) only the y-axis is logarithmic.}\label{ABCDE}
\end{figure}

The phase diagram in Fig.~\ref{phase_diagram_2Q}(d) is determined by analyzing the spatial decay patterns in correlation functions $\eta_{\theta/\phi_{\alpha}}$. For point A (Fig.~\ref{ABCDE}(a1)-(c3)), the distinct behaviors of $\eta_{\theta}$ (power-law decay) and $\eta_{\theta}$ (exponential decay) with respect to $\Delta r$, consistent with the 4e-SC phase. For point B (Fig.~\ref{ABCDE}(a2)-(c2)), the exponential decay observed in both $\eta_{\theta}$ and $\eta_{\theta}$ correlations confirms the MT phase. The correlation function analysis reveals more complex behavior at point C (Fig.~\ref{ABCDE}(a3)-(c3))), where both $\eta_{\theta}$ and $\eta_{\phi_{\alpha}}$ exhibit power-law decay, consistent with the C-PDW phase. Different signatures emerge at point D (Fig.~\ref{ABCDE}(a4)-(c4))), with $\eta_{\theta}$ showing power-law decay while $\eta_{\phi_{\alpha}}$ approaches a finite value at large $\Delta r$, consistent with the PDW phase. Finally, point E (Fig.~\ref{ABCDE}(a5)-(c5))): exponential decay in $\eta_{\theta}$ contrasts with power-law decay in $\eta_{\phi_{\alpha}}$, consistent with the C-CDW phase.

\begin{figure}[h]
	\centering
	\includegraphics[width=0.55\textwidth]{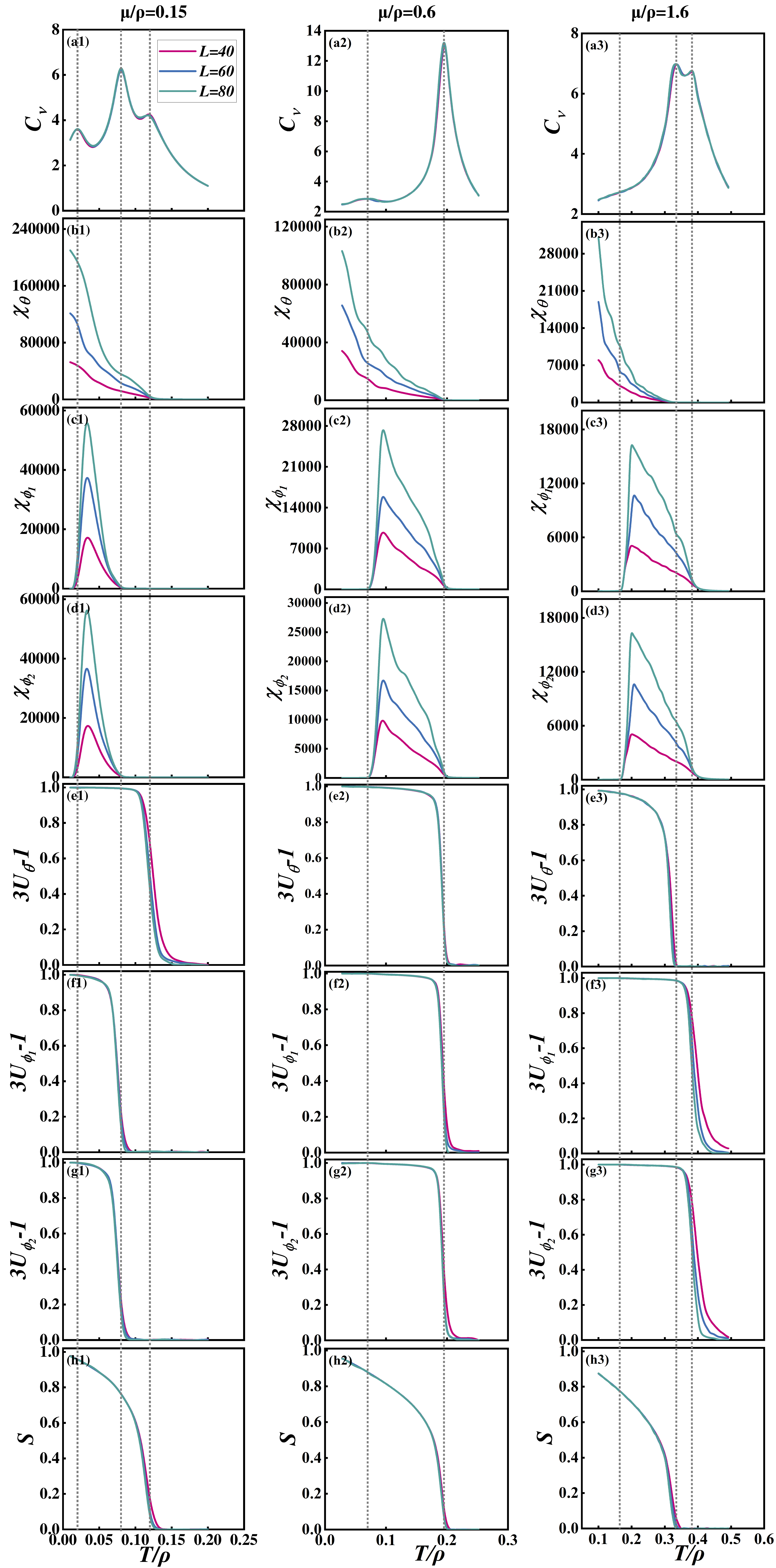}
	\caption{(Color online) The quantities as functions of temperature for $\mu/\rho=0.15$ (a1,b1,...,h1), $\mu/\rho=0.6$ (a2,b2,...,h2) and $\mu/\rho=1.6$ (a3,b3,...,h3) for $n=5$ (the $2Q$ PDW state). The scaling in all figures is $L=$ 40 (red line), 60 (blue line), and 80 (green line). (a1-a3) The specific heat $C_v$. (b1-b3) The susceptibilities $\chi_{\theta}$ of $\theta$. (c1-c3) The susceptibilities $\chi_{\phi_1}$ of $\phi_1$. (d1-d3) The susceptibilities $\chi_{\phi_2}$ of $\phi_2$. (e1-e3) $3U_{\theta}-1$, where $U_{\theta}$ is the Binder cumulant of the $\theta$-field. (f1-f3) $3U_{\phi_1}-1$, where $U_{\phi_1}$ is the Binder cumulant  of the $\phi_1$-field. (g1-g3) $3U_{\phi_2}-1$, where $U_{\phi_2}$ is the Binder cumulant  of the $\phi_2$-field. (h1-h3) The phase stiffness $S$ of $\theta$. The grey dotted lines represent the phase transitions in (a1)-(h3).}\label{ob_2Q_5}
\end{figure}

Figure~\ref{ob_2Q_5} displays thermodynamic data for varying lattice sizes at $\mu/\rho = 0.15$, $0.6$, and $1.6$. The specific heat $C_v$ is depicted in (a1–a3), while susceptibilities $\chi_\theta$ and $\chi_{\phi_{\alpha}}$ are shown in (b1–b3), (c1–c3), and (d1–d3). The Binder cumulants $3U_\theta-1$ and $3U_{\phi_{\alpha}}-1$ are plotted in (e1–e3), (f1–f3), and (g1–g3), with the stiffness $S$ of $\theta$ illustrated in (h1–h3).

For $\mu/\rho = 0.15$ (Fig.~\ref{ob_2Q_5}(a1–h1)): Three distinct phase transitions are observed. Near $T/\rho \approx 0.02$, the first transition is signaled by a broad finite peak in specific heat and a shift in $\chi_{\phi_{\alpha}}$ from finite to divergent values—a hallmark of a BKT transition that establishes quasi-long-range order in the $\phi_{\alpha}$-field. The system enters into the C-PDW phase. At $T/\rho \approx 0.08$, the second transition features another broad specific heat peak. The $\chi_{\phi_{\alpha}}$ becomes finite and $3U_{\phi_{\alpha}}-1$ drops to zero, marking the disordering of the $\phi_{\alpha}$-field and the system enters the 4e-SC phase. The third transition ($T/\rho \approx 0.12$) is characterized by a specific heat peak, $\chi_\theta$ transitioning from divergence to finiteness, and the collapse of $3U_\theta-1$ and $S$ to zero, indicating a BKT-driven disordering of the $\theta$-field and entry into the normal MT phase.

For $\mu/\rho = 0.6$ (Fig.~\ref{ob_2Q_5}(a2–h2)): Two transitions emerge. The first, at $T/\rho \approx 0.07$, exhibits a smooth specific heat curve and a divergent $\chi_{\phi_{\alpha}}$, consistent with a BKT transition where the $\phi_{\alpha}$-field decays from long-range to quasi-long-range order, stabilizing the C-PDW phase. The second transition ($T/\rho \approx 0.19$) shows a broad specific heat peak, finite $\chi_\theta$ and $\chi_{\phi_{\alpha}}$, and $3U_\theta-1$, $3U_{\phi_{\alpha}}-1$, and $S$ drops to zero, reflecting a BKT transition that disorders both $\theta$-and $\phi_{\alpha}$-fields, driving the system into the normal MT phase.

For $\mu/\rho = 1.6$ (Fig.~\ref{ob_2Q_5}(a3–h3)): Three transitions are identified. At $T/\rho \approx 0.16$, a smooth specific heat and divergent $\chi_{\phi_{\alpha}}$ suggest a BKT transition, with the $\phi_{\alpha}$-field losing long-range order to form the C-PDW phase. Near $T/\rho \approx 0.34$, a specific heat peak, finite $\chi_\theta$, and vanishing $3U_\theta-1$ and $S$ denote a BKT transition disordering the $\theta$-field, leading to the C-CDW phase. Finally, at $T/\rho \approx 0.38$, a specific heat peak, finite $\chi_{\phi_{\alpha}}$, and the decay of $3U_{\phi_{\alpha}}-1$ signify the $\phi_{\alpha}$-field’s disordering via BKT, transitioning the system to the normal MT phase.

\section{More details Results about the MC study for the unidirectional PDW}

The unidirectional commensurate PDW, which is described by two complex gap functions $\Delta_{\pm \mathbf{Q}_1}(\mathbf{r})$~\cite{Berg2009}. To simplify notation, we denote $\phi_1$ by $\phi$ in the following. To account for the anisotropy between x- and y-directions, \begin{eqnarray}\label{Hamiltonian_rs}
H_{0}=\int d^{2}\mathbf{r}\Big( \frac{\rho}{2}(|\nabla_{x}\theta|^{2}+\alpha_{1}|\nabla_{y}\theta|^{2}) +\frac{\mu}{2}(|\nabla_{x}\phi|^{2}+\alpha_{2}|\nabla_{y}\phi|^{2})+A \cos(2n\phi) \Big) .
\end{eqnarray}
Here $\alpha_{1}/\alpha_{2}$ are anisotropy parameters. To conduct the MC simulations, we begin with the discretized Hamiltonian (\ref{Hamiltonian_rs}) on the square lattice:
\begin{eqnarray}\label{Hamiltonian_ds}
H &=&-\kappa ( \sum_{\langle ij\rangle_x}\cos[\theta_{1}(\mathbf{r}_i)+\theta^{'}_{1}(\mathbf{r}_i)-\theta_{1}(\mathbf{r}_j)-\theta^{'}_{1}(\mathbf{r}_j)]	 
+\tau_{1}\sum_{\langle ij\rangle_y}\cos[\theta_{1}(\mathbf{r}_i)+\theta^{'}_{1}(\mathbf{r}_i)-\theta_{1}(\mathbf{r}_j)-\theta^{'}_{1}(\mathbf{r}_j)] )\nonumber\\
&-&\lambda (\sum_{\langle ij\rangle_x}\cos[\theta_{1}(\mathbf{r}_i)-\theta^{'}_{1}(\mathbf{r}_i)-\theta_{1}(\mathbf{r}_j)+\theta^{'}_{1}(\mathbf{r}_j)] 
+\tau_{2}\sum_{\langle ij\rangle_y}\cos[\theta_{1}(\mathbf{r}_i)-\theta^{'}_{1}(\mathbf{r}_i)-\theta_{1}(\mathbf{r}_j)+\theta^{'}_{1}(\mathbf{r}_j)] )\nonumber\\
&-&\gamma(\sum_{\langle ij\rangle_x}\cos[\theta_{1}(\mathbf{r}_i)-\theta_{1}(\mathbf{r}_j)]+\cos[\theta^{'}_{1}(\mathbf{r}_i)-\theta^{'}_{1}(\mathbf{r}_j)] 
+\sum_{\langle ij\rangle_y}\cos[\theta_{1}(\mathbf{r}_i)-\theta_{1}(\mathbf{r}_j)]+\cos[\theta^{'}_{1}(\mathbf{r}_i)-\theta^{'}_{1}(\mathbf{r}_j)]) \nonumber\\
&+& A\sum_{i}\cos[n\theta_{1}(\mathbf{r}_i)-n\theta^{'}_{1}(\mathbf{r}_i)].
\end{eqnarray}
Here $\langle ij\rangle$ represents nearest-neighbor bonding, and the positive coefficients $\kappa$, $\lambda$, $\tau_{1}$, $\tau_{2}$, and $\gamma$ satisfy
\begin{eqnarray}\label{relation}
\kappa &=&\frac{\rho-2\gamma}{4},~~~~~~~~~~~ \lambda = \frac{\mu-2\gamma}{4},~~~~~~ \nonumber\\
\tau_{1}&=&\frac{\rho\alpha_{1}-2\gamma}{4\kappa},~~~~~~~  \tau_{2} = \frac{\mu\alpha_{2}-2\gamma}{4\lambda}.
\end{eqnarray}
These coefficients ensure the discretized Hamiltonian~(\ref{Hamiltonian_ds}) matches the continuous Hamiltonian~(\ref{Hamiltonian_rs}) in the thermodynamic limit. The physical $\theta_{1}(\mathbf{r})$ and $\theta^{'}_{1}(\mathbf{r})$ phase fields should host only integer vortices, suggests that the $\theta$ and $\phi$ each can host integer or half-integer vortices in the x direction and y direction, respectively. This is the ``kinematics constraint'' between the $\theta$ and $\phi$ fields~\cite{Yu_Bo_Liu2023,Yu_Bo_Liu2024}. In the MC calculations, we set $\alpha_1=1.5$, $\alpha_2=1.6$, $\gamma=\frac{1}{4}\rho\mu/(\rho+\mu),~A=0.02\rho$, and slight adjustments of the parameters will not qualitatively change the structure of the phase diagram.

Considering the anisotropy along the x and y directions, we calculate the phase stiffness of the total-phase for the x-direction defined by

\begin{eqnarray}\label{ss}
	S_{x}=\frac{1}{N}(<H_{x}>-\beta<I_{x}^2>),
\end{eqnarray}
with
\begin{eqnarray}\label{Hamiltonian_p1x}
H_x &=& 4\kappa\sum_{<ij>_x}\cos[\theta_{1}(\mathbf{r}_i)+\theta^{'}_{1}(\mathbf{r}_i)-\theta_{1}(\mathbf{r}_j)-\theta^{'}_{1}(\mathbf{r}_j)]\nonumber\\
&&+\gamma\sum_{<ij>_x}\cos[\theta_{1}(\mathbf{r}_{i})-\theta_{1}(\mathbf{r}_{j})]+\cos[\theta^{'}_{1}(\mathbf{r}_{i})-\theta^{'}_{1}(\mathbf{r}_{j})],\nonumber\\
I_x &=& 2\kappa\sum_{<ij>_x}\sin[\theta_{1}(\mathbf{r}_i)+\theta^{'}_{1}(\mathbf{r}_i)-\theta_{1}(\mathbf{r}_j)-\theta^{'}_{1}(\mathbf{r}_j)]\nonumber\\
&&+\gamma\sum_{<ij>_x}\sin[\theta_{1}(\mathbf{r}_{i})-\theta_{1}(\mathbf{r}_{j})]+\sin[\theta^{'}_{1}(\mathbf{r}_{i})-\theta^{'}_{1}(\mathbf{r}_{j})],\nonumber\\
\end{eqnarray}
and for the y-direction is
\begin{eqnarray}\label{ss}
	S_{y}=\frac{1}{N}(<H_{y}>-\beta<I_{y}^2>),
\end{eqnarray}
with 
\begin{eqnarray}\label{Hamiltonian_p1y}
H_y &=& 4\kappa\tau_{1} \sum_{<ij>_y}\cos[\theta_{1}(\mathbf{r}_i)+\theta^{'}_{1}(\mathbf{r}_i)-\theta_{1}(\mathbf{r}_j)-\theta^{'}_{1}(\mathbf{r}_j)]\nonumber\\
&&+\gamma\sum_{<ij>_y}\cos[\theta_{1}(\mathbf{r}_{i})-\theta_{1}(\mathbf{r}_{j})]+\cos[\theta^{'}_{1}(\mathbf{r}_{i})-\theta^{'}_{1}(\mathbf{r}_{j})],\nonumber\\
I_y &=& 2\kappa\tau_{1} \sum_{<ij>_y}\sin[\theta_{1}(\mathbf{r}_i)+\theta^{'}_{1}(\mathbf{r}_i)-\theta_{1}(\mathbf{r}_j)-\theta^{'}_{1}(\mathbf{r}_j)]\nonumber\\
&&+\gamma\sum_{<ij>_y}\sin[\theta_{1}(\mathbf{r}_{i})-\theta_{1}(\mathbf{r}_{j})]+\sin[\theta^{'}_{1}(\mathbf{r}_{i})-\theta^{'}_{1}(\mathbf{r}_{j})].\nonumber\\
\end{eqnarray}

The $\theta$ and $\phi$ fields correlation functions along the x direction are defined as
\begin{eqnarray}
\eta_{\theta/\phi}(\Delta \mathbf{x})\equiv\frac{1}{N}\sum_{\mathbf{x}}\left\langle e^{i[\theta_1(\mathbf{x}) \pm \theta^{'}_1(\mathbf{x})-\theta_1(\mathbf{x}+\Delta \mathbf{x}) \mp\theta^{'}_1(\mathbf{x}+\Delta \mathbf{x})]}\right\rangle,  \nonumber\\
\end{eqnarray}
and the y direction correlation functions $\eta_{\theta/\phi}(\Delta \mathbf{y})$ defined similarly.

\begin{figure}[h]
	\centering
	\includegraphics[width=0.55\textwidth]{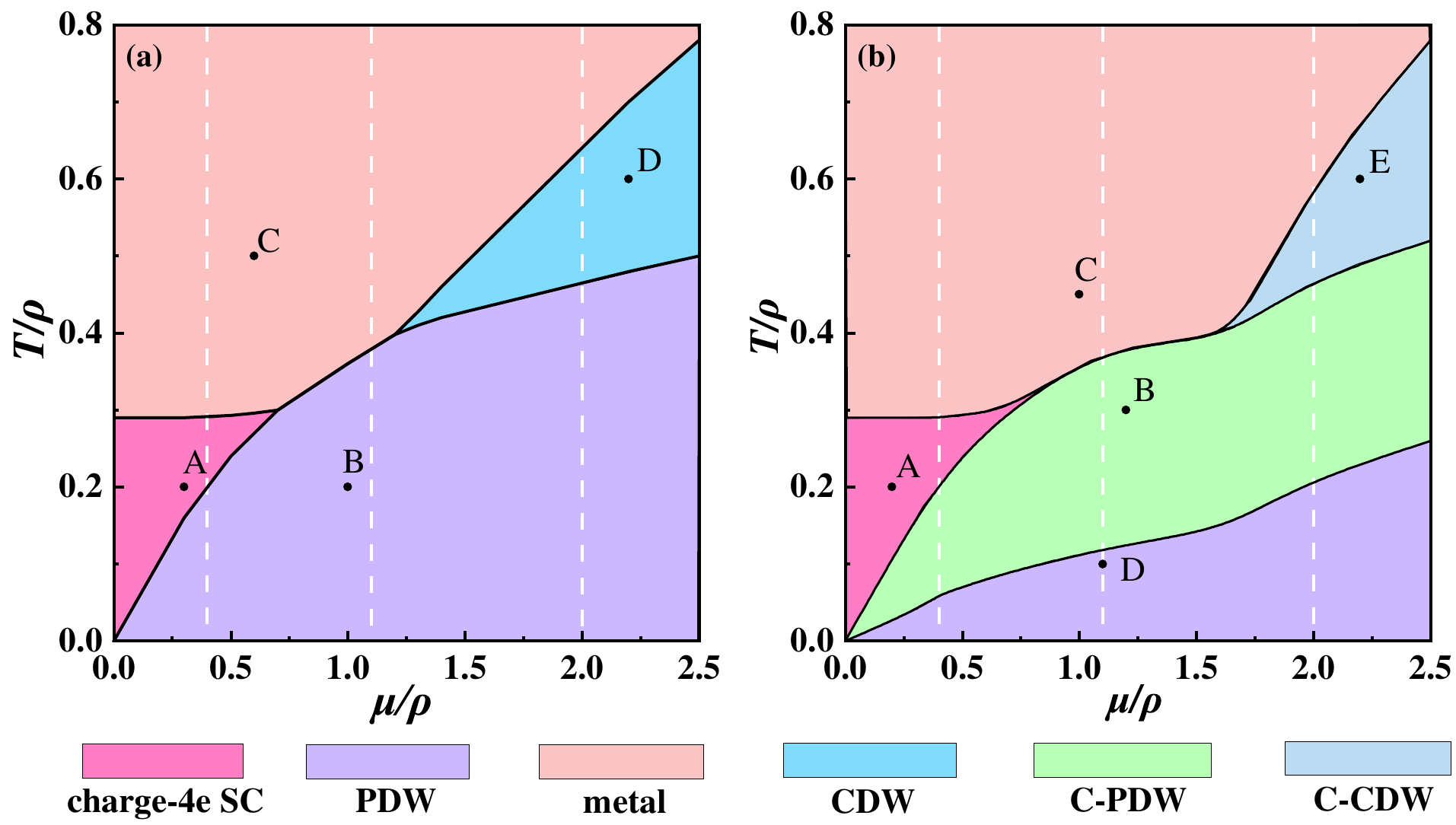}
	\caption{(Color online) The phase diagrams obtained from MC simulations for (a) $n=4$ and (b) $n=6$ (the $1Q$ PDW state) and $A=0.02\rho$ in Eq. (\ref{Hamiltonian_ds}).}\label{phase_diagram3}
\end{figure}

\subsection{$n=2$}

\begin{figure}[h]
	\centering
	\includegraphics[width=0.55\textwidth]{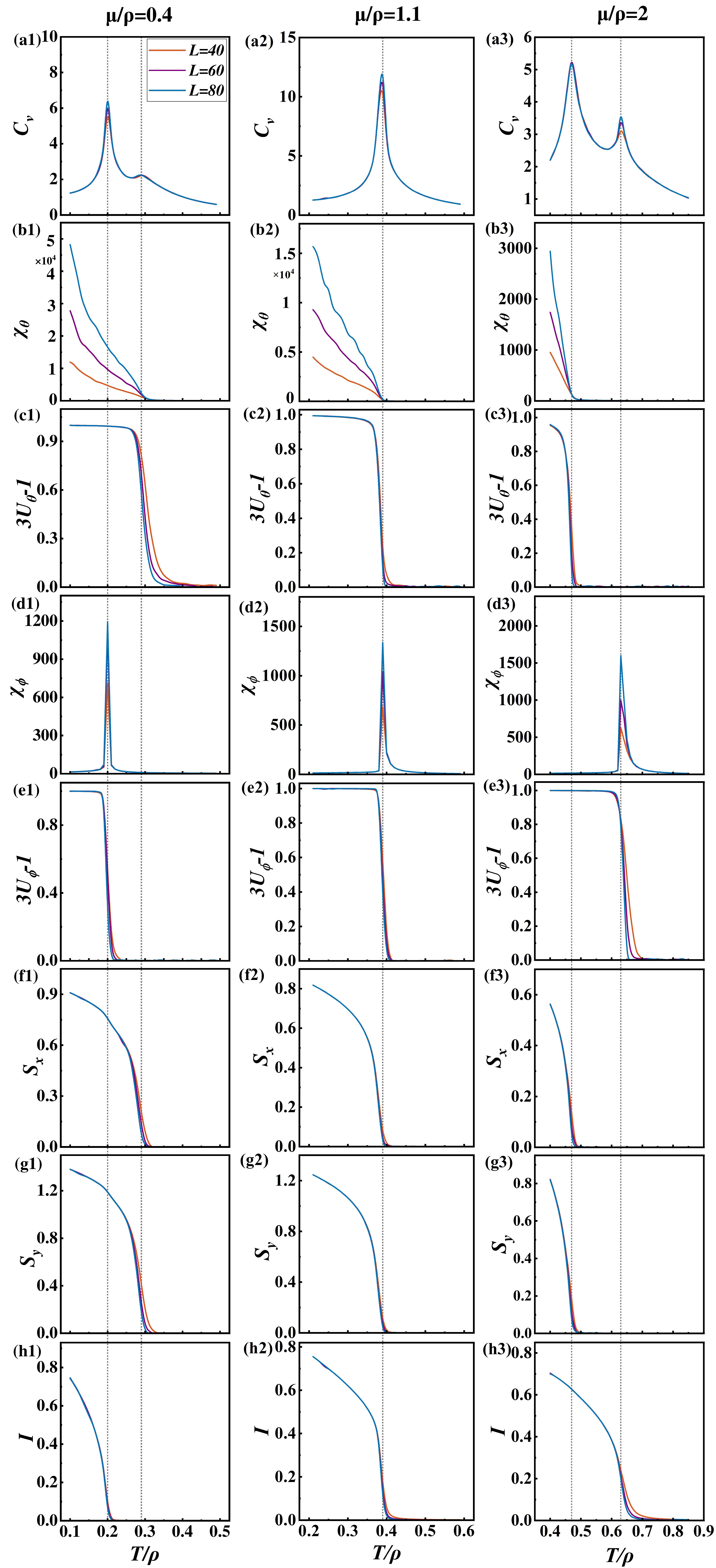}
	\caption{(Color online) The temperature dependence of thermodynamic quantities for $\mu/\rho=0.4$ (a1,b1,...,h1), $\mu/\rho=1.1$ (a2,b2,...,h2) and $\mu/\rho=2$ (a3,b3,...,h3) for $n=2$ (the $1Q$ PDW state). The scaling in all figures is $L=$ 40(chocolate line), 60(purple line), and 80(darkcyan line). (a1-a3) The specific heat $C_v$. (b1-b3) The susceptibilities $\chi_{\theta}$ of $\theta$. (c1-c3) $3U_{\theta}-1$, where $U_{\theta}$ is the Binder cumulant of the $\theta$-field. (d1-d3) The susceptibilities $\chi_{\phi}$ of $\phi$. (e1-e3) $3U_{\phi}-1$, where $U_{\phi}$ is the Binder cumulant  of the $\phi$-field. (f1-f3) The phase stiffness $S_{x}$ of $\theta$ in the x-direction. (g1-g3) The phase stiffness $S_{y}$ of $\theta$ in the y-direction. (h1-h3) Ising order parameter $\mathit{I}$.}\label{odps4}
\end{figure}

We carry out a numerical calculation of a MC study on a discretized Hamiltonian (\ref{Hamiltonian_ds})(we take $n=2$ without loss of generality).  The derived global phase diagram is displayed in Fig.~\ref{phase_diagram3}(a). In Fig.~\ref{odps4}, we show the thermodynamic quantities as functions of temperature for different lattice sizes ($L=40,~60,~80$) at $\mu/\rho=0.4,1.1$ and $2$. More detailedly, Fig.~\ref{odps4}(a1-a3) show the specific heat $C_v$, Fig.~\ref{odps4}(b1-b3) and (d1-d3) show the susceptibility $\chi_\theta$ and $\chi_{\phi}$, and Fig.~\ref{odps4}(c1-c3) and (e1-e3) show the Binder cumulant $3U_\theta-1$ and $3U_{\phi}-1$, Fig.~\ref{odps4}(f1-f3) show the phase stiffness $S_{x}$ in the x-direction, Fig.~\ref{odps4}(g1-g3) show the phase stiffness $S_{y}$ in the y-direction, Fig.~\ref{odps4}(h1-h3) show the Ising order parameter $I$, respectively. The grey dotted lines in (a1-h3) mark the phase transitions.

For $\mu/\rho=0.4$, Fig.~\ref{odps4}(a1,b1,...,h1) reveals two phase transitions. When the temperature $T/\rho$ increases to about 0.2, the specific heat $C_v$ exhibits a divergence. The susceptibility $\phi$-field is also divergence, and the cumulant $3U_{\phi}-1$ rapidly drops to zero, which suggests that the $\phi$-field experiences an Ising phase transition from long-range order to disorder at $T/\rho\approx 0.2$. The Ising order parameter $I$ characterizing the relative-phase order, which emerges at the critical temperature. The system enters the charge-4e SC since proliferating the dislocation charge. Next, when $T/\rho$ increases to about 0.29, the specific heat exhibits a small bump. The susceptibility $\theta$-field transitions form divergence to finit, and the cumulant $3U_{\theta}-1$ rapidly drops to zero, which suggesting that the $\theta$-field experiences a BKT phase transition from quasi-long-range order to disorder at $T/\rho\approx 0.29$. The system enters the normal metal (MT) phase upon this BKT transition.

For $\mu/\rho=1.1$, Fig.~\ref{odps4}(a2,b2,...,h2) reveals one phase transition. When the temperature $T/\rho$ increases to about 0.39, the specific heat $C_v$ exhibits a peak. The susceptibility $\chi _{\theta}$ rapidly drops to finite while the $\chi _{\phi}$ is divergence, the cumulant $3U_{\theta}-1$ and $3U_{\phi}-1$ rapidly drops to zero, the phase stiffness $S$ characterizing the SC in the x and y directions rapidly drops to zero. The system enters the normal MT phase from the PDW state, since proliferating the superconducting half-vortex and the dislocation half charge.

For $\mu/\rho=2$,  Fig.~\ref{odps4}(a3,b3,...,h3) reveals two phase transition. When the temperature $T/\rho$ increases to about 0.47, the specific heat $C_v$ exhibits a broad hump. The susceptibility $\chi _{\theta}$ changes from divergence to finite and the cumulant $3U_{\theta}-1$ rapidly drops to zero, which suggests that the $\theta$-field experiences a BKT phase transition from quasi-long-range order to disorder at $T/\rho\approx 0.47$. The phase stiffness $S$ characterizing the SC in the x and y directions rapidly drops to zero. The system enters the CDW phase upon this BKT transition. The CDW phase emerges since proliferating the superconducting phase vortex is favored. Next, when $T/\rho$ increases to about 0.63, the specific heat exhibits a divergence. The susceptibility $\chi _{\phi}$ is divergence and the cumulant $3U_{\phi}-1$ rapidly drops to zero, the Ising order parameter $I$ rapidly drops to zero, which suggesting that the $\phi$-field experiences an Ising phase transition from long-range order to disorder at $T/\rho\approx 0.63$. The system enters the metal upon this Ising transition.

We can determine the phase diagram Fig.~\ref{phase_diagram3}(a) based on the decaying behavior of the correlation functions $\eta_{\theta/\phi}$ in the x and y directions. The results are summarized in Table~\ref{tab:3}. The $\theta$-field and $\phi$-field  correlation functions are shown in Fig.~\ref{ABC} and Fig.~\ref{ABC4}. Fig.~\ref{ABC}(a1-a4) show that for the representative point A marked in Fig.~\ref{phase_diagram3}(a), while $\eta_{\phi_{x/y}}$ power-law decays with $\Delta x(y)$ suggesting quasi-long-range order of the $\theta$ field, $\eta_{\phi_{x/y}}$ decays exponentially with $\Delta x(y)$, suggesting disorder of the $\phi$ field. Such a phenomenon is the characteristics of the charge-4e SC phase. Fig.~\ref{ABC}(b1-b4) show that for the point D marked in Fig.~\ref{phase_diagram3}(a), while $\eta_{\phi_{x/y}}$ decays exponentially with $\Delta x(y)$ suggesting disorder of the $\theta$ field, $\eta_{\phi_{x/y}}$ saturates to a constant number for large enough $\Delta x(y)$ suggesting long-range order of the $\phi$ field, consistent with the CDW phase. The properties of the correlation function of the parameter point B and C in phase diagram Fig.~\ref{phase_diagram3}(a) is shown in Fig.~\ref{ABC4}. For the parameter point B, Fig.~\ref{ABC4}(a1-a4) show the correlation functions $\eta_{\theta_{x/y}}$ and $\eta_{{\phi}_{x/y}}$, respectively. The correlation function $\eta_{\theta_{x/y}}$ is power law decay but the correlation function $\eta_{\phi_{x/y}}$ saturates to a nonzero value when $\Delta x(y)\to \infty$, which proves that point parameter B is the PDW. For the parameter point C, Fig.~\ref{ABC4}(b1-b4) show the correlation functions $\eta_{\phi_{x/y}}$ and $\eta_{\theta_{x/y}}$, respectively. Both the correlation function $\eta_{\theta_{x/y}}$ and $\eta_{\phi_{x/y}}$ are exponentially decay, which proves that point parameter C is the metal state.

\begin{figure}[h]
	\centering
	\includegraphics[width=0.45\textwidth]{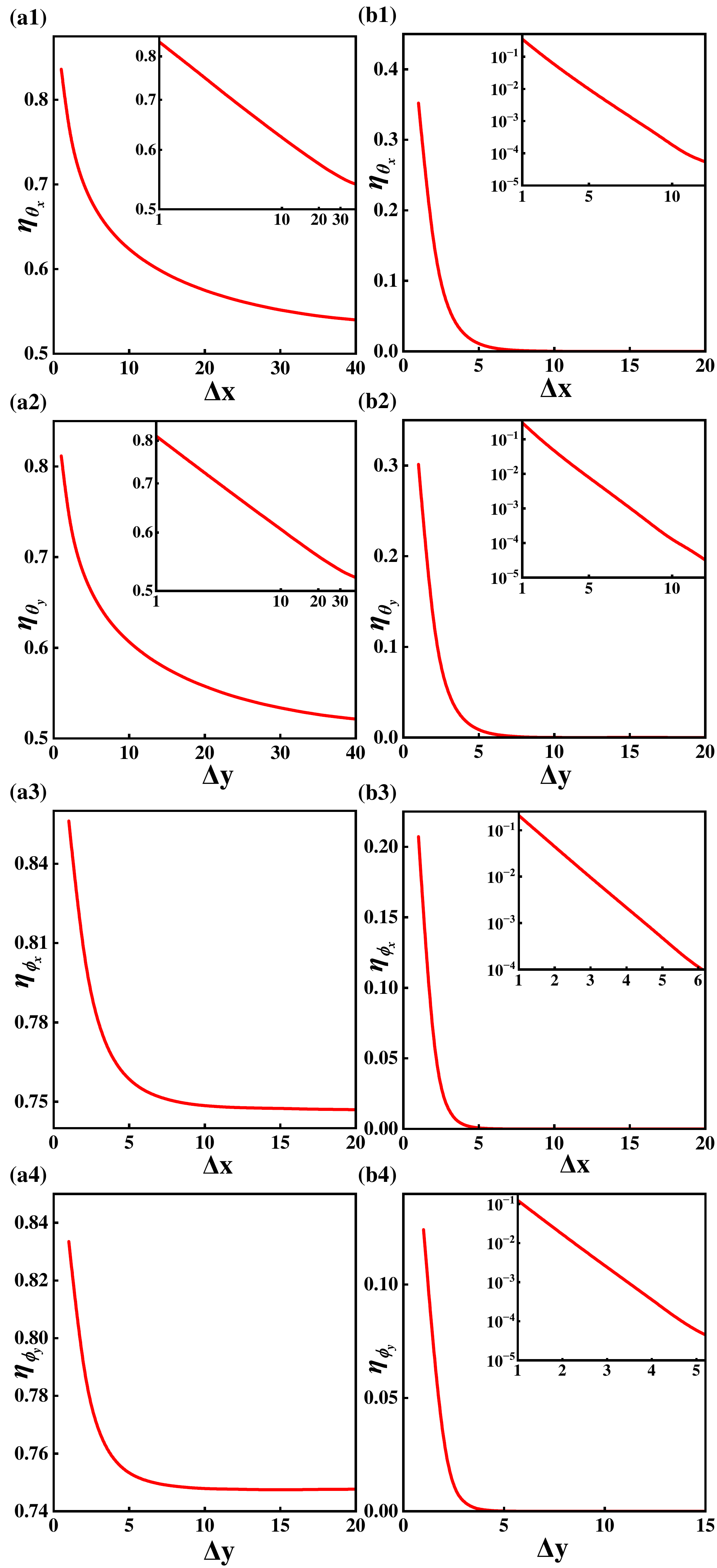}
	\caption{(Color online) The correlation function $\eta_{\theta/\phi}$ for (a1-a4) for the point $\mathbf{A}$ ($\mu=0.3\rho, T=0.2\rho$), for (b1-b4) for the point $\mathbf{D}$ ($\mu=2.2\rho, T=0.6\rho$) marked in Fig.~\ref{phase_diagram3}(a). Insets: (a1-a2) are the log-log plot, (a3-a4), (b1-b2) only the y-axis are logarithmic.}\label{ABC}
\end{figure}

\begin{figure}[h]
	\centering
	\includegraphics[width=0.55\textwidth]{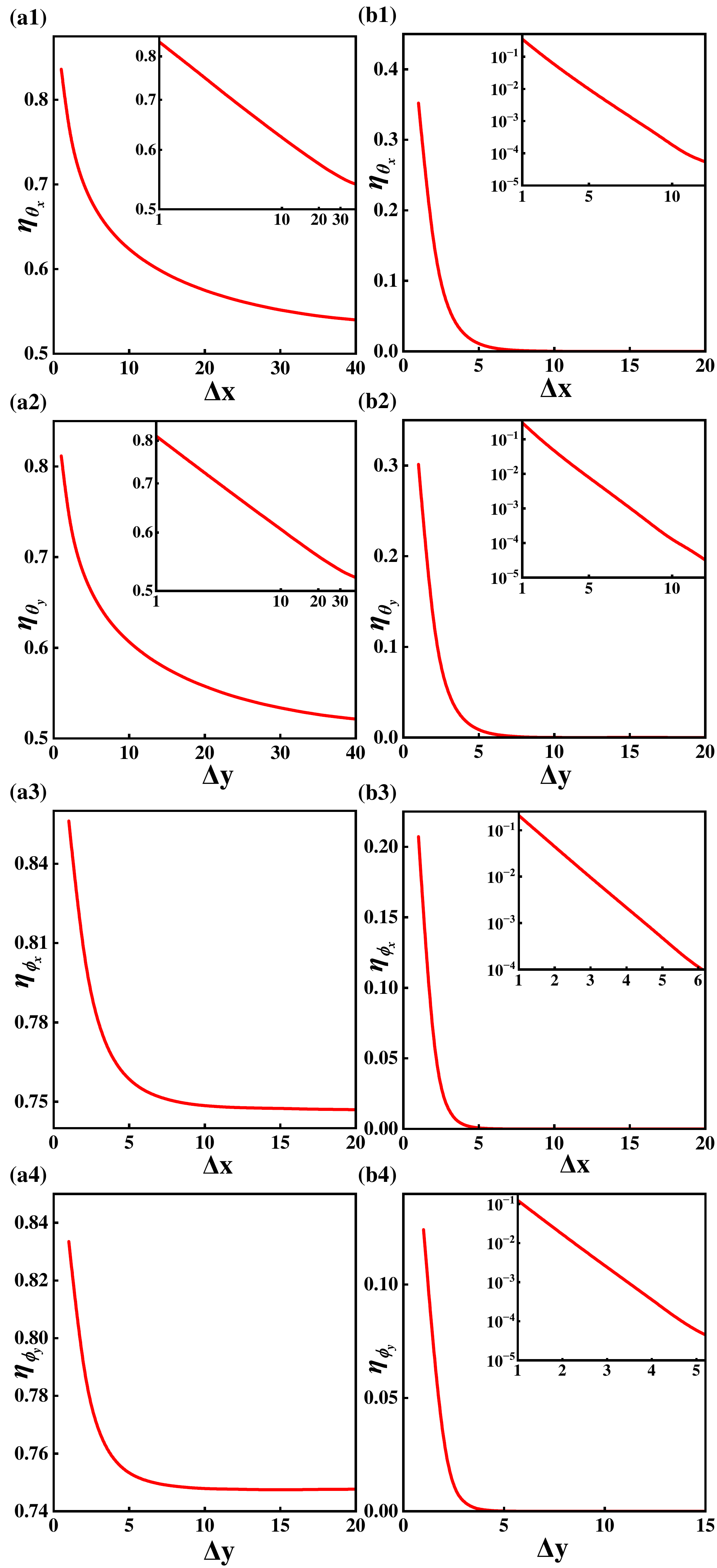}
	\caption{(Color online) The correlation function $\eta_{\theta/\phi}$ for (a) and (b) for the point $\mathbf{B}$ ($\mu=1\rho, T=0.2\rho$), for (c) and (d) for  the point $\mathbf{C}$ ($\mu=0.6\rho, T=0.5\rho$) marked in Fig.~\ref{phase_diagram3}(a). Insets: (b1-b4) only the y-axis are logarithmic, (a1-a2) are the log-log plot.}\label{ABC4}
\end{figure}

\subsection{$n=6$}

The derived global phase diagram is displayed in Fig.~\ref{phase_diagram3}(b) and the decaying behavior of the correlation functions $\eta_{\theta/\phi}$ in Table~\ref{tab:3}. However, the charge-4e SC phase is not the most interesting phase. The competition between the CDW elastic constant and the superfluid stiffness results two critical phases: the C-PDW and C-CDW. In order to gain insight into the essential physics of different phases, we perform a MC study on a discretized Hamiltonian (\ref{Hamiltonian_ds})(we take $n=6$ without loss of generality). Various $T$ dependent the above quantities on different lattice sizes are shown in Fig.~\ref{odps6} for $\mu/\rho=0.4,~1.1,~2$ marked in Fig.~\ref{phase_diagram3}(b). The grey dotted lines in (a1-g3) mark the phase transitions.

\begin{figure}[h]
	\centering
	\includegraphics[width=0.6\textwidth]{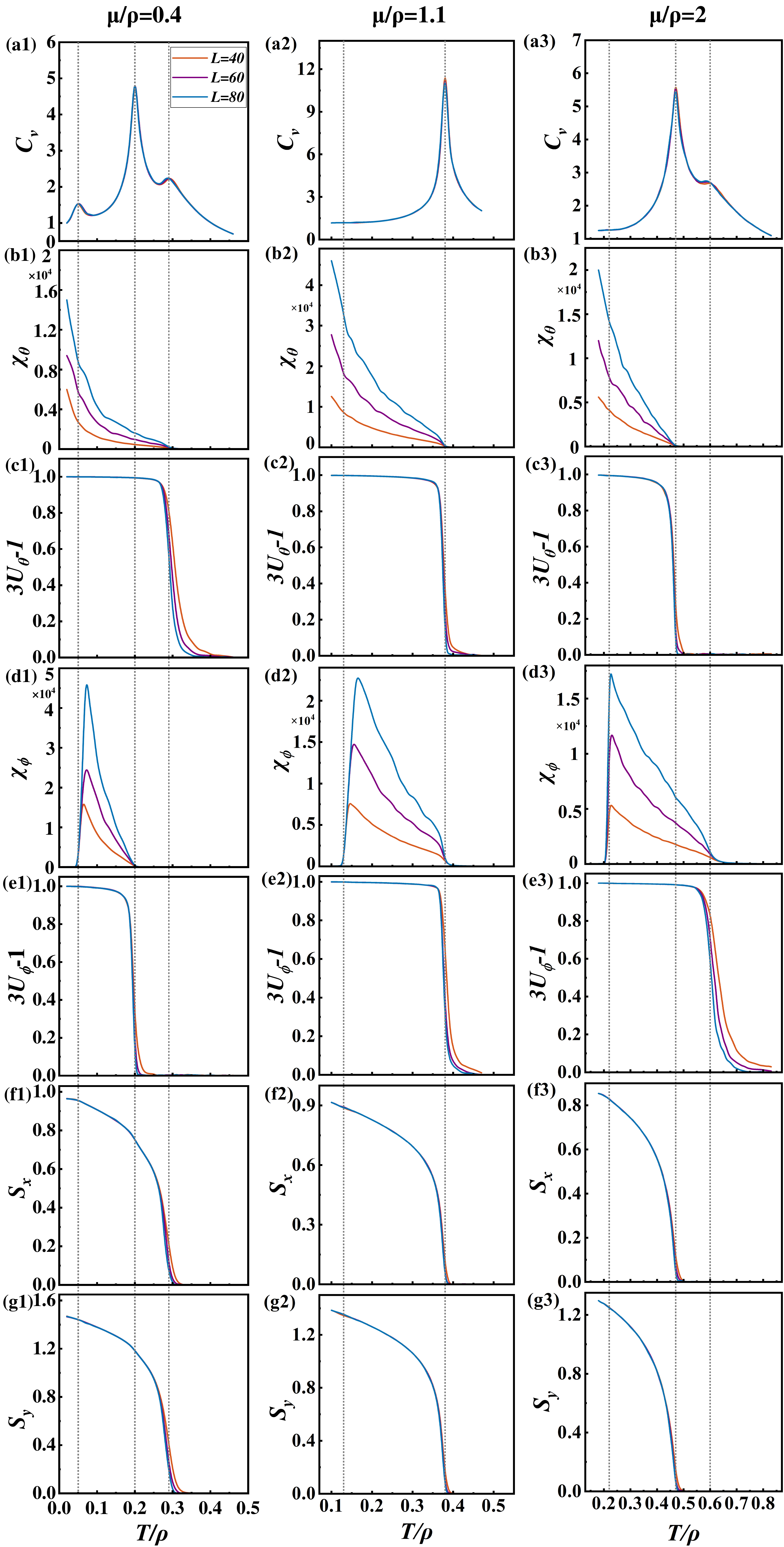}
	\caption{(Color online) The temperature dependence of thermodynamic quantities for $\mu/\rho=0.4$ (a1,b1,...,g1), $\mu/\rho=1.1$ (a2,b2,...,g2) and $\mu/\rho=2$ (a3,b3,...,g3) for $n=6$ (the $1Q$ PDW state). The scaling in all figures is $L=$ 40(chocolate line), 60(purple line), and 80(darkcyan line). (a1-a3) The specific heat $C_v$. (b1-b3) The susceptibilities $\chi_{\theta}$ of $\theta$. (c1-c3) $3U_{\theta}-1$, where $U_{\theta}$ is the Binder cumulant of the $\theta$-field. (d1-d3) The susceptibilities $\chi_{\phi}$ of $\phi$. (e1-e3) $3U_{\phi}-1$, where $U_{\phi}$ is the Binder cumulant  of the $\phi$-field. (f1-f3) The phase stiffness $S_{x}$ of $\theta$ in the x-direction. (g1-g3) The phase stiffness $S_{y}$ of $\theta$ in the y-direction.}\label{odps6}
\end{figure}

For $\mu/\rho=0.4$, the results are shown in Fig.~\ref{odps6} (a1,b1,...,g1). The four phases are separated by three phase transitions at $T/\rho\approx 0.05, 0.2, 0.29$. For $\mu/\rho=1.1$, the results are shown in Fig.~\ref{odps6} (a2,b2,...,g2). The three phases are separated by two phase transitions at $T/\rho \approx 0.12, 0.38$. For $\mu/\rho=2$, the results are shown in Fig.~\ref{odps6} (a3,b3,...,g3). The four phases are separated by three phase transitions at $T/\rho \approx 0.22, 0.47, 0.6$. 

The specific heat $C_v$, the susceptibility $\chi_{\phi}$ and the cumulant $3U_{\phi}-1$ suggest that all the transitions exhibits BKT behavior: for $C_v$ (a1-a3), it shows as broad humps or featureless at phase transition temperatures, which are insensitive to $L$; for $\chi_{\phi}$ (d1-d3), in the low-$T$ PDW phase and high-$T$ charge-4e SC and MT phases, it is finite and small and in the intermediate-$T$ critical phases (C-PDW and C-CDW), it starts to diverge in the thermodynamic limit; such a result is also reflected by the cumulant $3U_{\phi_1}-1$ (e1-e3). Here these characteristic features of the BKT transition coincide with the quasi-long-range order, corresponding to the ``quasi-broken'' translation symmetry. The susceptibility $\chi_{\theta}$ (b1-b3) changes from divergent to finite, and the cumulant $3U_{\theta}-1$ (c1-c3) rapidly drops to zero, suggesting the $\theta$-field becomes disorder. For (f1-f3) and (g1-g3), the numerical phase stiffness S characterizing the SC in the x and y directions as a function of temperature is shown. It can be seen that the phase stiffness $S_x$ and $S_y$ start to dramatically drop to zero at the same critical temperatures. These characteristic features suggest the superconducting phase is disappear.

\begin{figure}[h]
	\centering
	\includegraphics[width=0.4\textwidth]{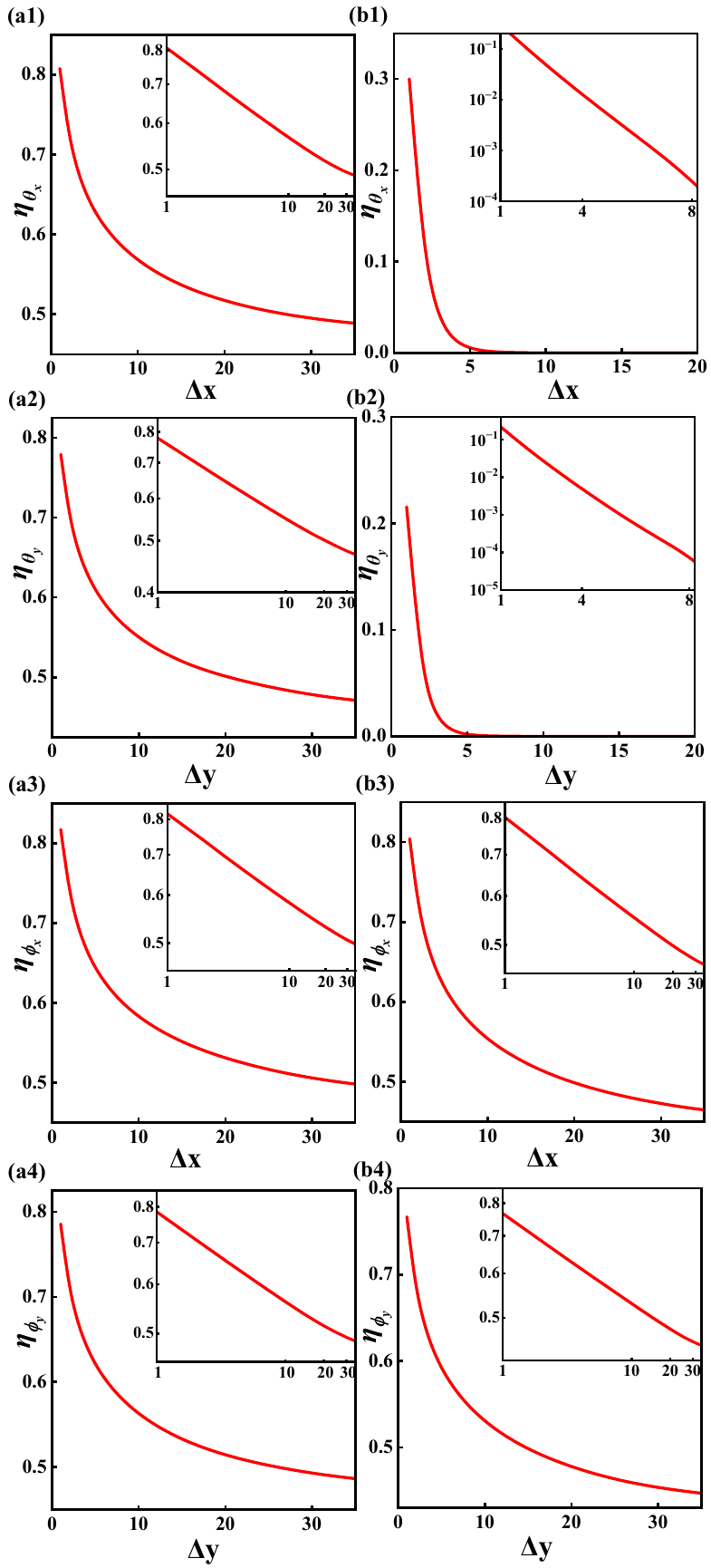}
	\caption{(Color online) The correlation function $\eta_{\theta/\phi}$ for (a1-a4) for the point $\mathbf{B}$ ($\mu=1.2\rho, T=0.3\rho$), for (b1-b4) for  the point $\mathbf{E}$ ($\mu=2.2\rho, T=0.6\rho$) marked in Fig.~\ref{phase_diagram3}(b). Insets: (a1-a4) and (b3-b4) are the log-log plot, and (b1-b2) only the y-axis are logarithmic.}\label{ABC2}
\end{figure}

When we focus on the two critical phases, the correlation functions Fig.~\ref{ABC2} (a1-a4) show both $\eta_{\theta}$ and $\eta_{\phi}$ power-law decay with $\Delta x(y)$  for the typical point B marked in Fig.~\ref{phase_diagram3}(b), reflecting the C-PDW. While for the typical point E marked in Fig.~\ref{phase_diagram3}(b), Fig.~\ref{ABC2} (b1-b4) show that $\eta_{\theta}$ decays exponentially with $\Delta x(y)$, $\eta_{\phi}$ power-law decays with $\Delta x(y)$, reflecting the C-CDW. The properties of the correlation function for the parameter point A, C and D in phase diagram Fig.~\ref{phase_diagram3}(b) is shown in Fig.~\ref{ABC3}. For the parameter point A marked in Fig.~\ref{phase_diagram3}(b), Fig.~\ref{ABC3}(a1-a4) show the correlation functions $\eta_{\theta_{x/y}}$ and $\eta_{\phi_{x/y}}$, respectively. The correlation function $\eta_{\theta_{x/y}}$ is power law decay but the correlation function $\eta_{\phi_{x/y}}$ is exponentially decay, which proves that parameter point A is the charge-4e SC. For the parameter point C marked in Fig.~\ref{phase_diagram3}(b), Fig.~\ref{ABC3}(b1-b4) show the correlation functions $\eta_{\phi_{x/y}}$ and $\eta_{\theta_{x/y}}$, respectively. Both the correlation function $\eta_{\theta_{x/y}}$ and $\eta_{\phi_{x/y}}$ are exponentially decay, which proves that point parameter C is the metal state. For the parameter point D marked in Fig.~\ref{phase_diagram3}(b), Fig.~\ref{ABC3}(c1-c4) show the correlation functions $\eta_{\theta_{x/y}}$ and $\eta_{\phi_{x/y}}$, respectively. The correlation function $\eta_{\theta_{x/y}}$ is power law decay but the correlation function $\eta_{\phi_{x/y}}$ saturates to a nonzero value when $\Delta x(y)\to \infty$, which proves that point parameter D is the PDW.

\begin{figure}[h]
	\centering
	\includegraphics[width=0.7\textwidth]{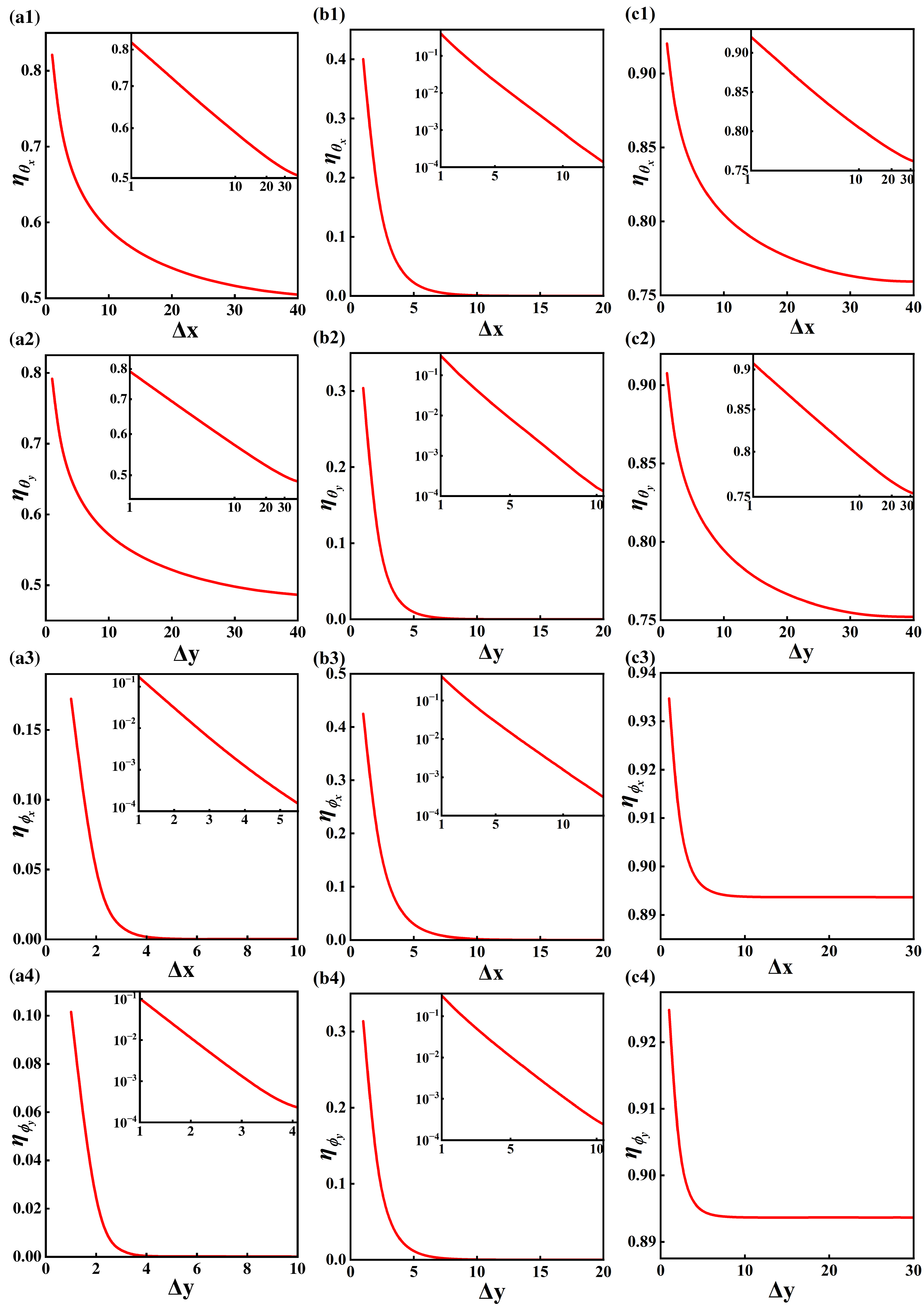}
	\caption{(Color online) The correlation function $\eta_{\theta/\phi}$ for (a1-a4) for the point $\mathbf{A}$ ($\mu=0.2\rho, T=0.2\rho $), for (b1-b4) for the point $\mathbf{C}$ ($\mu=1\rho, T=0.45\rho$), for (c1-c4) for the point $\mathbf{D}$ ($\mu=1.1\rho, T=0.1\rho$) marked in Fig.~\ref{phase_diagram3}(b). Insets: (a1-a2) and (c1-c2) are the log-log plot, (a3-a4) and (b1-b4) only the y-axis are logarithmic.}\label{ABC3}
\end{figure}

\end{widetext}
\end{document}